\documentclass{aa}  

\usepackage{xcolor}
\usepackage{graphicx}
\usepackage{txfonts}
\usepackage{natbib,twoopt}

\bibpunct{(}{)}{;}{a}{}{,} 
\makeatletter
\newcommandtwoopt{\citeads}[3][][]{\href{http://adsabs.harvard.edu/abs/#3}%
{\def\hyper@linkstart##1##2{}%
\let\hyper@linkend\@empty\citealp[#1][#2]{#3}}}
\newcommandtwoopt{\citepads}[3][][]{\href{http://adsabs.harvard.edu/abs/#3}%
{\def\hyper@linkstart##1##2{}%
\let\hyper@linkend\@empty\citep[#1][#2]{#3}}}
\newcommandtwoopt{\citetads}[3][][]{\href{http://adsabs.harvard.edu/abs/#3}%
{\def\hyper@linkstart##1##2{}%
\let\hyper@linkend\@empty\citet[#1][#2]{#3}}}
\newcommandtwoopt{\citeyearads}[3][][]%
{\href{http://adsabs.harvard.edu/abs/#3}
{\def\hyper@linkstart##1##2{}%
\let\hyper@linkend\@empty\citeyear[#1][#2]{#3}}}
\makeatother

\def\fd#1#2{{{\rm d} #1 \over {\rm d} #2}}

\newcommand{\celltspace}{\rule{0pt}{2.5ex}}
\newcommand{\cellbspace}{\rule[-1.4ex]{0pt}{0pt}}

\begin{document} 

\title{Gamma-ray emission from internal shocks in novae}
\author{P. Martin\inst{1}
            \and
            G. Dubus\inst{2}
            \and
            P. Jean\inst{1}
            \and
            V. Tatischeff\inst{3}
            \and 
            C. Dosne\inst{1,4}
            }
\institute{Univ. Paul Sabatier, CNRS, Institut de Recherche en Astrophysique et Plan\'etologie (IRAP), F-31028, Toulouse Cedex, France
              \and
              Univ. Grenoble Alpes, CNRS, Institut de Plan\'etologie et d'Astrophysique de Grenoble (IPAG), F-38000, Grenoble, France
              \and
              Centre de Sciences Nucl\'eaires et de Sciences de la Mati\`ere, IN2P3-CNRS and Univ. Paris-Sud, F-91405 Orsay Cedex, France
              \and
              Institut Sup\'erieur de l'A\'eronautique et de l'Espace (ISAE-Supaero), 10 avenue Edouard Belin, F-31055 Toulouse Cedex, France
              }
\date{Received ; accepted ; in original form \today}

\abstract
   {Gamma-ray emission at energies $\ge 100\rm\,MeV$ has been detected from nine novae using the {\em Fermi} Large Area Telescope (LAT), and it can be explained by particle acceleration at shocks in these systems. Eight out of these nine objects are classical novae in which interaction of the ejecta with a tenuous circumbinary material is not expected to generate detectable gamma-ray emission.}
   {We examine whether particle acceleration at internal shocks can account for the gamma-ray emission from these novae. The shocks result from the interaction of a fast wind radiatively-driven by nuclear burning on the white dwarf with material ejected in the initial runaway stage of the nova outburst.}
   {We present a one-dimensional model for the dynamics of a forward and reverse shock system in a nova ejecta, and for the associated time-dependent particle acceleration and high-energy gamma-ray emission. Non-thermal proton and electron spectra are calculated by solving a time-dependent transport equation for particle injection, acceleration, losses, and escape from the shock region. The predicted emission is compared to LAT observations of V407 Cyg, V1324 Sco, V959 Mon, V339 Del, V1369 Cen, and V5668 Sgr.}
   {The $\ge 100$\,MeV gamma-ray emission arises predominantly from particles accelerated up to $\sim100$\,GeV at the reverse shock and undergoing hadronic interactions in the dense cooling layer downstream of the shock. The emission rises within days after the onset of the wind, quickly reaches a maximum, and its subsequent decrease reflects mostly the time evolution of the wind properties. Comparison to gamma-ray data points to a typical scenario where an ejecta of mass $10^{-5} - 10^{-4}$\,M$_\odot$ expands in a homologous way with a maximum velocity 1000-2000\,km\,s$^{-1}$, followed within a day by a wind with mass-loss rate $10^{-4} - 10^{-3}$\,M$_\odot$\, yr$^{-1}$ and velocity $< 2000$\,km\,s$^{-1}$ and declining over a time scale of a few days. Because of the large uncertainties in the measurements, many parameters of the problem are degenerate and/or poorly constrained except for the wind velocity, the relatively low values of which result in the majority of best-fit models having gamma-ray spectra that turn down abruptly below $\sim10$\,GeV. Our typical model is able to account for the main features in the observations of the recent gamma-ray nova ASASSN-16ma.}
   {The internal shock model can account for the gamma-ray emission of the novae detected by {\em Fermi} LAT. Gamma-ray observations hold potential for probing the mechanism of mass ejection in novae, but should be combined to diagnostics of the thermal emission at lower energies to be more constraining.}


\maketitle

\section{Introduction}

Gamma-ray emission at energies $\ga 100\rm\,MeV$ has now been detected from nine novae \citep{Ackermann:2014a,Cheung:2015a,Cheung:2016a,Cheung:2016b,Li:2016a,Li:2016b}, and two candidates at the $2 \sigma$ significance level were recently found in a re-analysis of the first 7.4 years of {\em Fermi}-LAT observations using the enhanced instrument performances provided by the Pass 8 event reconstruction scheme \citep{Franckowiak:2017a}. Whereas the first detection was associated with a nova eruption in a symbiotic system (V407 Cyg), all other detections have been associated with classical binary systems. Novae eruptions are known to eject part of the white dwarf envelope at velocities $\sim 10^{3}-10^{4}\rm\,km\,s^{-1}$. In V407 Cyg, gamma-ray emission is thought to have occurred when this high-velocity material shocked the dense stellar wind from the red giant companion, accelerating a fraction of the swept-up particles to high energies by diffusive shock acceleration \citep{Abdo:2010a}. However, the companion in classical novae is a main sequence star with a very tenuous stellar wind, similar to the solar one, so, although a strong shock may be produced, no detectable gamma-ray emission is expected. Furthermore, the binaries are much tighter with orbital periods of order hours or less compared to years in the case of symbiotic systems. The rapid $\la 1\rm\,day$ expansion of the white dwarf envelope to $\sim 100\rm\,R_\odot$ engulfs the binary system in a common envelope during the eruption. Hence, interactions with the companion star also seem unlikely to explain high-energy emission on the observed timescales of weeks. The origin of the high-energy gamma-ray emission from classical novae is thus puzzling.

Although all novae models agree that mass loss is an important ingredient of the eruption, the exact sequence is not well constrained by theory. Shock ejection, envelope expansion and/or  an outflow are all possible depending upon the rate of energy deposition in the optically thick envelope \citep{Quataert:2016a}. \citet{Prialnik:1986a} found that the thermonuclear runaway at the onset of the eruption leads to the ejection of a fraction of the envelope, followed by the launch of a wind driven by radiation pressure due to continued nuclear burning at (super-)Eddington luminosities at the base of the envelope. This nova wind plays an important role in setting the timescale of the eruption by removing mass from the envelope \citep{Kato:1994a}.  The interaction of faster material with slower material ejected during various phases of the nova eruption might thus generate an internal shock liable to accelerate particles. 

Strong observational evidence exists for internal shocks, as thoroughly reviewed by \citet{Metzger:2014a}, who explored the physics of these shocks and their consequences on the emission of classical novae. The typical nova wind outflow rate is in the range $10^{-6}-10^{-3}\rm\,M_\odot\,yr^{-1}$ for a total ejected mass in the range $10^{-7}-10^{-3}\rm\,M_\odot$ \citep{Kato:1994a}. \citet{Metzger:2014a} pointed out that the resulting densities at the shocks are high enough that X-ray emission from the shock-heated material is absorbed and reprocessed to optical wavelengths, providing a novel interpretation of the thermal emission from novae. The ionization structure associated with the reprocessing impacts the thermal bremsstrahlung emission in the radio, possibly supplemented by non-thermal synchrotron emission from shock-accelerated electrons \citep{Vlasov:2016a}. High densities favor non-thermal gamma-ray emission, via $\pi^0$ production and decay in interactions of shock-accelerated protons with thermal protons \citep{Metzger:2015a}, in which case a weak neutrino signal is expected to be associated with the nova eruption \citep{Metzger:2016a}.

Here, inspired by the nova mass loss sequence described by \citet{Prialnik:1986a}, we compute the gamma-ray emission resulting from shock acceleration when a nova wind propagates through slower material ejected at the onset of the nova eruption. This assumption allows us to derive the detailed hydrodynamical evolution of the shock and to associate it with a model for particle acceleration, along the lines we set out in \citet{Martin:2013a} for the case of V407 Cyg. We are then able to compute gamma-ray light curves and spectra. The comparison to {\em Fermi}-LAT observations of V407 Cyg, V1324 Sco, V959 Mon, V339 Del, V1369 Cen, and V5668 Sgr allows us to identify the regions of parameter space that are relevant to the gamma-ray emission of classical novae\footnote{The other three novae detected by the LAT in 2016 were not included in this work because their gamma-ray emission properties were not published, and their distances were still poorly determined.}.

\section{Hydrodynamics of the interaction}
\label{hydro}

\begin{figure*}
\begin{center}
\includegraphics[width=\linewidth]{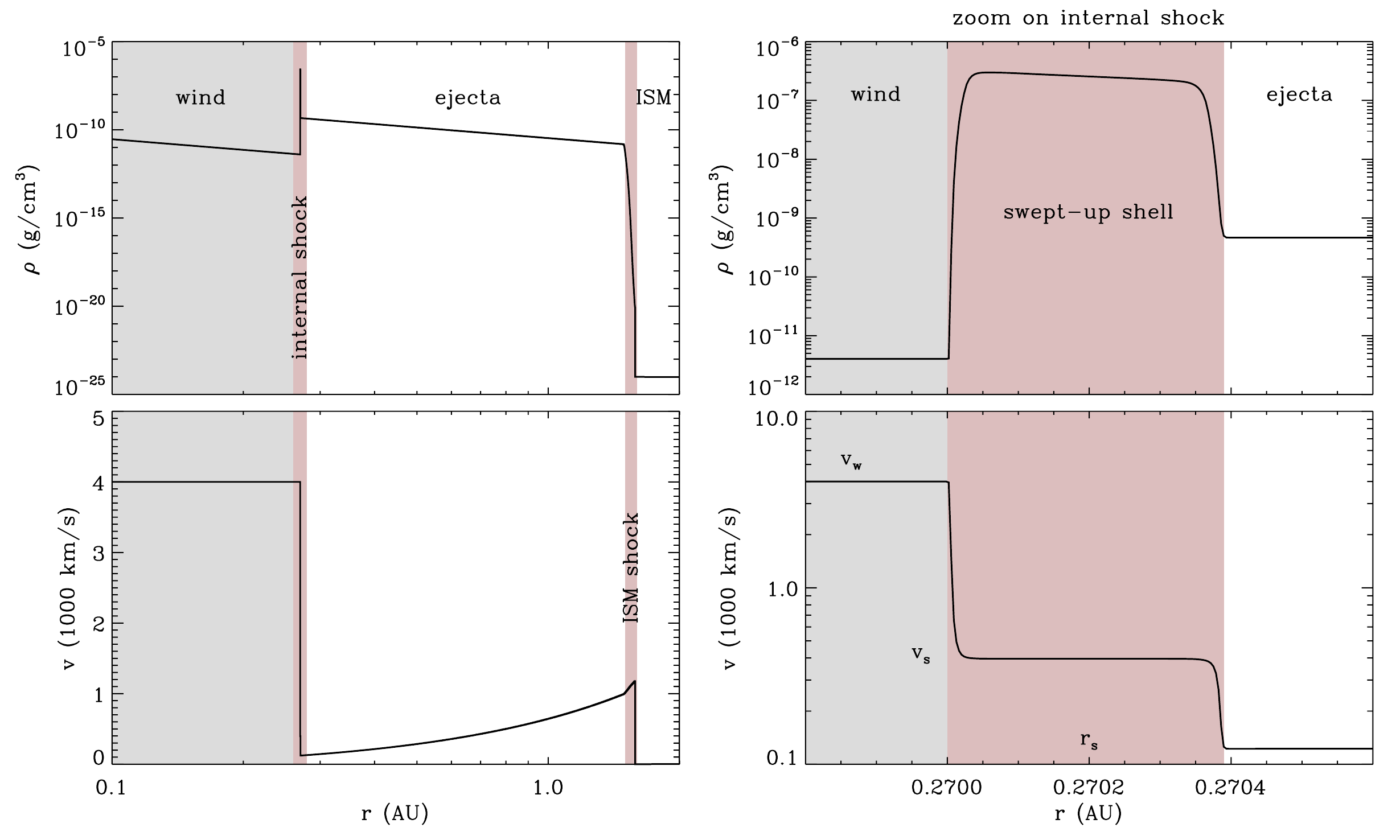} 
\caption{Structure of the interaction between a wind and a homologous ejecta with $\rho\propto r^{-2}$ from a computation using PLUTO with an isothermal equation of state. The left panels show the density and the velocity of the material. The right panels zoom in on the internal shock structure at the interface between the wind and the ejecta. Another shock structure appears at the interface between the ejecta and the ISM at $r \approx 1.6$\,AU.}
\label{fig_shock}
\end{center}
\end{figure*}

\subsection{Ejecta and nova wind}
\label{hydro_ejwind}

We assume that part of the white dwarf envelope is ejected at the onset of the thermonuclear runaway. By analogy with supernova explosions, we assume that this ejecta is in homologous expansion with a velocity  $v\propto r$ and a density distribution $\rho\propto r^{-n_{\rm ej}}$. Neglecting pressure effects (adiabatic expansion quickly cools the ejecta) and assuming spherical symmetry, the velocity and density of material in the ejecta follow from  conservation of mass and  momentum,
\begin{align}
v(r,t)={}& v_{\rm ej} \left(\frac{r -r_{\rm in}}{r_{\rm ej}+v_{\rm ej} t-r_{\rm in}}\right)\label{vej}\\
\rho(r,t) = {}& \rho_{i}\left(\frac{r}{r_{\rm ej}}\right)^{-n_{\rm ej}}\left(1+\frac{r_{\rm in}}{r}\frac{v_{\rm ej}t}{r_{\rm ej}-r_{\rm in}}\right)^{2-n_{\rm ej}}\left(1+\frac{v_{\rm ej}t}{r_{\rm ej}-r_{\rm in}}\right)^{n_{\rm ej}-3}\label{rhoej}
\end{align}
where  $r_{\rm ej}$ (resp. $r_{\rm in}$) is the outer (resp. inner) boundary of the ejected shell at $t=0$, $v_{\rm ej}$ is the velocity of the outer boundary of the ejecta, and the normalization $\rho_i$ is set by the total mass in the ejecta
\begin{equation}
M_{\rm ej}\equiv \int_{r_{\rm in}}^{r_{\rm ej}+v_{\rm ej}t}4\pi \rho(r,t) r^{2}dr
\end{equation}
In the following, we will assume that  $r_{\rm in}\ll r_{\rm ej}$. The ejecta is then characterized by four parameters: $M_{\rm ej}$, $v_{\rm ej}$, $r_{\rm ej}$ and $n_{\rm ej}$. 

The nova wind is characterized by a constant velocity $v_{\rm w}$ and an exponentially decreasing mass loss rate $\dot{M}_{\rm w}$ with typical timescale of $\sim$ weeks. We identify the timescale 
\begin{equation}
t_{\rm w}=\frac{r_{\rm ej}}{v_{\rm ej}}\approx 1.7\,\left(\frac{r_{\rm ej}}{\rm 1\,AU}\right)\left(\frac{1000\rm\,km\,s^{-1}}{v_{\rm ej}}\right)\rm\, day
\end{equation}
as the delay between the ejection of the envelope and the launch of this wind. Hence, $t=0$ in Eq.~\ref{vej}-\ref{rhoej} corresponds to the onset of the interaction between the nova wind and the homologous ejecta (neglecting the time for the wind to reach the inner boundary of the ejected envelope, since we have set  $r_{\rm in} \sim 0$).

\subsection{Shell structure of the interaction}
\label{hydro_shstruc}

The propagation of the nova wind through the ejecta results in the formation of a forward shock in the ejecta and of a reverse shock in the nova wind, separated by a contact discontinuity. \citet{Metzger:2014a} found that these shocks are highly radiative for the typical parameters under consideration, in the sense that the timescale on which shock-heated material cools due to emission is fast compared to the timescale on which the shock propagates. We confirm this is indeed the case for our setup by computing the shock  evolution in the adiabatic limit and deriving the associated radiative timescale, assuming a simple cooling function (Appendix A). We implicitly assume that radiation can indeed escape to cool the shock region. Preliminary calculations using a hydrodynamical code including radiative transfer (Dessart, priv. comm.) indicate nearly isothermal shock conditions are reached, as expected for efficient radiative cooling. In the following, we will assume that the interaction is isothermal, and that radiative transfer and ionization balance set $T\approx 10^4\rm\,K$.

The structure of the interaction region is greatly simplified in the isothermal limit (but see Sect. \ref{model_downstream}). The nova wind pushes through the ejecta a narrow shell of material composed of shocked wind and ejecta material (Fig.~\ref{fig_shock}). This shell of swept-up material propagates at a velocity $v_{\rm s}={\rm d}r_{\rm s}/{\rm d}t$. At the leading edge of the shell, the Mach number of the shock is 
\begin{equation}
{\cal M}_{\rm fs}=\frac{v_{\rm s}-v}{c_s}
\label{machfs}
\end{equation}
where $c_s\approx 10\rm\,km\,s^{-1}$ is the sound speed in the ejecta, $v\equiv v(r_{\rm s}(t),t)$ is given by Eq.~\ref{vej}, and $r_s(t)$ is the location of the shell at time $t$. Similarly, the Mach number of the shock at the trailing edge of the shell is 
\begin{equation}
{\cal M}_{\rm rs}=\frac{v_{\rm w}-v_{\rm s}}{c_s}
\end{equation}
If maximum compression of the shocked gas is achieved (but see Sect. \ref{model_gammaray}), the density behind the forward shock is $\rho_{\rm fs}={\cal M}^2_{\rm fs}\rho$ with $\rho\equiv \rho(r_{\rm s}(t),t)$ given by Eq.~\ref{rhoej}, and the density behind the reverse shock is $\rho_{\rm rs}={\cal M}^2_{\rm rs}\rho_{\rm w}$ with $\rho_{\rm w}$ defined by $\dot{M}_{\rm w}=4\pi r_{\rm s}^2 \rho_{\rm w}  v_{\rm w}$. In this case, the width of the shell $\Delta r_{\rm s}$ can be approximated as 
\begin{equation}
\Delta r_{\rm s} = \frac{M_{\rm s}}{4\pi r_{\rm s}^2}\frac{\ln\left(\rho_{\rm fs}/\rho_{\rm rs}\right)}{\rho_{\rm fs}-\rho_{\rm rs}}
\end{equation}
in the limit $\Delta r_{\rm s}\ll r_{\rm s}$, with $M_{\rm s}$ the mass of swept-up material. 

\subsection{Dynamical evolution of the shell}
\label{hydro_dynevol}

The shell dynamics are governed by conservation of momentum, pushed by the nova wind and slowed by the ejecta, neglecting gravity. The equation of motion is 
\begin{equation}
\fd{}{t}\left[M_{\rm s}v_{\rm s}\right]=\dot{M}_{\rm ej,s} v+\dot{M}_{\rm w,s}v_{\rm w}
\label{dynamics}
\end{equation}
where $\dot{M}_{\rm ej,s}$ (resp. $\dot{M}_{\rm w,s}$) is the rate at which swept-up ejecta (resp. wind) material is added to the shell and 
\begin{equation}
M_{\rm s}=M_{\rm ej,s}+M_{\rm w,s}=\int_0^t \left(\dot{M}_{\rm ej,s}+\dot{M}_{\rm w,s}\right)dt.
\end{equation}
The contribution to the shell mass  from the swept-up wind is
\begin{align}
M_{\rm w,s}= {} &\int_0^t \dot{M}_{\rm w,s}dt=\int_0^t 4\pi  r_s^2 \rho_w \left(v_{\rm w}-v_{\rm s}\right)dt \nonumber\\
 ={} &\dot{M}_{\rm w} \left(t - \frac{r_{\rm s}}{v_{\rm w}}\right)
\end{align}
and the contribution from the swept-up ejecta is 
\begin{equation} 
M_{\rm ej,s}=\int_0^{t}\dot{M}_{\rm ej, s}dt =\int_{0}^{t} 4\pi  r_{\rm s}^2 \rho  \left(v_{\rm s}-v\right)dt\leq M_{\rm ej}
\label{mejs}
\end{equation}
Eq.~\ref{dynamics} is an integro-differential equation on $r_{\rm s}$ that we solve numerically using a Runge-Kutta integration. Analytical solutions exist with additional assumptions (see Appendix B). We have verified the validity of our approach by comparing these analytical and numerical solutions to computations carried out with the hydrodynamical code PLUTO \citep{Mignone:2012a}.

We define $t_{\rm catch}$ as the time when the shell has propagated throughout the ejecta or, in other words, when $M_{\rm ej,s}=M_{\rm ej}$ in Eq.~\ref{mejs}. In Appendix~\ref{analyticalISO}, we show that $t_{\rm catch}\sim t_{\rm M}$ where  
\begin{equation}
t_{\rm M}=\frac{M_{\rm ej}v_{\rm ej}}{\dot{M}_{\rm w} v_{\rm w}}=7\,\left(\frac{M_{\rm ej}}{10^{-5}\rm\, M_\odot}\right)\left(\frac{10^{-5}\rm\,M_\odot\,week^{-1}}{\dot{M}_{\rm w}}\right)\left(\frac{v_{\rm ej}}{v_{\rm w}}\right)\,{\rm days}
\label{tM}
\end{equation}
is the timescale over which the integrated thrust of the nova wind matches the momentum of the ejecta, as could be expected by  inspecting the equation of motion (Eq.~\ref{dynamics}).

\subsection{Interaction with the external medium}
\label{hydro_intext}

We have neglected the interaction of the ejecta with the surrounding external medium. For the interstellar medium, the swept-up material is indeed negligible until the transition to the Sedov-Taylor regime at 
\begin{align}
 t_{\rm ISM}={}&\left(\frac{3 M_{\rm ej}}{4\pi \rho_a v_{\rm ej}^3}\right)^{1/3}\nonumber\\
\approx {}& 45\,\left(\frac{M_{\rm ej}}{10^{-5}\rm\,M_\odot}\right)^{1/3}\left(\frac{1\rm\,cm^{-3}}{\rho_{\rm a}}\right)^{1/3}\left(\frac{10^3\rm\,km\,s^{-1}}{v_{\rm ej}}\right)\rm\,yr,
\end{align}
which is much longer than the timescales under consideration here. 

\section{Modelling of particle acceleration and radiation}
\label{model}

\subsection{Particle acceleration}
\label{model_partacc}

Particle acceleration at the shocks is computed separately from the hydrodynamics, in the test particle approximation, i.e. assuming no retroaction of the acceleration process on the dynamics of the shocks. We used the two-zone model introduced in \citet{Martin:2013a}, which works as follows: a small fraction of particles entering the shock undergo diffusive shock acceleration (DSA) in a layer at the shock front, the acceleration region, and their evolution depends on a competition between inflow of material into the shock, advection away from it, energy gain through DSA, and energy loss mostly from radiative processes; at any time, an energy-dependent fraction of particles in the acceleration region are advected downstream of the shock, and they accumulate in a layer, the cooling region, where they can only lose energy and radiate. 

In each zone, particle distributions are computed over time by solving a transport equation in momentum space, using as input the shock trajectory (position and velocity) and environmental conditions (upstream gas density and velocity, radiation field density) computed from the equations given in Sect. \ref{hydro_dynevol}. Namely, the evolution of the non-thermal particle distribution $N(p)$ in each zone is computed through the following general equation (with subscripts A or B for the acceleration zone and cooling zone, respectively):
\begin{align}
\label{eq_evol}
\frac{\partial N_{A,B}}{\partial t}  &= \frac{\partial}{\partial p} \left( \dot{p}_{A,B} N_{A,B} \right) - \frac{N_{A,B}}{\tau_{A,B}} + Q_{A,B}
\end{align}
where $\dot{p}$ is the momentum gain/loss rate, $\tau$ the characteristic escape time, and $Q$ the source term. In the acceleration zone, these quantities have the following expressions:
\begin{align}
& \dot{p}_A =  \left( \frac{dp}{dt} \right)_{\mathrm{DSA}} + \left( \frac{dp}{dt} \right)_{\mathrm{losses}} \label{eq_coeffa_1} \\
& \tau_A =  \frac{q \, (1+q) \, D_B(p)}{V_{\mathrm{X,s}}^2} \label{eq_coeffa_2} \\
& Q_A = \frac{\eta_{\mathrm{inj}}}{\mu m_p} \frac{d M_{\mathrm{X,s}}}{dt} \, \delta (p-p_{\mathrm{inj}}) \label{eq_coeffa_3}
\end{align}
where $m_p$ and $\mu$ are the proton mass and the mean molecular mass, respectively, $q$ is the shock compression ratio, and $M_{\mathrm{X,s}} = M_{\mathrm{ej,s}} \; \mathrm{or} \; M_{\mathrm{w,s}}$ and $V_{\mathrm{X,s}} = (v_{\rm s}-v) \; \mathrm{or} \; (v_{\rm w}-v_{\rm s})$ depending on whether we are treating the forward or reverse shock. The escape time is computed as the ratio of the average cycle time by the escape probability, and involves the Bohm diffusion coefficient $D_B(p)$. The source term is a Dirac at the momentum $p_{\mathrm{inj}}$ and its amplitude is controlled by the injection fraction $\eta_{\mathrm{inj}}$. In the cooling zone, the terms in the transport equation are:
\begin{align}
\label{eq_coeffb}
& \dot{p}_B = \left( \frac{dp}{dt} \right)_{\mathrm{losses}} \\
& \tau_B =  \infty \\
& Q_B = \frac{ N_A }{ \tau_A }
\end{align}
The source term is now the escape from zone A and there is no acceleration.

Both zones are treated in the thin shell approximation and are characterized at each time by a single set of hydrodynamical quantities. In particular, the code is not suited to track the downstream evolution of non-thermal particles in a spatially resolved way, with particles of different ages experiencing different physical conditions. Instead, all particles advected downstream up to a given time are assumed to experience the same physical conditions at that time.

In the equations above, the following hypotheses were made: (1) A fraction $\eta_{\mathrm{inj}}=10^{-4}$ of the particles crossing the shock front enter the acceleration process, and they do so with a fixed momentum $p_{\mathrm{inj}}=1$\,MeV/c, irrespective of the conditions at the shock; in the comparison to the \textit{Fermi}-LAT observations, the predicted spectra and light curves will be rescaled by a renormalisation factor that is related linearly to the particle injection fraction. (2) The electron-to-proton ratio at injection is set at $10^{-2}$; as will be illustrated further down, the gamma-ray emission is largely dominated by hadronic interactions and the exact value of the electron-to-proton ratio therefore is irrelevant in that context; it is more important in the discussion of the synchrotron emission at radio wavelengths or bremsstrahlung emission in hard X-rays. (3) The spatial diffusion of non-thermal particles back and forth across the shock occurs in the Bohm limit, where the particle scatter with a mean free path equal to the gyroradius and at a rate depending on the total amplified magnetic field; hybrid simulations of ion acceleration in non-relativistic shocks showed that this is the case for shocks with Mach number ${\cal M}_{\rm s} \ga 30$ \citep{Caprioli:2014c}. With the parameters we considered for the ejecta and wind properties, this condition is always satisfied at the reverse shock, which, as we will see, dominates the gamma-ray emission. At the forward shock, however, ${\cal M}_{\rm s}$ is closer to 20 and reaches above 30 only for light ejecta and strong winds. (4) The magnetic field is amplified by particles streaming ahead of the shock, up to a strength that corresponds to a magnetic energy density of 1\% of the upstream kinetic energy density \citep[the typical value inferred for shell-type supernova remnants, see][]{Volk:2005a}. We investigated the effect of higher conversion efficiencies, up to 5\%, and found it to be minor. 

As noted in Sect. \ref{hydro_shstruc}, ejecta densities in the first days-weeks after eruption are such that shock-heated material will cool very rapidly compared to the dynamical time scale of the double-shock system. Cooling actually is so efficient that the characteristic length of the radiative relaxation layer downstream of the shock can be comparable to the characteristic diffusion length of particles being accelerated. This could affect particle shock acceleration in two ways: first, the cooled medium may become neutral and strongly damp the magnetic turbulence needed for high-energy particle scattering; second, the strong compression in the radiative relaxation layer would dramatically increase energy losses for the highest-energy particles that propagate downstream into the cooled shocked gas layer. We implemented an exponential cutoff at particle momentum $p_{\rm max}$ to ensure that, at any time, the size of the acceleration region downstream is lower than the characteristic size of radiative relaxation layer, or
\begin{equation}
\frac{(n_e+n_i) k_B T_d}{\Lambda(T_d) n_e n_i} v_d \geq \frac{D_B(p_{\rm max})}{v_d}
\label{eq_rrl}
\end{equation}
where $n_e$, $n_i$, $T_d$, and $v_d$ are the electron density, ion density, temperature, and flow velocity immediately downstream of the shock, $D_B(p_{max})$ is the Bohm diffusion coefficient at the maximum particle momentum, and $\Lambda$ is the temperature-dependent cooling function. By default, we used the cooling function of \citet{Schure:2009a} that includes thermal bremsstrahlung and atomic line transitions for a solar metallicity plasma (but see the discussion in Sect. \ref{model_downstream}). For the typical internal shock parameters explored here, the implications of this effect on the forward and reverse shocks are very different: at the latter, the maximum particle energy is reduced by a factor $\la 2$ over the first days\footnote{Interestingly, over the first couple of weeks, the criterion of Eq. \ref{eq_rrl} is comparable to the criterion suggested in \citet{Metzger:2015a} that the Larmor radius of the highest energy particles should be lower than the size of the photo-ionized layer upstream, where magnetic turbulence can exist.}, whereas at the former it is reduced by a factor $\ga 10$ over the first weeks. This is due to the $\sim10$ times higher upstream densities and $\sim10$ times lower flow velocities at the forward shock, which makes it much more efficiently radiative and for a longer time than the reverse shock. Overall, the contribution of the forward shock to the gamma-ray emission is subdominant over most of the parameter space we explored, even without the above quenching of the particle acceleration, and we will restrict ourselves to the contribution of the reverse shock in the following. Only in the case of light ejecta and powerful wind will the forward shock have a comparable contribution to the emission, because this leads to higher velocity and mass inflow at the forward shock, and exactly the opposite at the reverse shock. Such parameters are, however, not favoured by the current observations (see Sect. \ref{pred_complat}).

\begin{table}[!t]
\caption{Ejecta and wind parameters considered in the simulations}
\label{tab_novapars}
\begin{tabular}{|c|c|}
\hline
\celltspace Parameter (unit) & Values \\
\hline
\celltspace $M_{\rm ej}$ (M$_\odot$) &  $3 \times 10^{-6}$, $10^{-5}$, $3 \times 10^{-5}$, $10^{-4}$, $3 \times 10^{-4}$ \\
\hline
\celltspace $v_{\rm ej}$ (km\,s$^{-1}$) & 1000, 2000, 3000 \\
\hline
\celltspace $\dot{M}_{\rm w}$ (M$_\odot$\,yr$^{-1}$) & $10^{-4}$, $3 \times 10^{-4}$, $10^{-3}$, $3 \times 10^{-3}$ \\
\hline
\celltspace $v_{\rm w}$ (km\,s$^{-1}$) & 1000, 1500, 2000, 2500, 3000 \\
\hline
\celltspace $D_{\rm w}$ (d) & 5, 10, 20 \\
\hline
\celltspace $t_{\rm w}$ (d) & 0.5, 1.0, 2.0 \\
\hline
\end{tabular}
\end{table}

\subsection{Gamma-ray emission from accelerated particles}
\label{model_gammaray}

From the particle distributions thus obtained, gamma-ray emission is computed as a function of time. The emission is largely dominated by the population of particles accumulated in the cooling region downstream of the shock. Protons and electrons radiate respectively via neutral pion production and bremsstrahlung in the downstream gas, the density of which will increase rapidly because of the radiative nature of the shocks. As already mentioned, our code does not allow to follow the downstream evolution of accelerated particles as they move away from the shock and experience increasing densities. Instead, a single density is adopted at each time, and we discuss in the following section our assumption on this point. 

In addition to the gas-related processes mentioned above, electrons will also radiate through inverse-Compton scattering, predominantly with the nova light because the companion star is several orders of magnitude dimmer than the nova. The radiation density at any position was computed assuming a constant luminosity of $10^{38}$\,erg\,s$^{-1}$ and an effective temperature of $10^4$\,K \citep[these are typical values for the early fireball stage, see][for the case of V339 Del]{Skopal:2014a}. The evolution of the nova light over the course of weeks would affect inverse-Compton losses and the corresponding emission spectrum, but we neglected it because inverse-Compton scattering was found to be a minor contribution to the gamma-ray output of the nova (in the context of internal shocks). We also found that secondary electrons and positrons from charged pions produced in hadronic interactions have a negligible contribution at gamma-ray energies $\geq$100\,MeV. Secondary leptons have an energy distribution comparable to that of gamma-rays from neutral pions, but most of this energy is used up in synchrotron losses. We therefore neglected the contribution of secondary leptons in the present context, focused on high-energy gamma-rays, but their contribution will be relevant in the soft gamma-ray / hard X-ray range \citep{Vurm:2016a}.

Given the very high gas densities at stake in the ejecta, especially in the first days after the nova eruption ($> 10^{10-11}$\,cm$^{-3}$), gamma-ray absorption from interactions with matter cannot be neglected. We implemented the absorption of gamma rays by interaction with nuclei using the cross-section formulae in \citet{Schlickeiser:2002}, and the attenuation from Compton scattering using the cross-section formulae in \citet{Rybicki:1986}, under the approximation of a simple exponential attenuation. The effects are weak except over the first few days and for the most massive ejecta with $M_{\rm ej} \ge$ a few $10^{-4}$\,M$_\odot$. We neglected the opacity from gamma-gamma interactions because, at the maximum gamma-ray energies observed, $\sim$10\,GeV, pair creation requires target photons with energies in the soft X-ray range, much more than those produced by a nova over the first few weeks, even in the nebular phase.

\subsection{On downstream density}
\label{model_downstream}

\begin{figure}
\begin{center}
\includegraphics[width=\columnwidth]{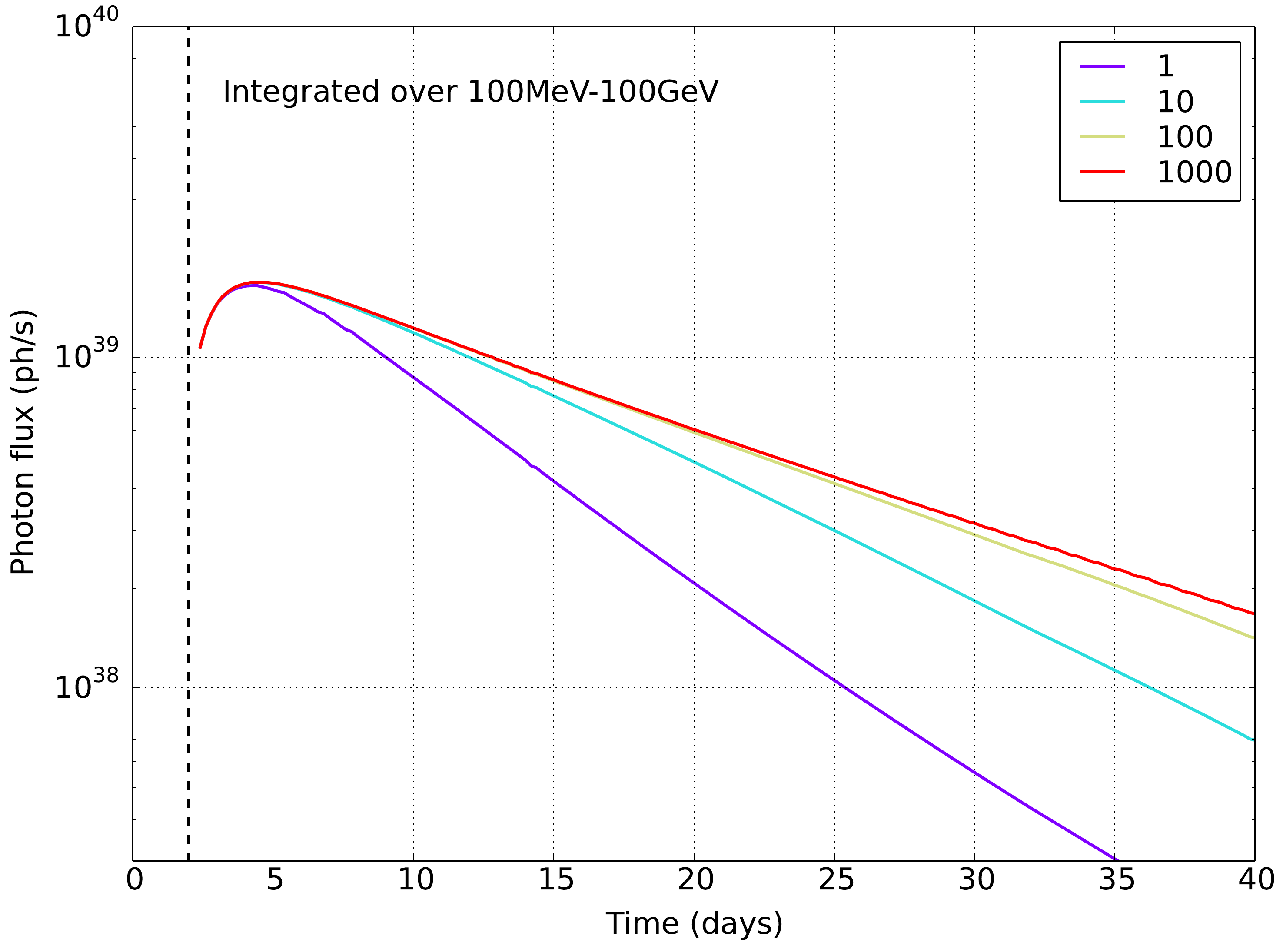} 
\caption{Photon light curves for 4 values of the gas compression factor in the cooling zone. The vertical dashed line marks the onset of the wind.}
\label{fig_cc}
\end{center}
\end{figure}

In the isothermal limit, gas can be compressed by up to a factor ${\cal M}_{\rm s}^2$ in radiative shocks, or about $\sim 10^4$ for the typical conditions we consider here. Yet, such high densities may not be reached because magnetic field and non-thermal particles may limit gas compressibility to more modest values of $\sim 10^2$. Even in the latter case, such high compression factors may have a major influence on the non-thermal energy that can be extracted and turned into radiation.

Depending on the actual thermal cooling rate (whether or not line cooling contributes and to what extent, depending on the ionization state and chemical composition of the plasma), the timescale for thermal cooling of the $\sim 10^{7-8}$\,K shocked gas is comparable to or lower than the timescale for hadronic interactions ($t_\pi \simeq 10^{15} \, \mathrm{s}/ n $, where $n$ is the gas number density in cm$^{-3}$). Over the thermal cooling time scale, relativistic particles advected away from the shock will gain energy by a factor of a few as a result of adiabatic heating (the energy gain is proportional to $n^{1/3}$). A proper treatment of non-thermal particle evolution downstream thus requires solving for the structure of the cooling layer. This was examined in \citet{Vurm:2016a}, who computed the downstream evolution of the thermal, non-thermal, and magnetic pressure components, but this is beyond the capabilities of our code. Yet, for the densities at stake in our nova internal shock scenario, immediate downstream densities of $\sim 10^{11-14}$\,cm$^{-3}$ over the first 10\,d, the timescale for hadronic interactions is of the order of minutes to hours; over a day, the typical timescale on which we want to predict gamma-ray emission properties, non-thermal protons therefore radiate most of their energy away, whatever the cooling layer structure.

In the framework of our code, we neglected the energy gained by particles as a result of adiabatic compression and adopted a density in the cooling region that is 10 times the immediate post-shock density; this provides a good approximation of proton energy losses over the first 2-3 weeks while avoiding the very small time steps that would be required by much larger densities and very short cooling time scales. Beyond $\sim 20$\,d, we underestimate the energy lost by protons in pion production. On the other hand, the radiative efficiency of the shocks diminishes with time because upstream densities decrease rapidly; for our reference set of parameters, the reverse shock transitions from radiative to adiabatic over the first few weeks, so our assumption of a modest compression factor may not be so inappropriate at such late times. In Fig. \ref{fig_cc}, we show the gamma-ray light curves obtained for various compression factors from 1 to 1000. The deviations are negligible over the first 10 days, and using a factor of 10 instead of a larger value introduces a downward bias of the order of 20\% after 20 days. We also tested the case of cooling in the average shell density, computed from the shell swept-up mass, radius, and thickness (see Sect. \ref{hydro_shstruc}), and it gives comparable results.

\section{Predicted spectra and light curves}
\label{pred_speclc}

\subsection{Grid of models}
\label{pred_grid}

The main ingredients of the simulations consist in: an ejecta with mass $M_{\rm ej}$, a power-law density profile with exponent $n_{\rm ej}$, in homologous expansion with maximum velocity $v_{\rm ej}$ and minimum velocity $v_{\rm in}$; a wind starting at time $t_{\rm w}$ after the eruption, with a constant velocity $v_{\rm w}$, and an exponentially decreasing mass-loss rate $\dot{M}_{\rm w}$ with characteristic time $D_{\rm w}$. We assessed how parameters of the internal shock scenario affect the gamma-ray output and considered the ranges of values listed in Table \ref{tab_novapars}. The complete grid of models consists in 2700 runs. Our reference set of parameters is $(M_{\rm ej},v_{\rm ej},v_{\rm in},\dot{M}_{\rm w},v_{\rm w},D_{\rm w},t_{\rm w})=(10^{-4},1000,10,10^{-3},2000,10,2)$, with the units given in Table \ref{tab_novapars}, and we discussed below variations from that reference case. A power-law radial density profile with index $n_{\rm ej}=2$ was assumed for the ejecta in all runs.

\subsection{Properties of the gamma-ray emission}
\label{pred_prop}

The principal features of the gamma-ray emission and their dependence on model parameters were assessed from the grid of models defined above.

\textit{Emission site}: As already mentioned, the emission is dominated by the reverse shock, by more than one order of magnitude over most of the parameter space. The reverse shock is able to accelerate particles up to higher energies, as could be expected from the discussion in Sect. \ref{hydro_shstruc}: the upstream flow velocity in the shock frame is much higher at the reverse shock than at the forward shock, which means particle acceleration up to higher energies because of the velocity-squared dependence of the energy gain rate through DSA (and the fact that magnetic fields are of comparable amplitude at the forward and reverse shocks).

\textit{Emission process}: The emission above 100\,MeV is dominated by pion decay. The strong inverse-Compton and synchrotron energy losses severely depletes the population of electrons and cause their spectral energy distribution to reach a maximum at a few 100\,MeV and to decrease steeply beyond that energy. As a consequence, pion decay completely dominates the gamma-ray spectrum, except in the 1-20\,MeV range, where bremsstrahlung takes over. 

\begin{figure}
\begin{center}
\includegraphics[width=\columnwidth]{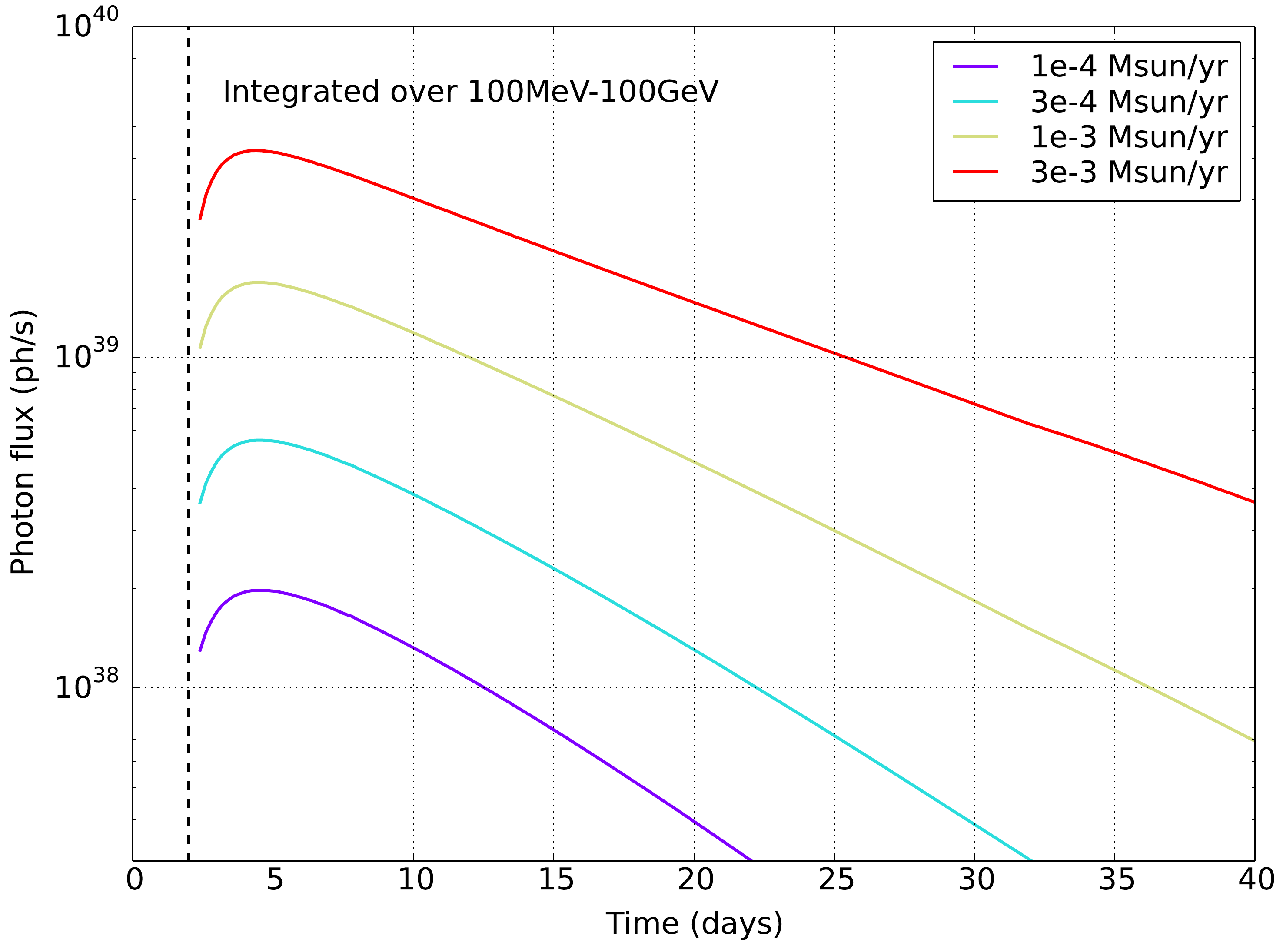}
\includegraphics[width=\columnwidth]{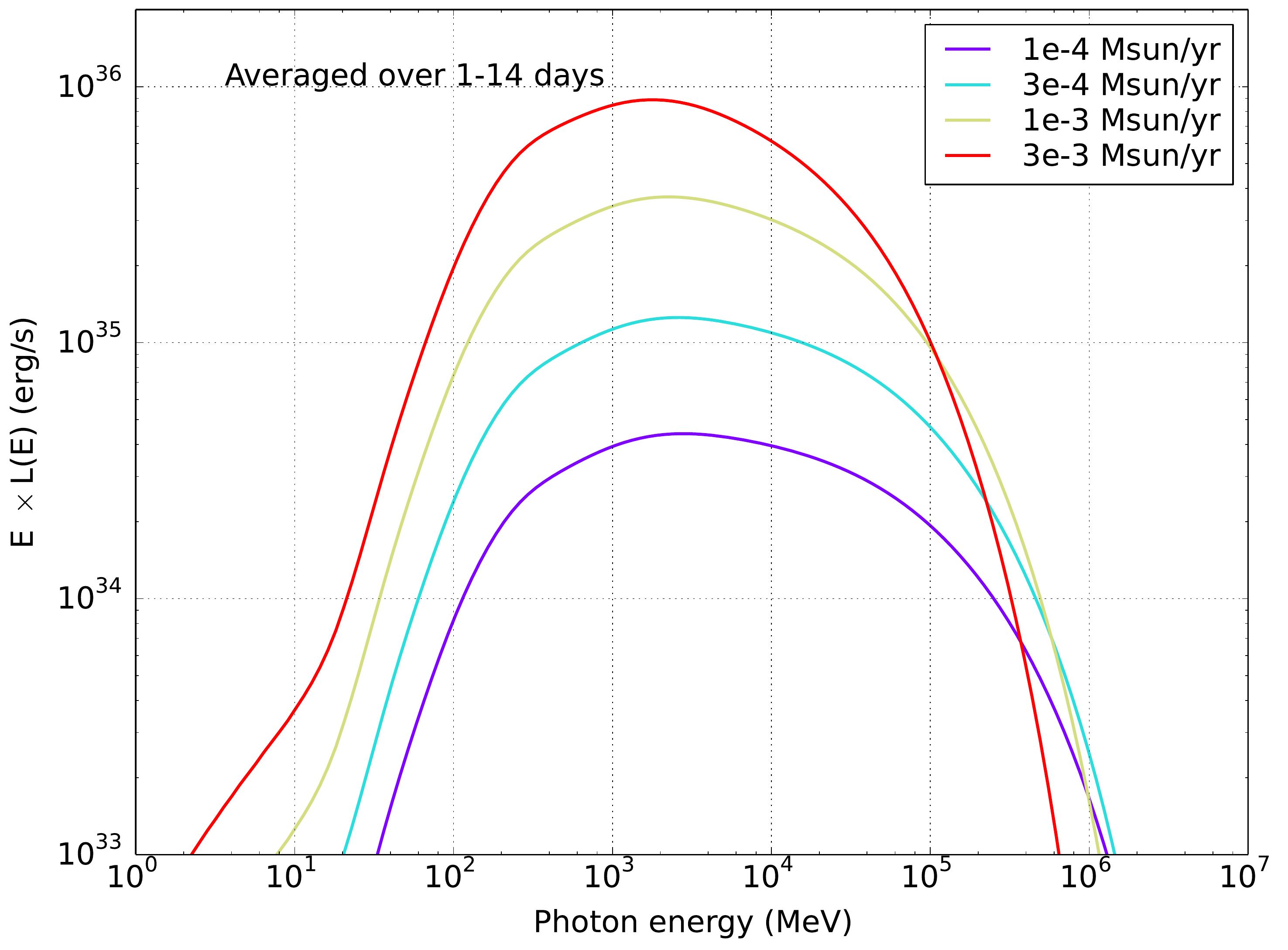}
\caption{Photon light curves and averaged spectra for 4 values of the wind mass loss rate $\dot{M}_{\rm w}$ from $10^{-4}$ to $3 \times 10^{-3}$\,M$_\odot$\,yr$^{-1}$. The vertical dashed line marks the onset of the wind.}
\label{fig_mout}
\end{center}
\end{figure}

\textit{Emission level}: The total luminosity is primarily set by the wind mass-loss rate. This is illustrated in Fig. \ref{fig_mout}, for 4 values of $\dot{M}_{\rm w}$ from $10^{-4}$ to $3 \times 10^{-3}$\,M$_\odot$\,yr$^{-1}$. The photon luminosity scales almost linearly with $\dot{M}_{\rm w}$. The latter parameter sets the rate at which particles enter the acceleration process, hence the total number of non-thermal particles radiating after a given amount of time (all other things being equal, especially the injection rate).

\begin{figure}
\begin{center}
\includegraphics[width=\columnwidth]{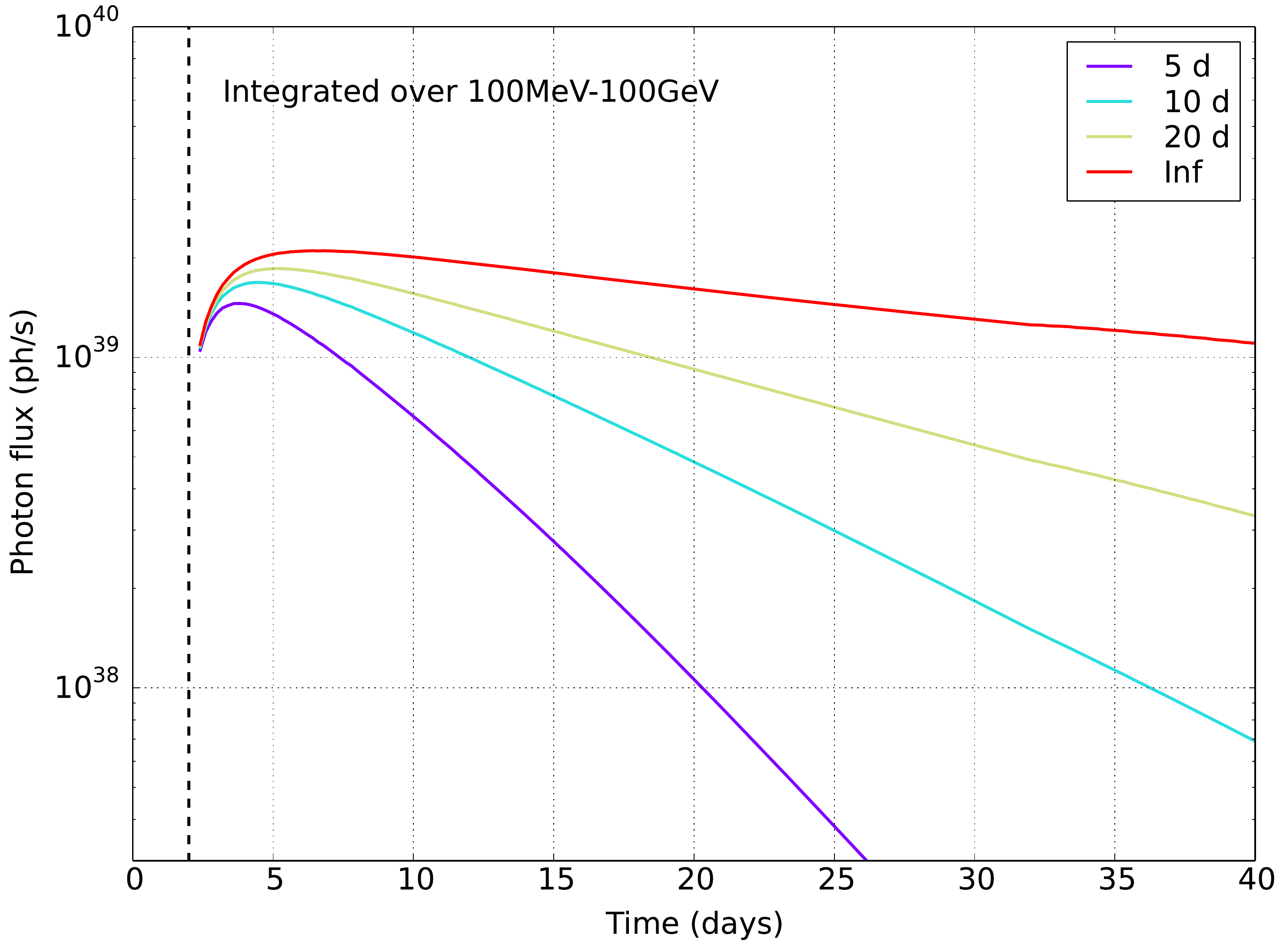}
\includegraphics[width=\columnwidth]{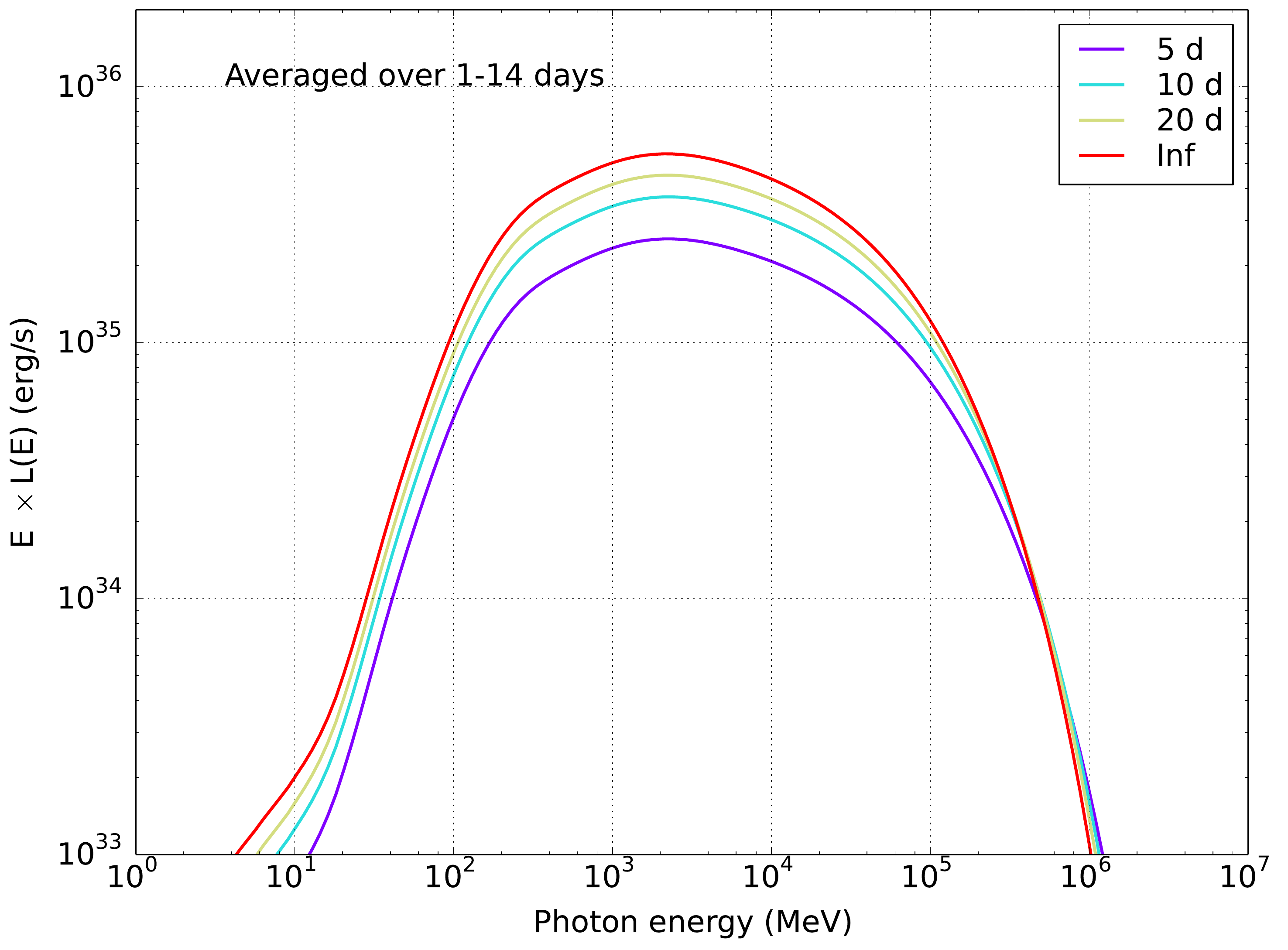}
\caption{Photon light curves and averaged spectra for 4 values of the wind characteristic duration $D_{\rm w}$ of 5, 10, 20 days and an infinite value to illustrate the case of a constant wind mass-loss rate. The vertical dashed line marks the onset of the wind.}
\label{fig_tout}
\end{center}
\end{figure}

\textit{Emission temporal profile}: The light curve profile is mostly driven by the wind properties. The emission rises after a time $t_{\rm w}$, the delay between the start of the nova eruption and the onset of the nova wind. Once started, the emission reaches a peak within days, up to about a week for the most massive ejecta $>10^{-4}$\,M$_\odot$ because it takes more time to the double shock system to move out and so gamma-ray opacity in the dense layers above the shock system prevents the radiation from escaping over a longer period. Beyond maximum, the emission decrease reflects mostly the time evolution of the wind mass-loss rate, as shown in Fig. \ref{fig_tout} for 4 values of $D_{\rm w}$: 5, 10, 20 days and an infinite value to illustrate the case of a steady wind. In theory, gamma-ray light curves could thus be a direct probe of this wind component of nova eruptions. In practice, however, as illustrated in Sect. \ref{pred_complat}, current uncertainties on the measurements are still too large to be constraining.

\begin{figure}
\begin{center}
\includegraphics[width=\columnwidth]{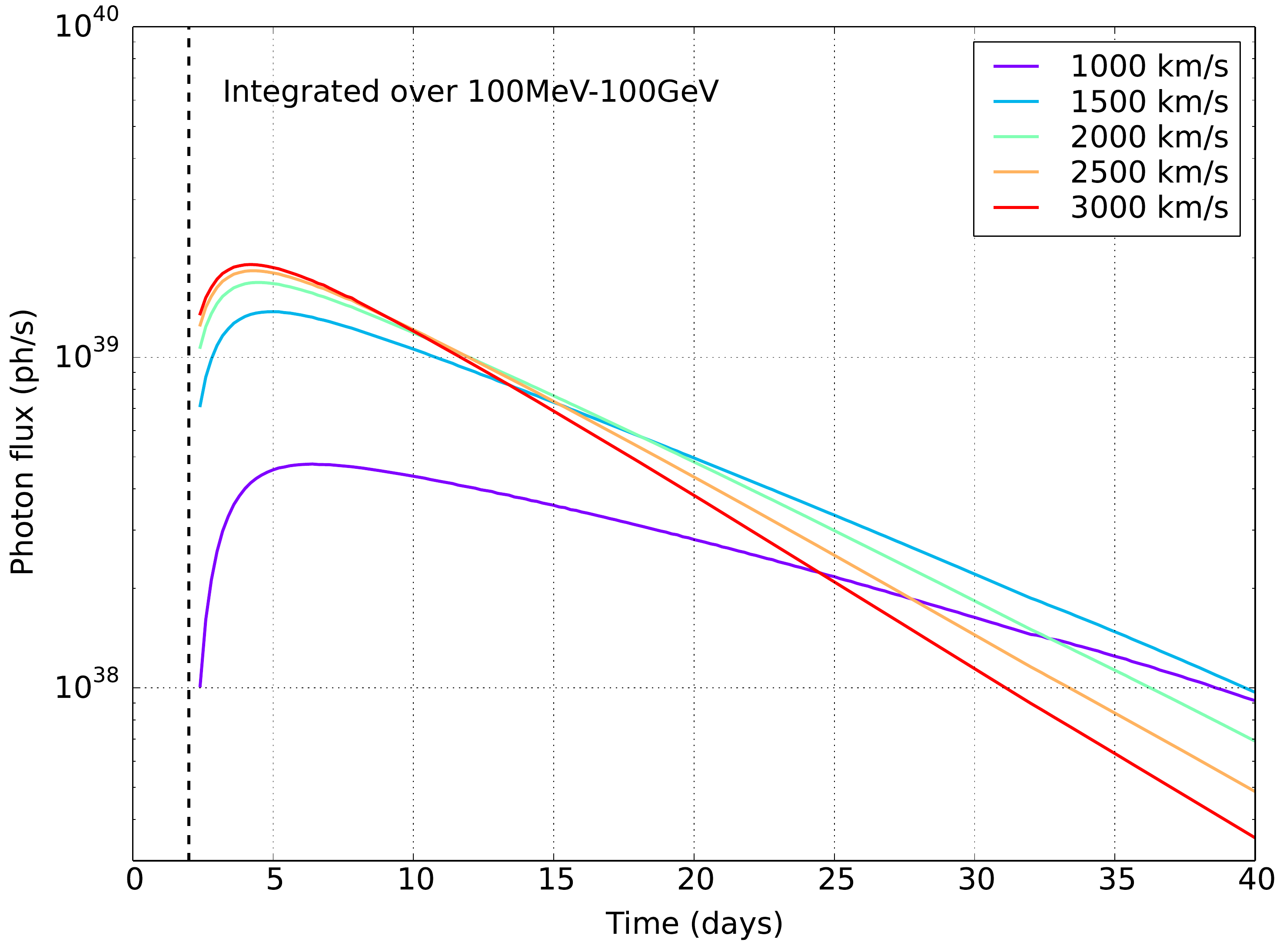}
\includegraphics[width=\columnwidth]{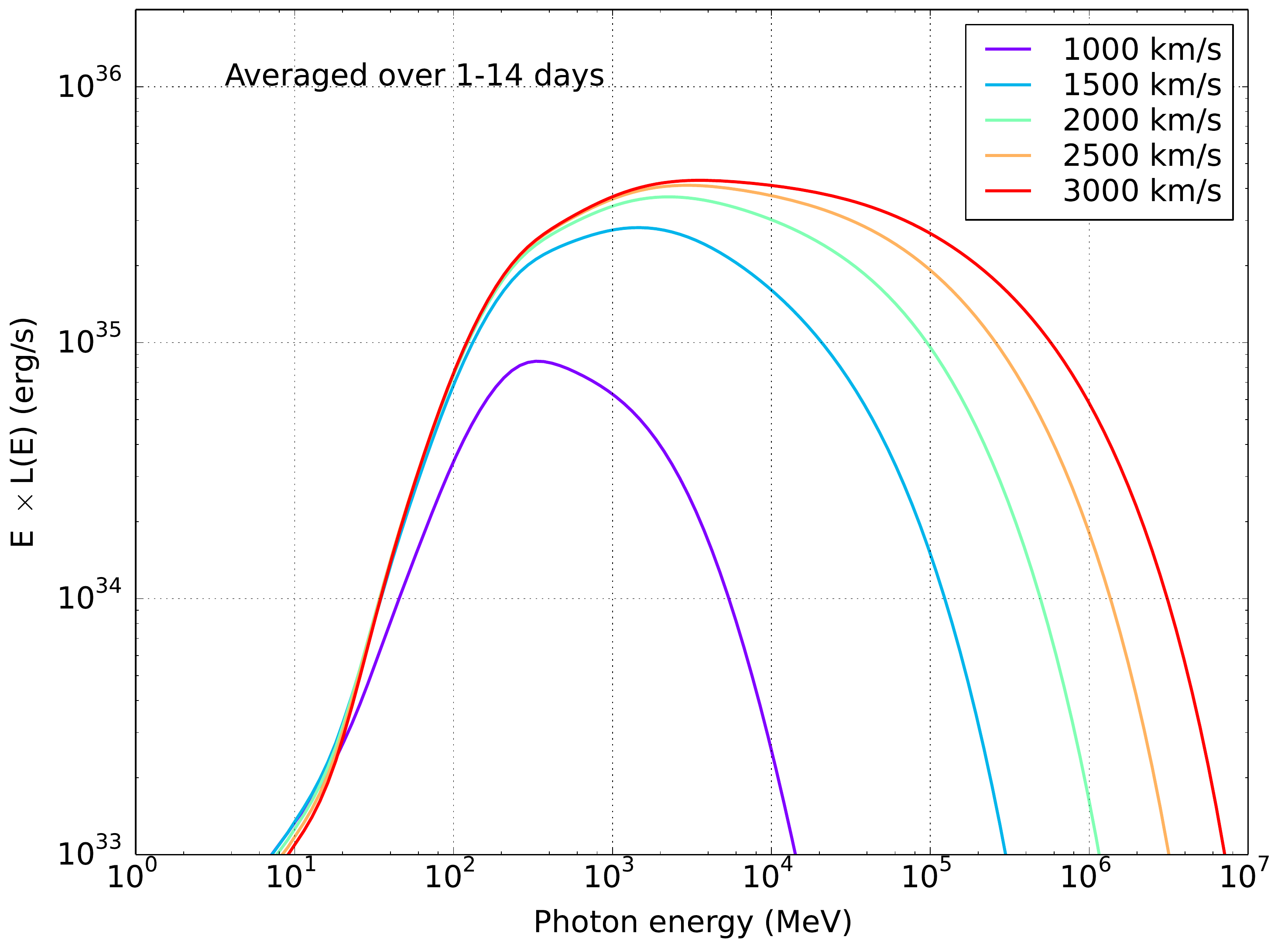}
\caption{Photon light curves and averaged spectra for 5 values of the wind velocity $v_{\rm w}$ from 1000 and 3000\,km\,s$^{-1}$. The vertical dashed line marks the onset of the wind.}
\label{fig_vout}
\end{center}
\end{figure}

\begin{figure}
\begin{center}
\includegraphics[width=\columnwidth]{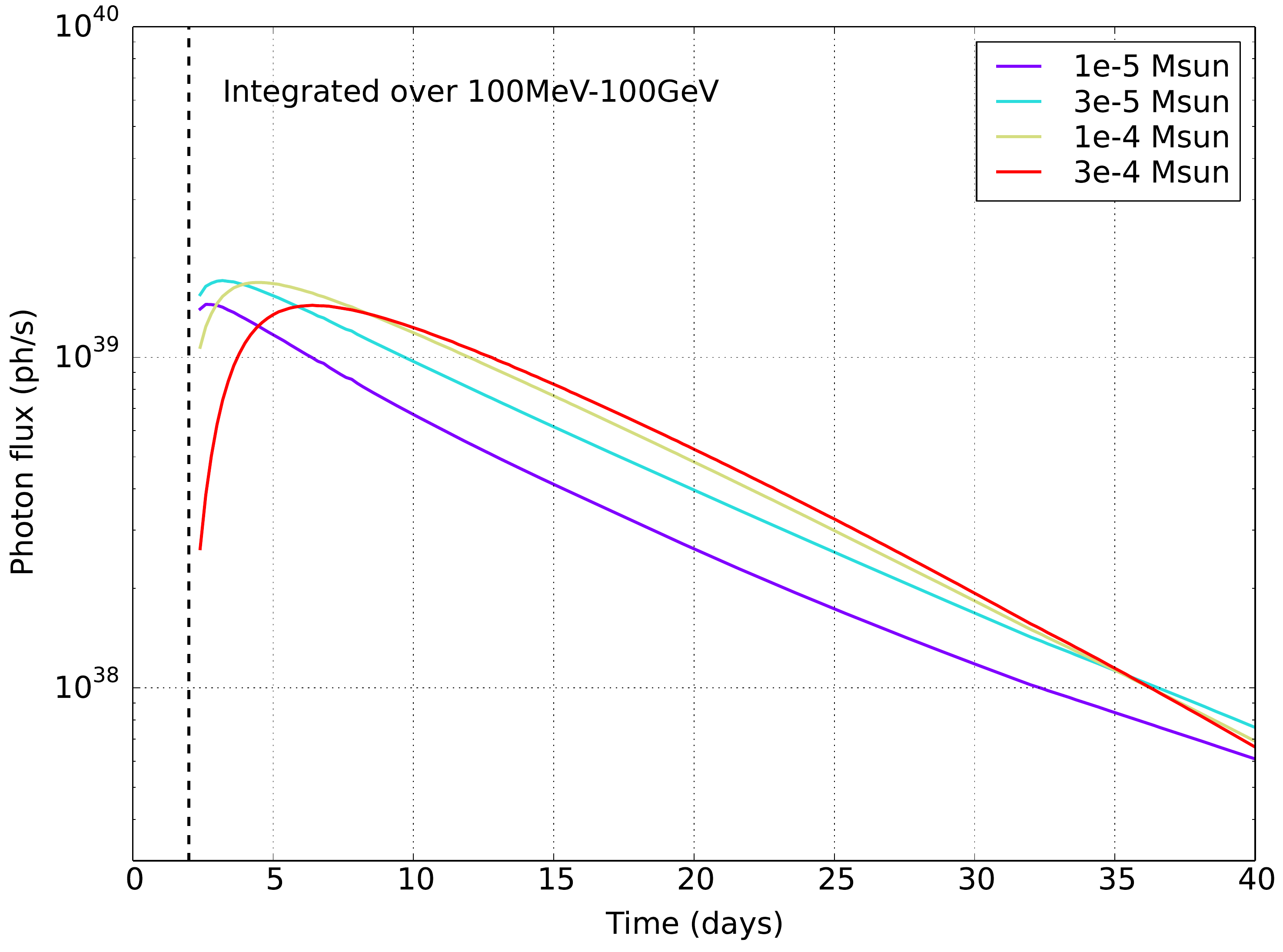}
\includegraphics[width=\columnwidth]{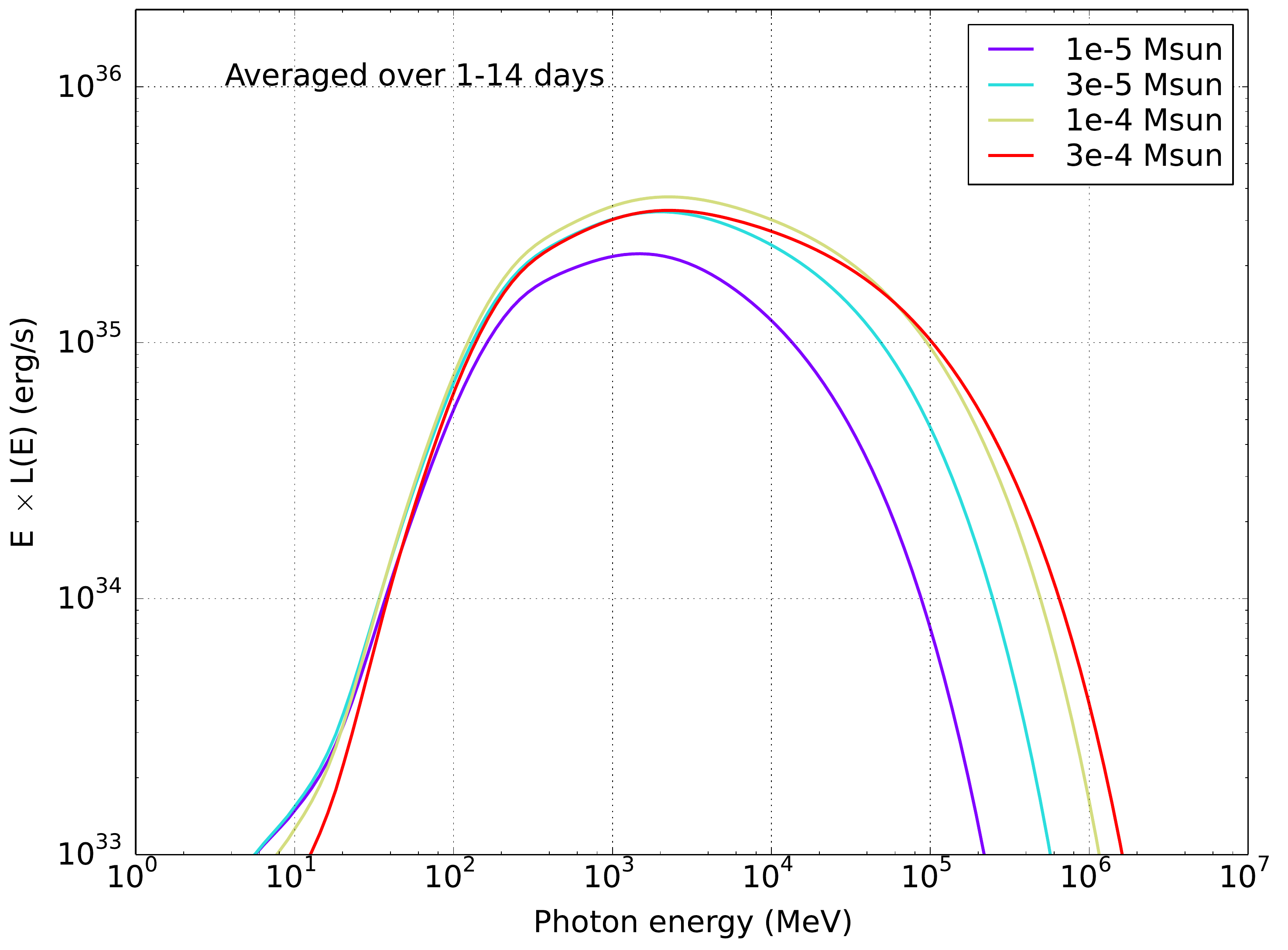}
\caption{Photon light curves and averaged spectra for 4 values of the ejecta mass $M_{\rm ej}$ from $10^{-5}$ and $3 \times 10^{-4}$\,M$_\odot$. The vertical dashed line marks the onset of the wind.}
\label{fig_mej}
\end{center}
\end{figure}

\textit{Emission spectral profile}: The spectrum of the emission is driven by the upstream flow velocity at the reverse shock. Because of the velocity-squared dependence of the energy gain rate through DSA, this has a strong impact the maximum particle (hence gamma-ray) energies that can be reached. The parameter of prime importance is the wind velocity $v_{\rm w}$, but other parameters affecting the dynamics of the double shock system, such as $\dot{M}_{\rm w}$ and $M_{\rm ej}$, will also have some influence on the gamma-ray spectrum. High values for $v_{\rm w}$ and $\dot{M}_{\rm w}$ and low values for $M_{\rm ej}$ will allow the shock system to move away from the WD more rapidly, and thus reduce the upstream flow velocity in the reverse shock frame. In contrast, $v_{\rm ej}$ and $v_{\rm in}$ have a negligible influence on the spectrum. The effects of $v_{\rm w}$ and $M_{\rm ej}$ are illustrated in Figs. \ref{fig_vout} and \ref{fig_mej}, while the effect of $\dot{M}_{\rm w}$ was presented in Fig. \ref{fig_mout}. Wind velocities $\ge 3000$\,km\,s$^{-1}$ result in gamma-ray emissions reaching more than 100\,GeV but, as we will see below, such high velocities are not favoured by the current \textit{Fermi}-LAT observations.

The parameters discussed above are those with the most noticeable impact on the predicted gamma-ray signal. We assessed the effects of several other quantities or assumptions and found differences of the order of $\la 10$\% over the first two weeks after nova eruption, which would thus be undetectable given the uncertainties in the \textit{Fermi}-LAT observations. In particular, we tested values from 0 to 2 for the density index $n_{ej}$; lower values result in a slightly flatter light curve and a spectrum reaching slightly lower energies, because the lower densities in the inner ejecta allow the shock system to move out more rapidly (hence lower upstream flow velocities in the reverse shock frame) and travel across denser material at late times (hence more particles being accelerated). We also compared the cases of continuum cooling and line cooling for the cooling function $\Lambda$ that sets the size of the radiative relaxation layer (see Eq. \ref{eq_rrl}; we refer the reader to \citet{Metzger:2014a} for a discussion on the uncertain ionization state of the cooling plasma in the presence of potentially strong photo-ionization); the more efficient line cooling results in a spectrum reaching slightly lower energies, because the smaller radiative relaxation layer imposes a smaller acceleration region, hence lower maximum particle energies $p_{\rm max}$, but the difference remains limited to a factor of a few because the criterion of Eq. \ref{eq_rrl} is naturally fulfilled after about a week for most conditions, allowing $p_{\rm max}$ to increase unconstrained afterwards.

\begin{figure}
\begin{center}
\includegraphics[width=\columnwidth]{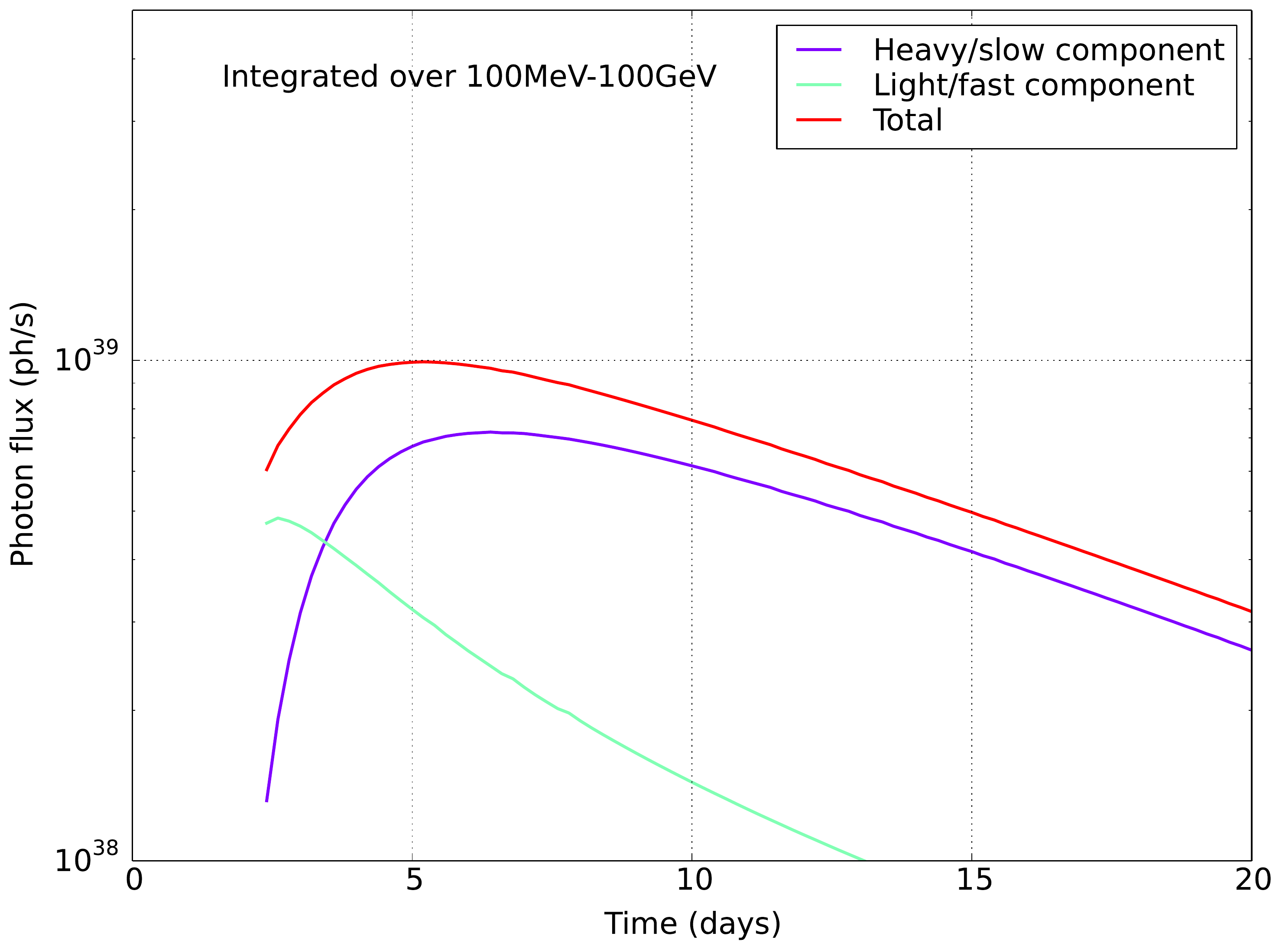}
\caption{Photon light curves for a non-spherical case, with heavy and slow ejecta in the equatorial plane and light and fast ejecta in the polar directions (obtained by linearly combining two runs and neglecting interactions at the boundaries; see text).}
\label{fig_combi}
\end{center}
\end{figure}

We also evaluated the impact of a departure from spherical symmetry such as, for instance, ejecta with different masses and velocities in the equatorial plane and polar directions \citep[as inferred for V959 Mon, see][]{Chomiuk:2014a}. We mimicked this scenario by a simple linear combination of two runs. We assumed that half of the solid angle consists in heavy and slow ejecta with $(M_{\rm ej},v_{\rm ej})=(3 \times 10^{-4}, 1000)$, while the other half consists in light and fast ejecta with $(M_{\rm ej},v_{\rm ej})=(10^{-5}, 3000)$ (we neglect interactions at the boundaries between the two regions); both components are then hit by a wind starting 2\,d after the eruption, with velocity 2000\,km\,s$^{-1}$ and mass-loss rate $10^{-4}$\,M$_\odot$\,yr$^{-1}$. The light curve thus obtained is plotted in Fig. \ref{fig_combi}. Combining the components results in some flattening of the light curve, and the plot suggests that multiple peaks may be produced for specific combinations of parameters. This could explain the double peak in the light curve of the bright nova ASASSN-16ma detected by the LAT last year \citep{Li:2016a}.
 
\subsection{Comparison to \textit{Fermi}-LAT observations}
\label{pred_complat}

\begin{table*}[!t]
\begin{minipage}[][7cm][c]{\textwidth}
\begin{center}
\caption{Novae assumed distances and best-fit eruption parameters}
\label{tab_novafit}
\begin{tabular}{|c|c|c|c|c|c|c|c|c|c|}
\hline
\celltspace Nova & Distance & $M_{\rm ej}$ & $v_{\rm ej}$ & $\dot{M}_{\rm w}$ & $v_{\rm w}$ & $D_{\rm w}$ & $t_{\rm w}$ & $f$ & $\chi^2_r$ \cellbspace \\
\hline
\celltspace V407 Cyg & 2.7$^a$ & $3 \times 10^{-5}$ & $3000$ & $10^{-4}$ & $1000$ & $10$ & $0.5$ & 6.25 & 0.73 \cellbspace \\
\hline
\celltspace V1324 Sco & 6.5$^b$ & $10^{-4}$ & $2000$ & $3 \times 10^{-4}$ & $1000$ & $5$ & $0.5$ & 15.6 & 0.71 \cellbspace \\
\hline
\celltspace V959 Mon & 1.4$^c$ & $3 \times 10^{-5}$ & $3000$ & $3 \times 10^{-4}$ & $1000$ & $5$ & $1.0$ & 1.33 & 0.82 \cellbspace \\
\hline
\celltspace V339 Del & 4.5$^d$ & $3 \times 10^{-5}$ & $1000$ & $3 \times 10^{-3}$ & $1500$ & $20$ & $2.0$ & 0.52 & 0.50 \cellbspace \\
\hline
\celltspace V1369 Cen & 2.5$^e$ & $10^{-4}$ & $2000$ & $10^{-3}$ & $1000$ & $10$ & $2.0$ & 0.78 & 0.45 \cellbspace \\
\hline
\celltspace V5668 Sgr & 2.0$^f$ & $10^{-4}$ & $2000$ & $10^{-3}$ & $1000$ & $20$ & $1.0$ & 0.18 & 0.32 \cellbspace \\
\hline
\end{tabular}
\end{center}
Distances are given in kpc and the references for them are: $(a)$ \citet{Munari:1990a}; $(b)$ \citet{Finzell:2015a}; $(c)$ \citet{Linford:2015a}; $(d)$ \citet{Schaefer:2014a}; $(e)$ \citet{Shore:2014a}; $(f)$ \citet{Banerjee:2016a}. Columns 3 to 8 are the best-fit parameters, with the units given in Table \ref{tab_novapars}. Columns 9 and 10 are the renormalization factors $f$ in the fit to the data and the reduced chi-square of the fit, respectively. 
\end{minipage}
\end{table*}

We have compared the \textit{Fermi}-LAT observed spectra and light curves for the six novae presented in \citet{Ackermann:2014a} and \citet{Cheung:2016a} to our grid of 2700 models. For each model, an averaged spectrum and a photon light curve were jointly fitted to the data using a $\chi^2$ criterion. The only fit parameter is a renormalization factor $f$. This factor $f$ is related linearly to the particle injection fraction and can therefore be interpreted as a correction to the initially assumed value of $10^{-4}$ (see Sect. \ref{model_partacc}). For each nova, we identify the best-fit set of parameters as well as a $1\sigma$ uncertainty domain. The latter is defined as the ensemble of model fits that are in the 68.3\% confidence interval, i.e. fits yielding $\chi^2 \leq \chi^2_{\rm opt}+8.18$ for 7 degrees of freedom, where $\chi^2_{\rm opt}$ is the best-fit criterion\footnote{Note that our $1\sigma$ uncertainty domain is underestimated because it hits the boundaries of our grid our models for some parameters.}. We excluded from this range all models that, after renormalization, yielded non-thermal efficiencies $>10$\%, to be consistent with our initial test particle approximation (see Sect. \ref{model_partacc}; in practice, only a small number of fits were filtered out by this criterion).

\begin{figure}[h]
\begin{center}
\includegraphics[width=\columnwidth]{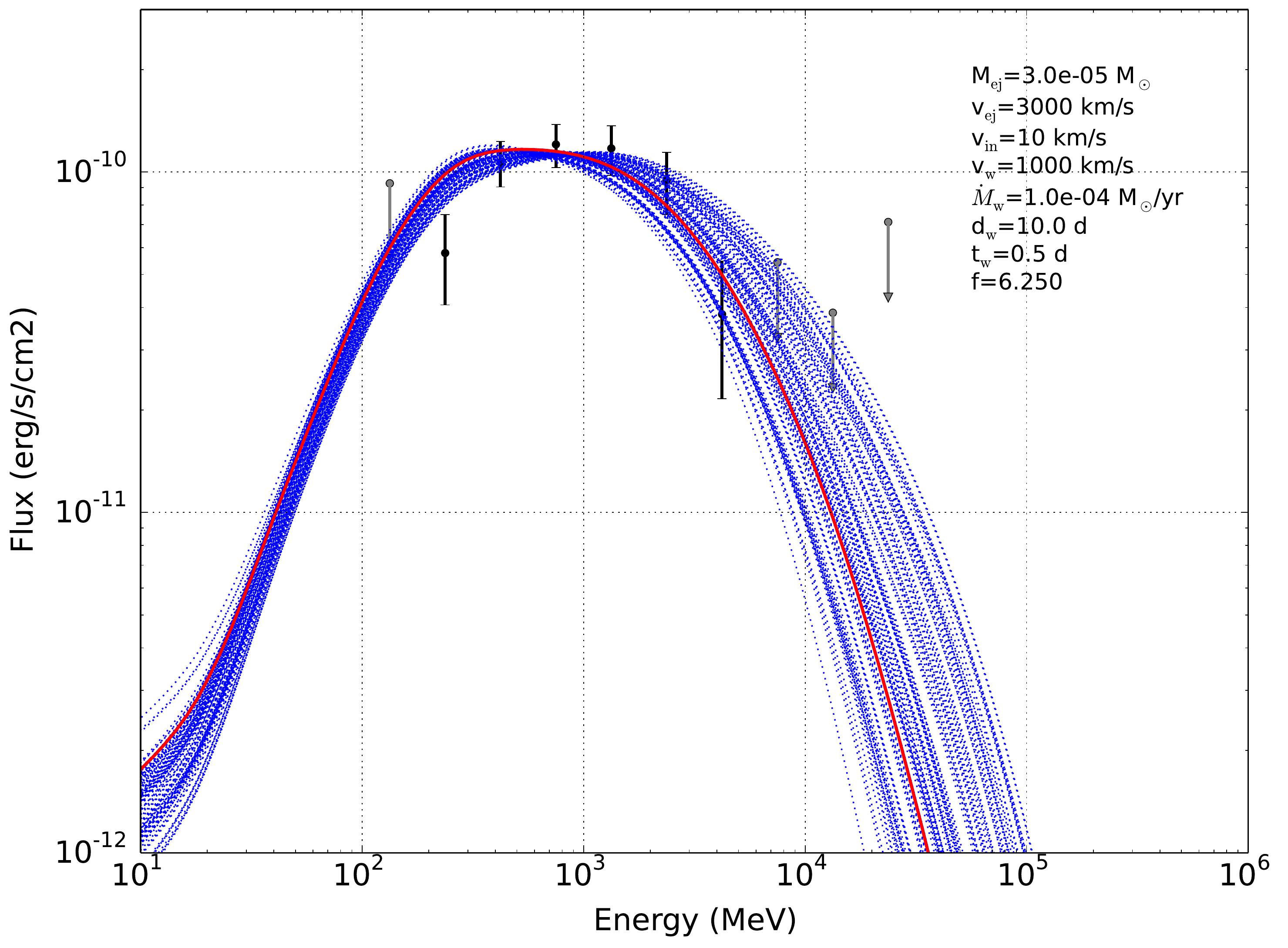}
\includegraphics[width=\columnwidth]{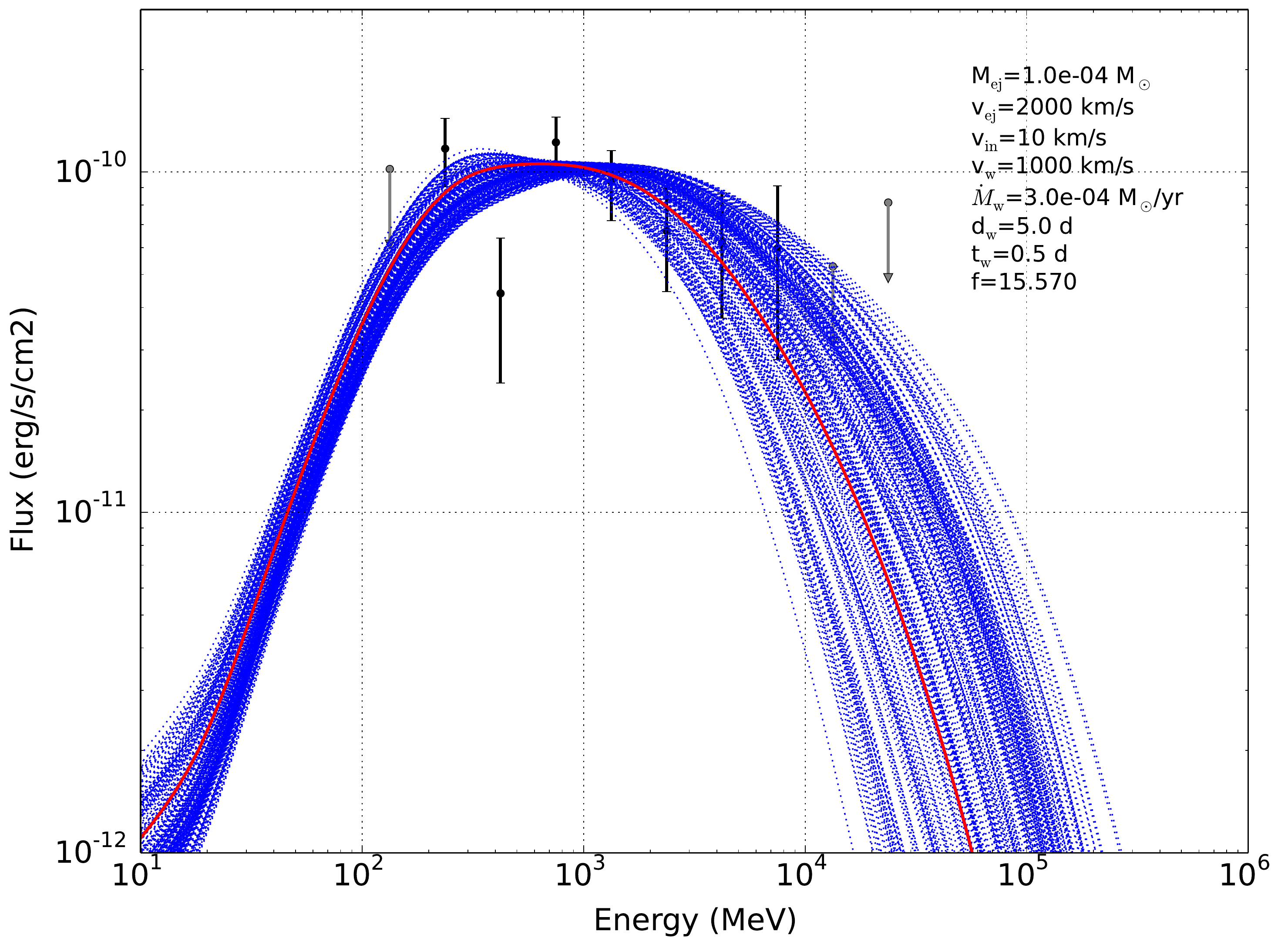}
\includegraphics[width=\columnwidth]{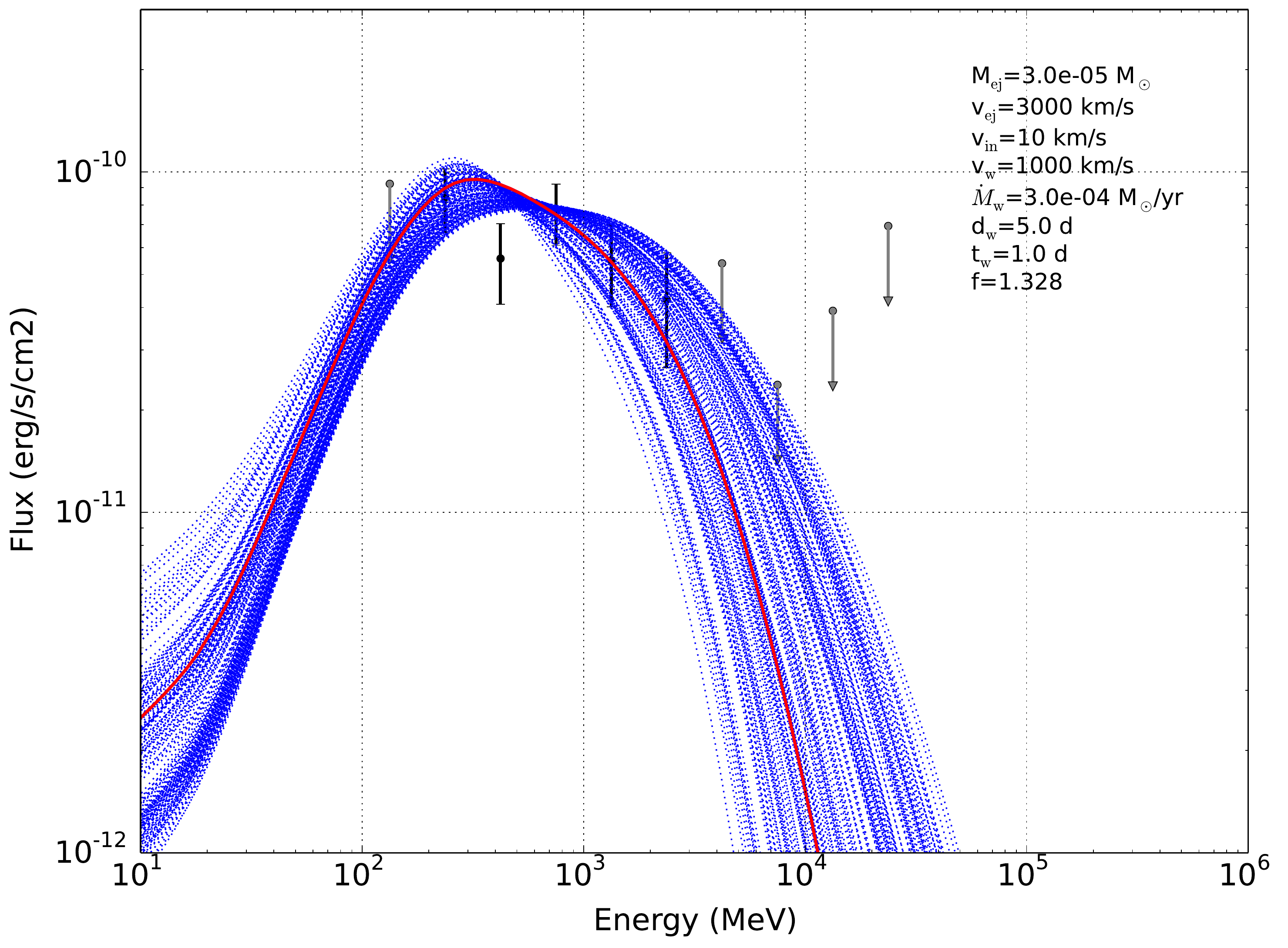}
\caption{Model spectra fitted to \textit{Fermi}-LAT observations for V407 Cyg, V1324 Sco, and V959 Mon (from top to bottom). The red solid curve correspond to the best-fit model, whose parameters are listed in the top-right corner ($f$ being the renormalization factor). The blue dotted curves correspond to model fits that are in the 68.3\% confidence interval.}
\label{fig_specfit_1}
\end{center}
\end{figure}

\begin{figure}[h]
\begin{center}
\includegraphics[width=\columnwidth]{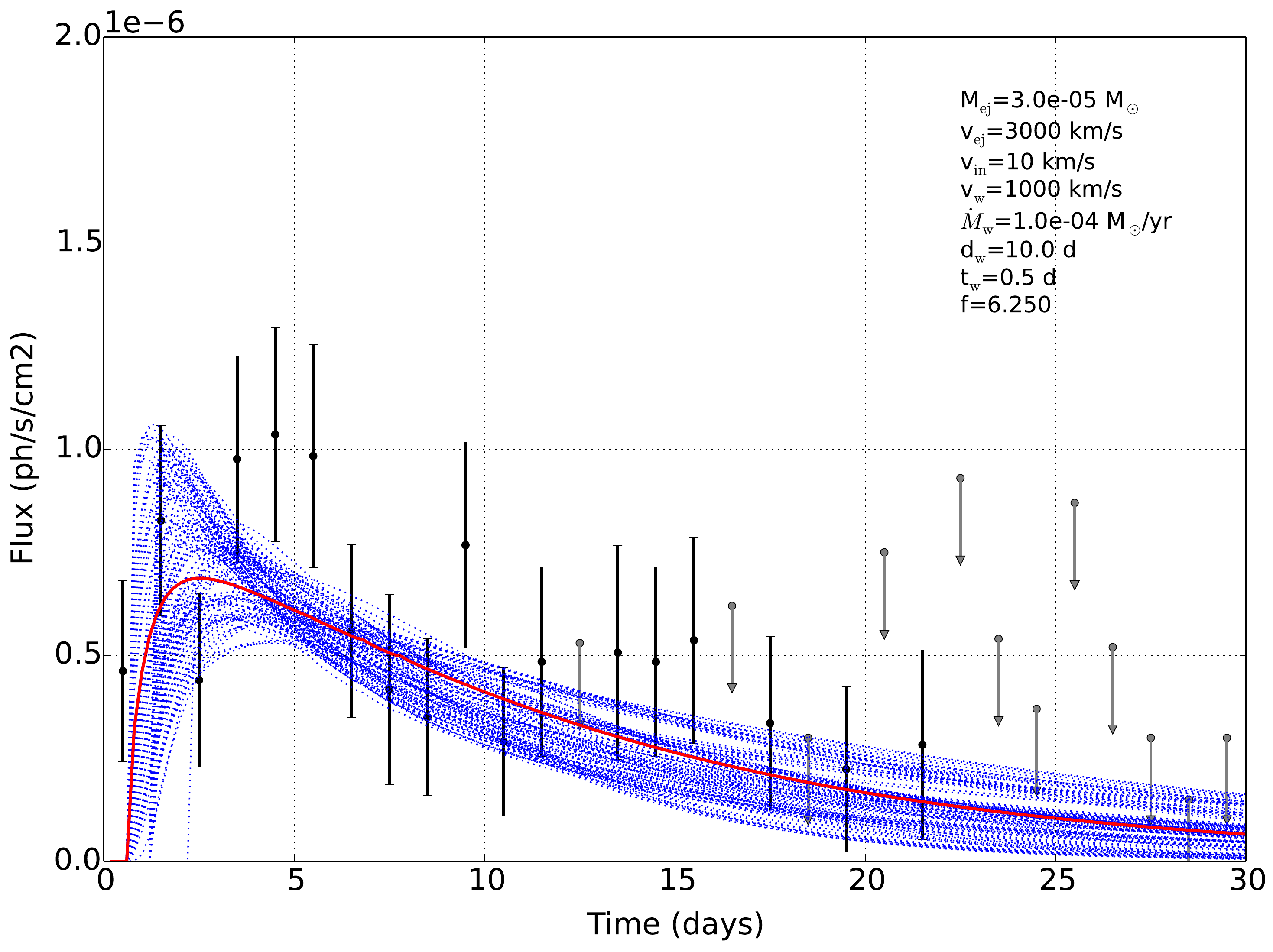}
\includegraphics[width=\columnwidth]{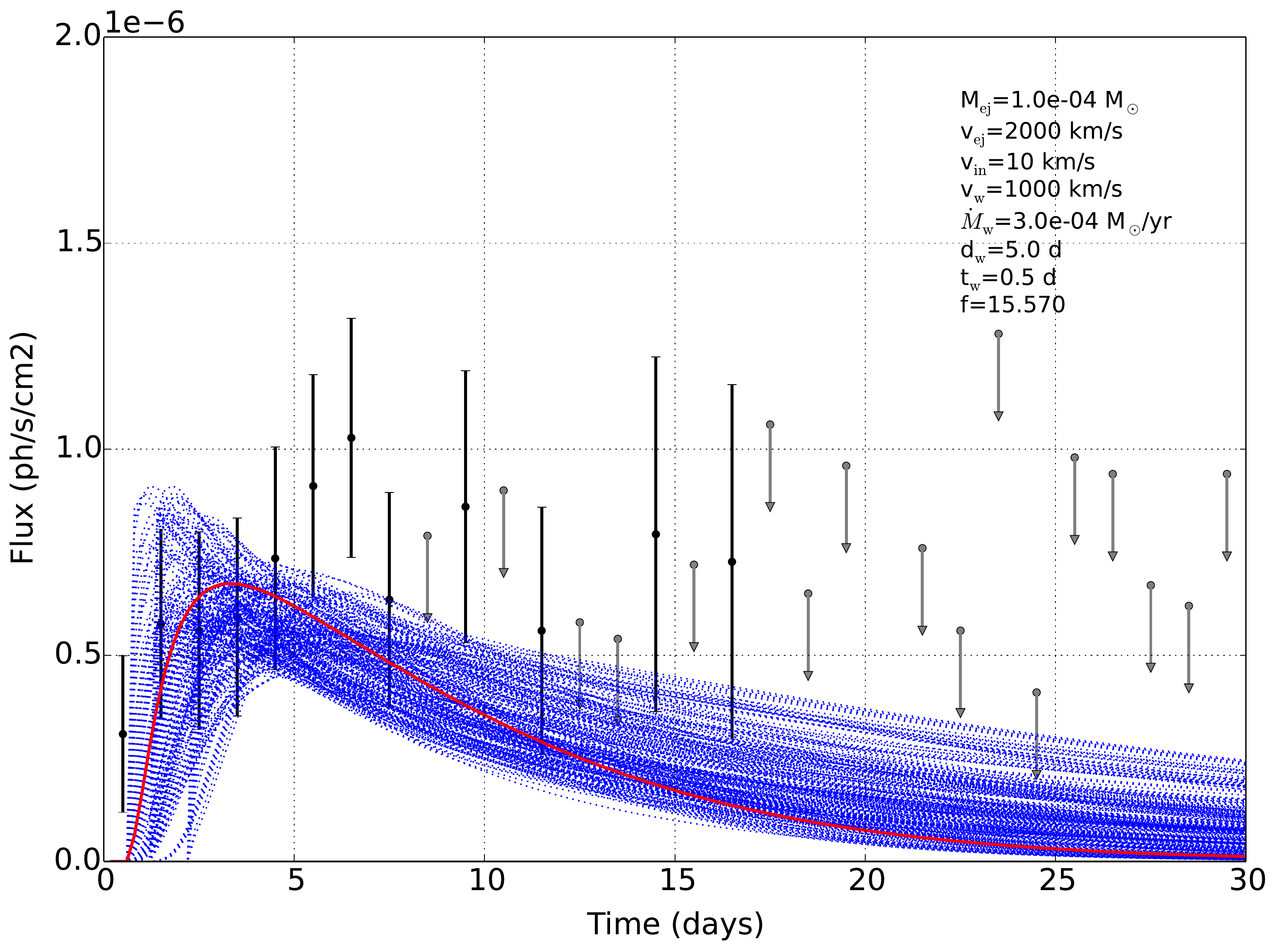}
\includegraphics[width=\columnwidth]{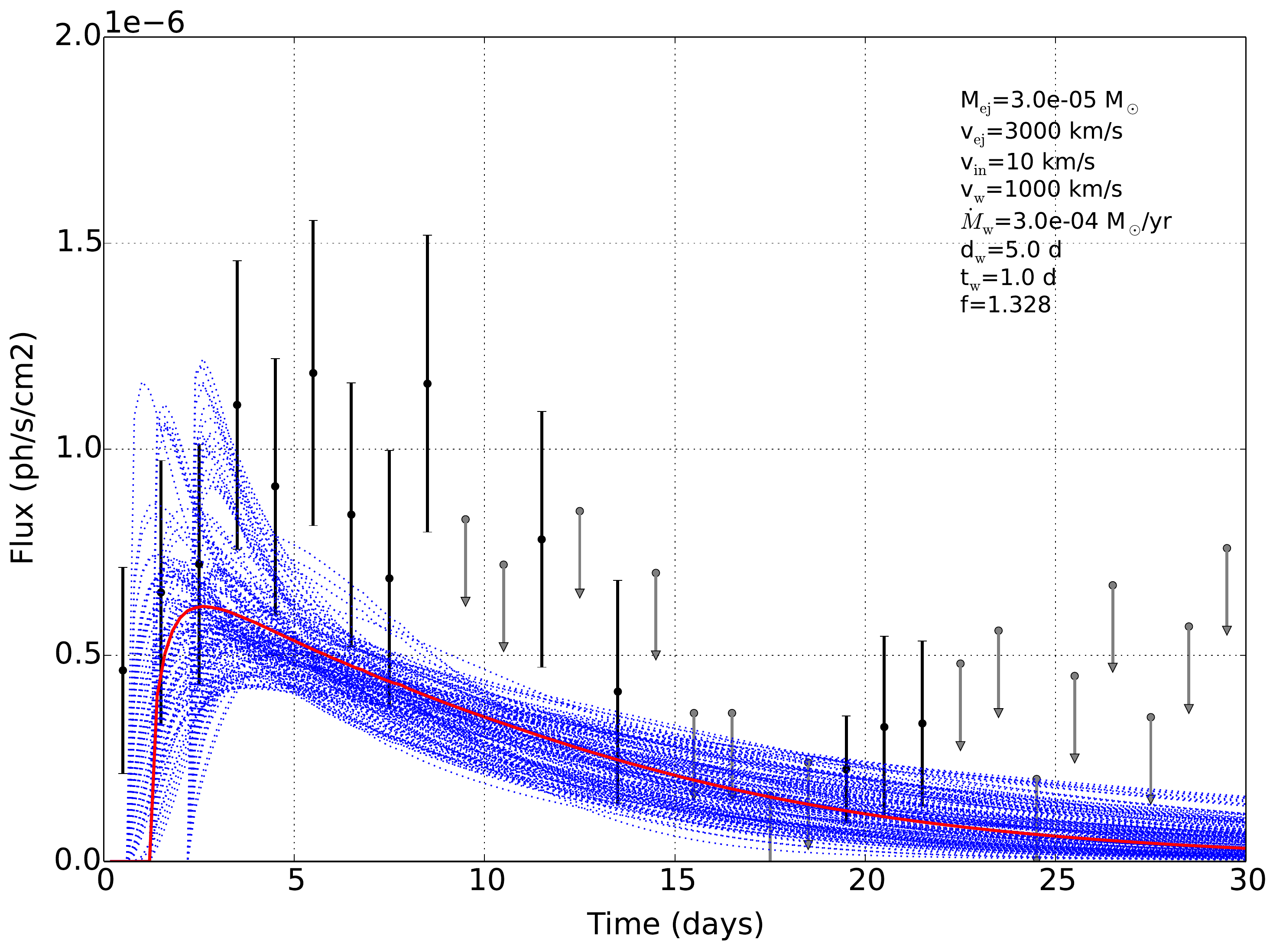}
\caption{Model light curves fitted to \textit{Fermi}-LAT observations for V407 Cyg, V1324 Sco, and V959 Mon (from top to bottom). The red solid curve correspond to the best-fit model, whose parameters are listed in the top-right corner ($f$ being the renormalization factor). The blue dotted curves correspond to model fits that are in the 68.3\% confidence interval.}
\label{fig_lcfit_1}
\end{center}
\end{figure}

\begin{figure}[h]
\begin{center}
\includegraphics[width=\columnwidth]{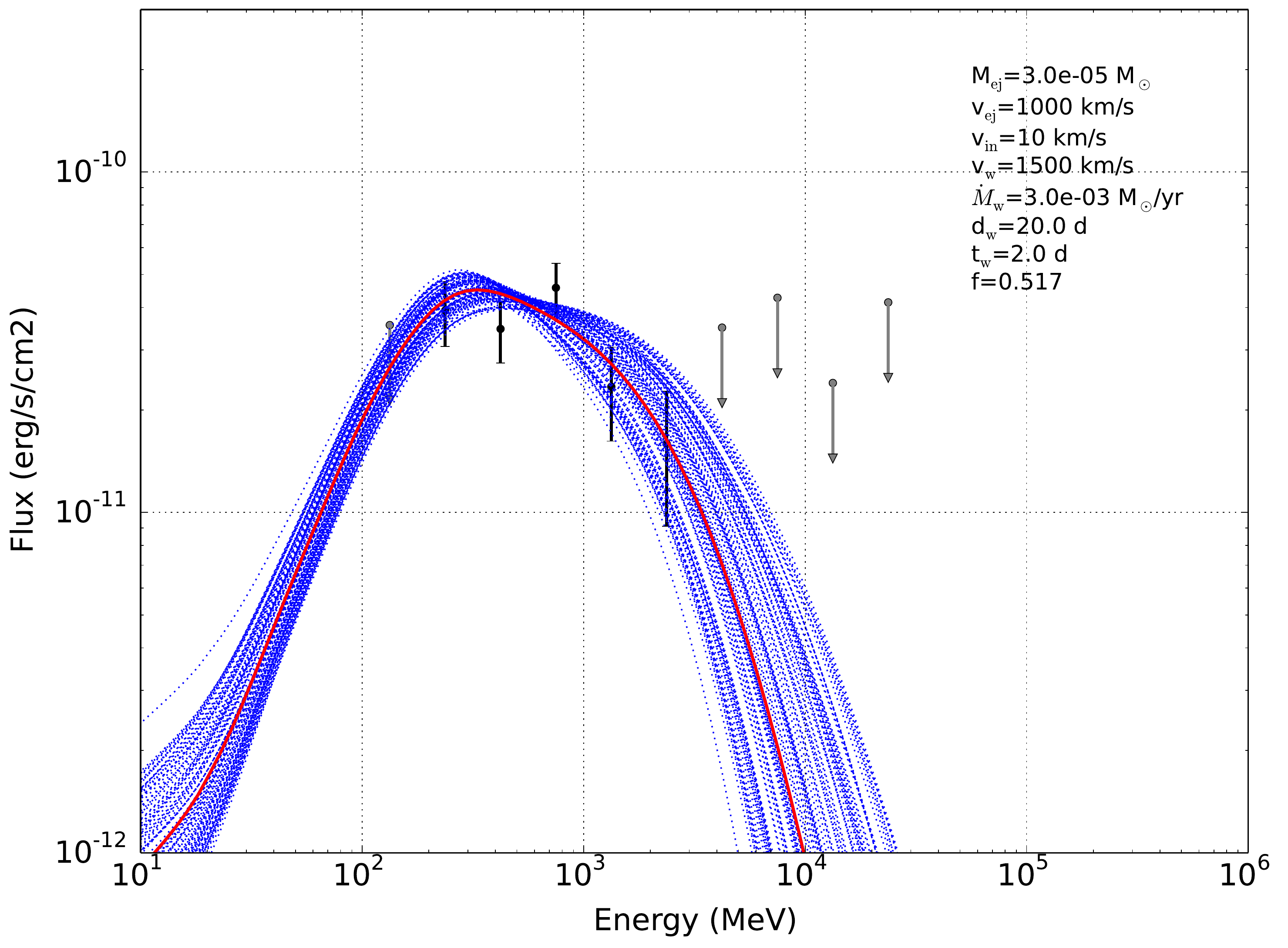}
\includegraphics[width=\columnwidth]{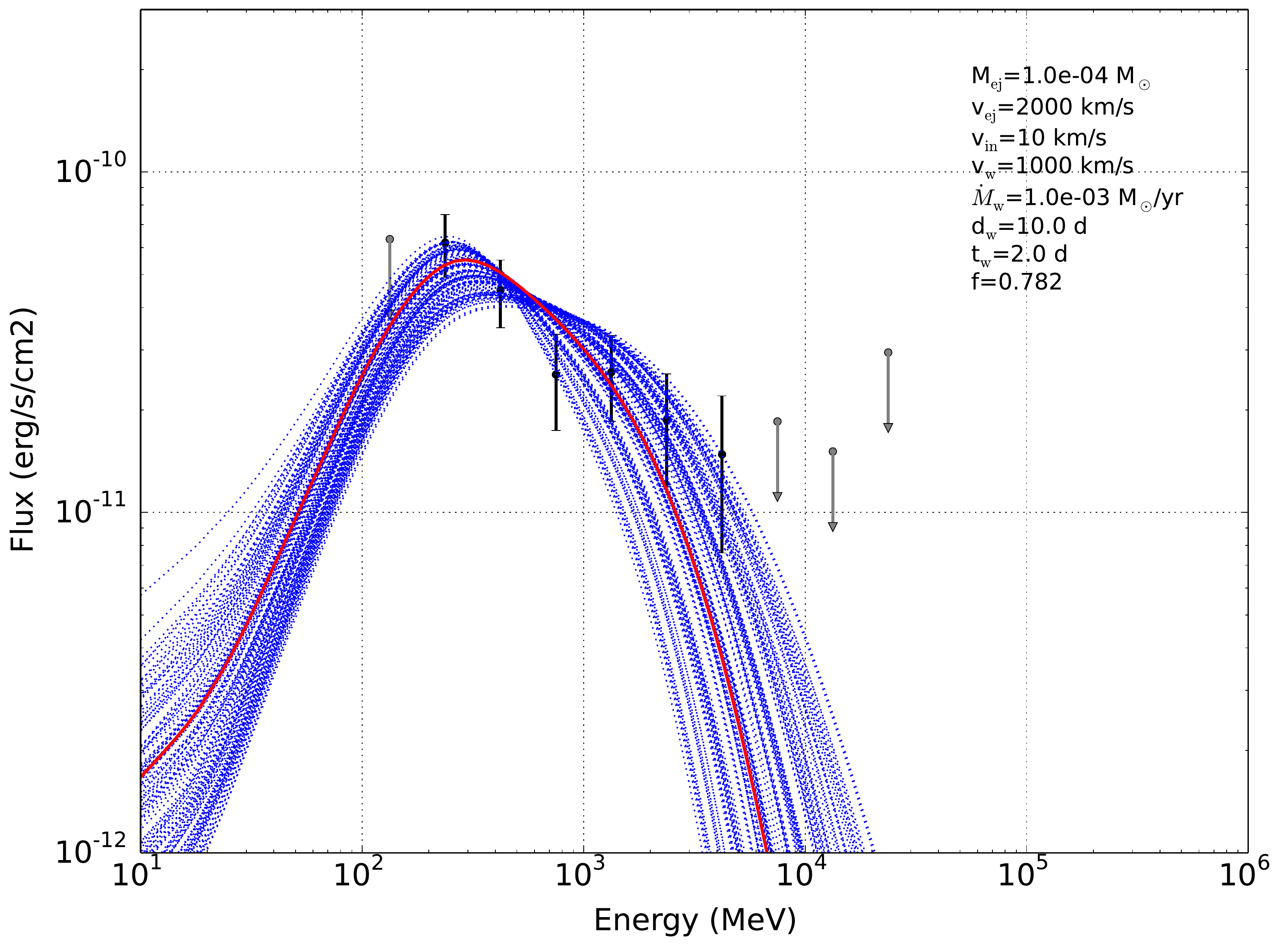}
\includegraphics[width=\columnwidth]{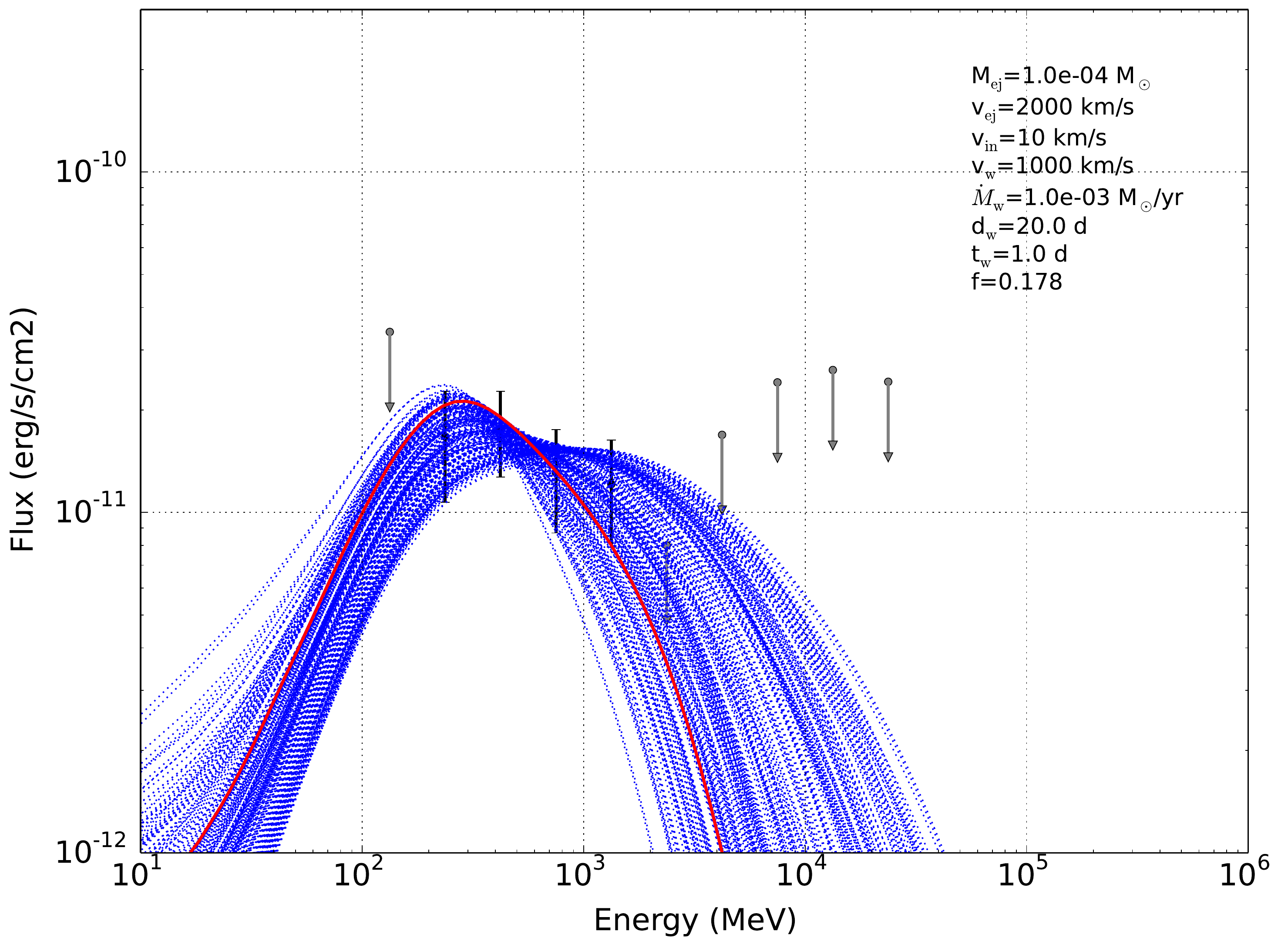}
\caption{Same as Fig. \ref{fig_specfit_1} for V339 Del, V1369 Cen, and V5668 Sgr (from top to bottom).}
\label{fig_specfit_2}
\end{center}
\end{figure}

\begin{figure}[h]
\begin{center}
\includegraphics[width=\columnwidth]{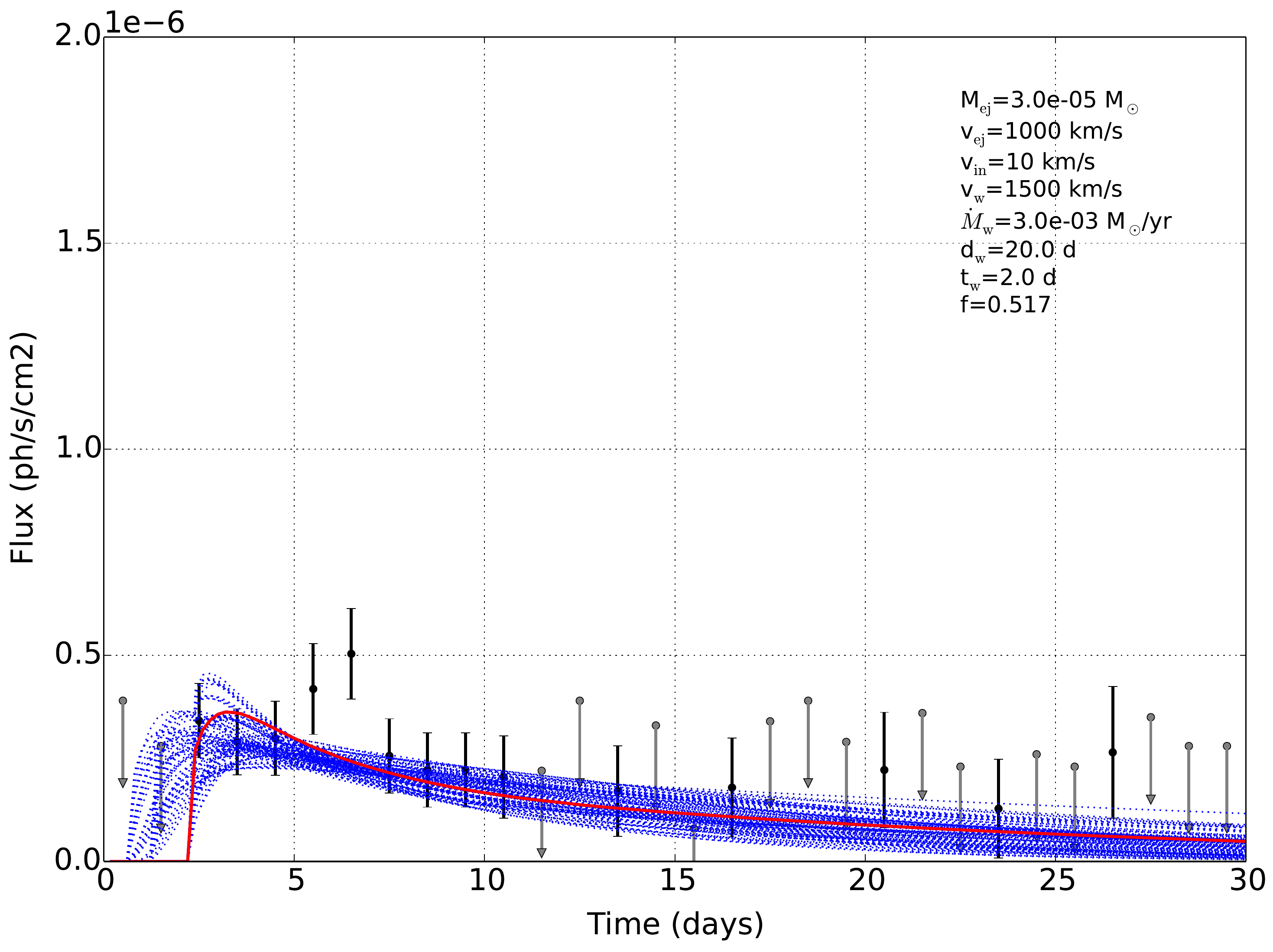}
\includegraphics[width=\columnwidth]{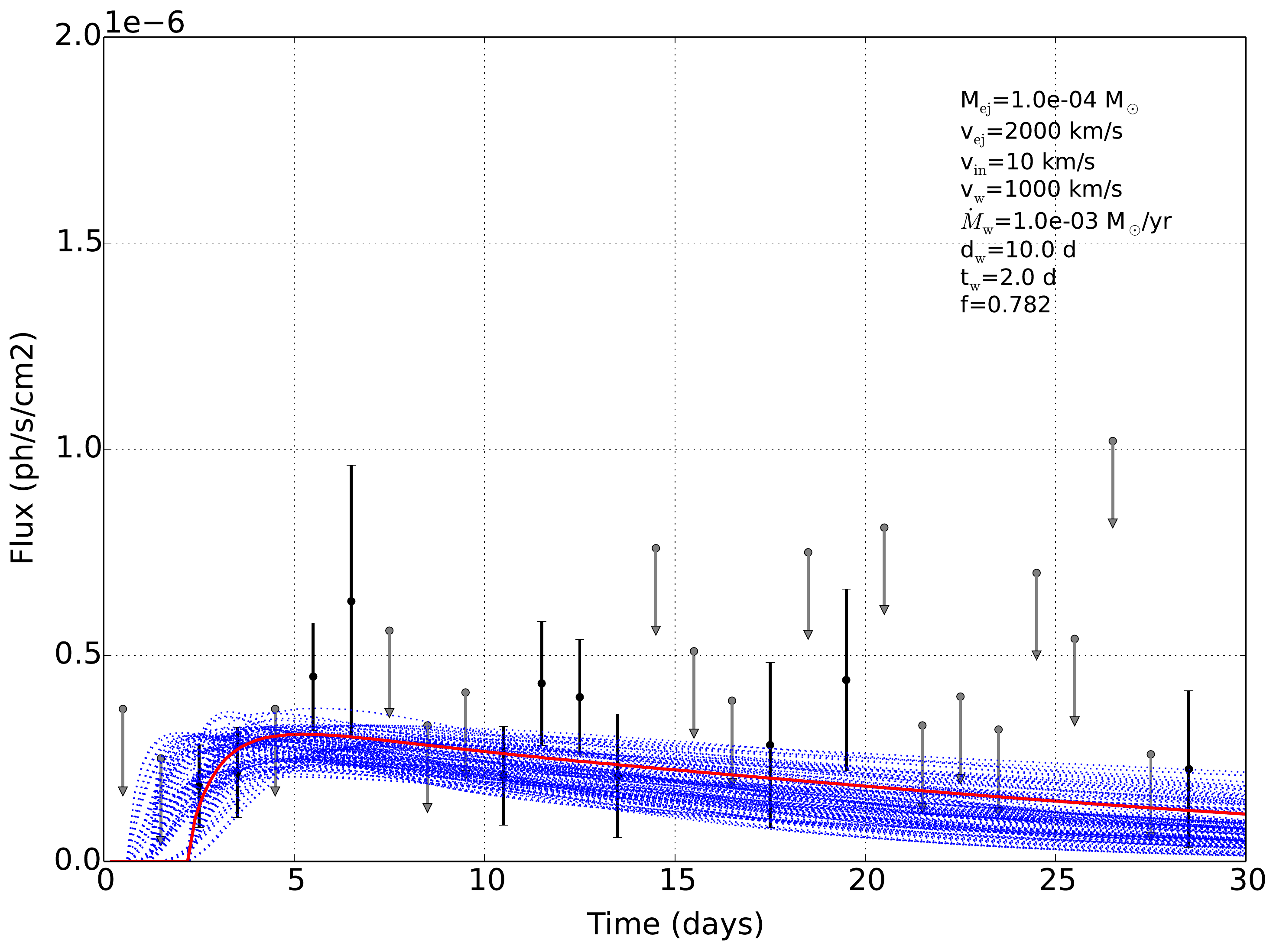}
\includegraphics[width=\columnwidth]{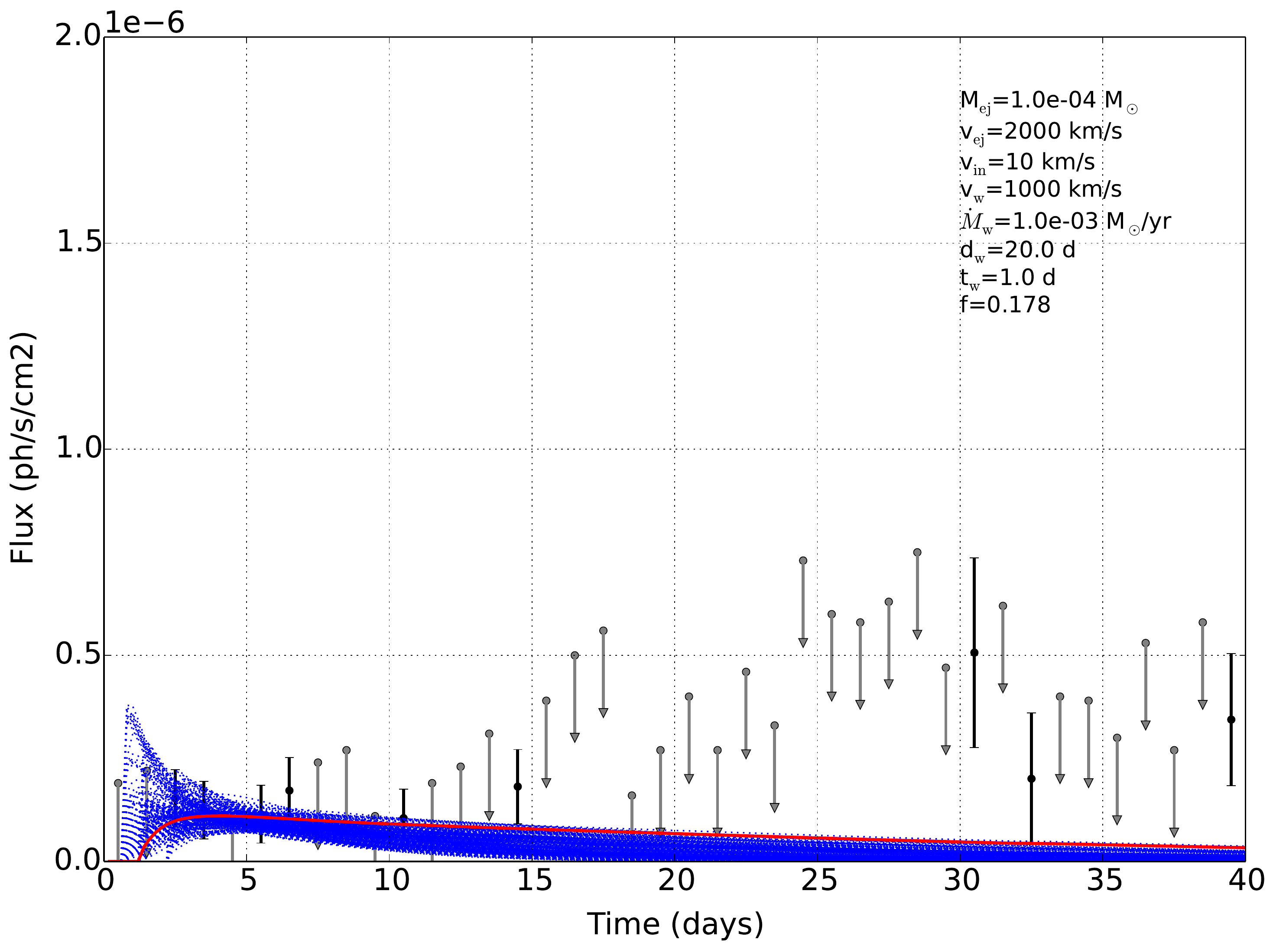}
\caption{Same as Fig. \ref{fig_lcfit_1} for V339 Del, V1369 Cen, and V5668 Sgr (from top to bottom).}
\label{fig_lcfit_2}
\end{center}
\end{figure}

The distances assumed for the novae and the fit results are summarized in Table \ref{tab_novafit}. Spectra and light curves of all models in the 68.3\% confidence interval are overplotted to the \textit{Fermi}-LAT data points and upper limits in Figs. \ref{fig_specfit_1} to \ref{fig_lcfit_2}, with the best-fit profiles marked as bold red lines. In the light curves, the \textit{Fermi}-LAT data points were scaled down compared to the plots published in \citet{Ackermann:2014a} and \citet{Cheung:2016a}. The original data points were computed under the assumption of a power-law spectrum with fixed index 2.2 or 2.3 over time, which biases the comparison with the strongly curved spectra from our model. We have therefore rescaled the data points by the ratio of integrated photon fluxes from the best-fit exponentially cutoff power-law and simple power-law models provided in \citet{Ackermann:2014a} and \citet{Cheung:2016a} (which results in corrections by factors of $\sim0.7-0.9$).

The best fits obtained are quite satisfactory, with reduced $\chi^2$ below 1 for all six novae. The renormalization factors $f$ are all of order unity, except for V1324 Sco, which is the most distant and most luminous one in gamma rays, and thus requires a larger population of non-thermal particles. The observed averaged spectra are well reproduced, both regarding data points around 1\,GeV and upper limits at $\sim$0.1 and $\sim$10\,GeV. Fits to the light curves are also satisfactory, although one can note that the model tends to slightly underestimate the flux level between day $\sim$5 and $\sim$10. This is a consequence of the time profile assumed for the wind mass-loss rate: in the absence of solid constraints, we considered an exponential decrease but the comparison to the data suggests that a plateau phase lasting about a week and followed by a decrease would have provided a slightly better fit.

\begin{figure}[h]
\begin{center}
\includegraphics[width=\columnwidth]{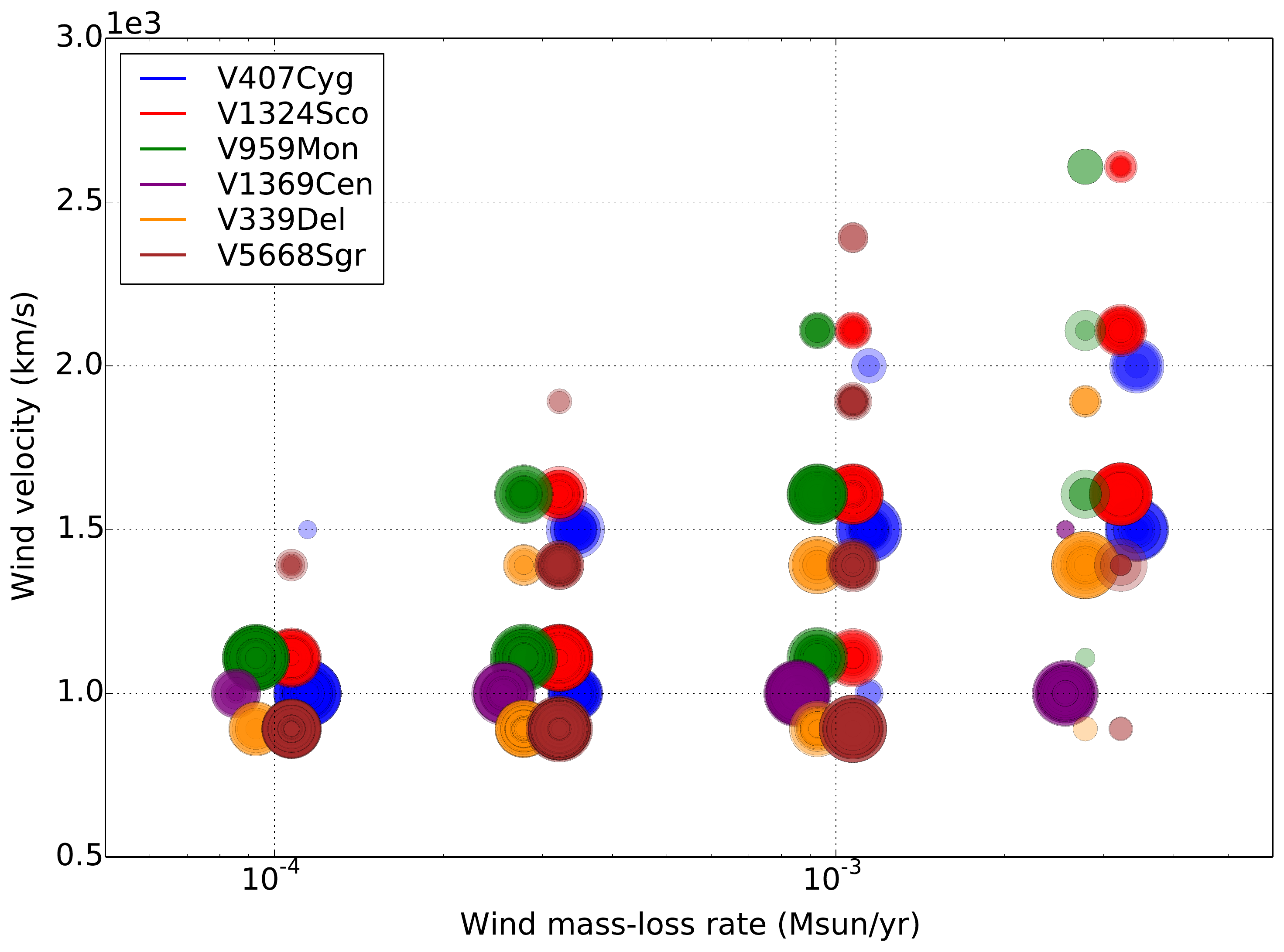}
\includegraphics[width=\columnwidth]{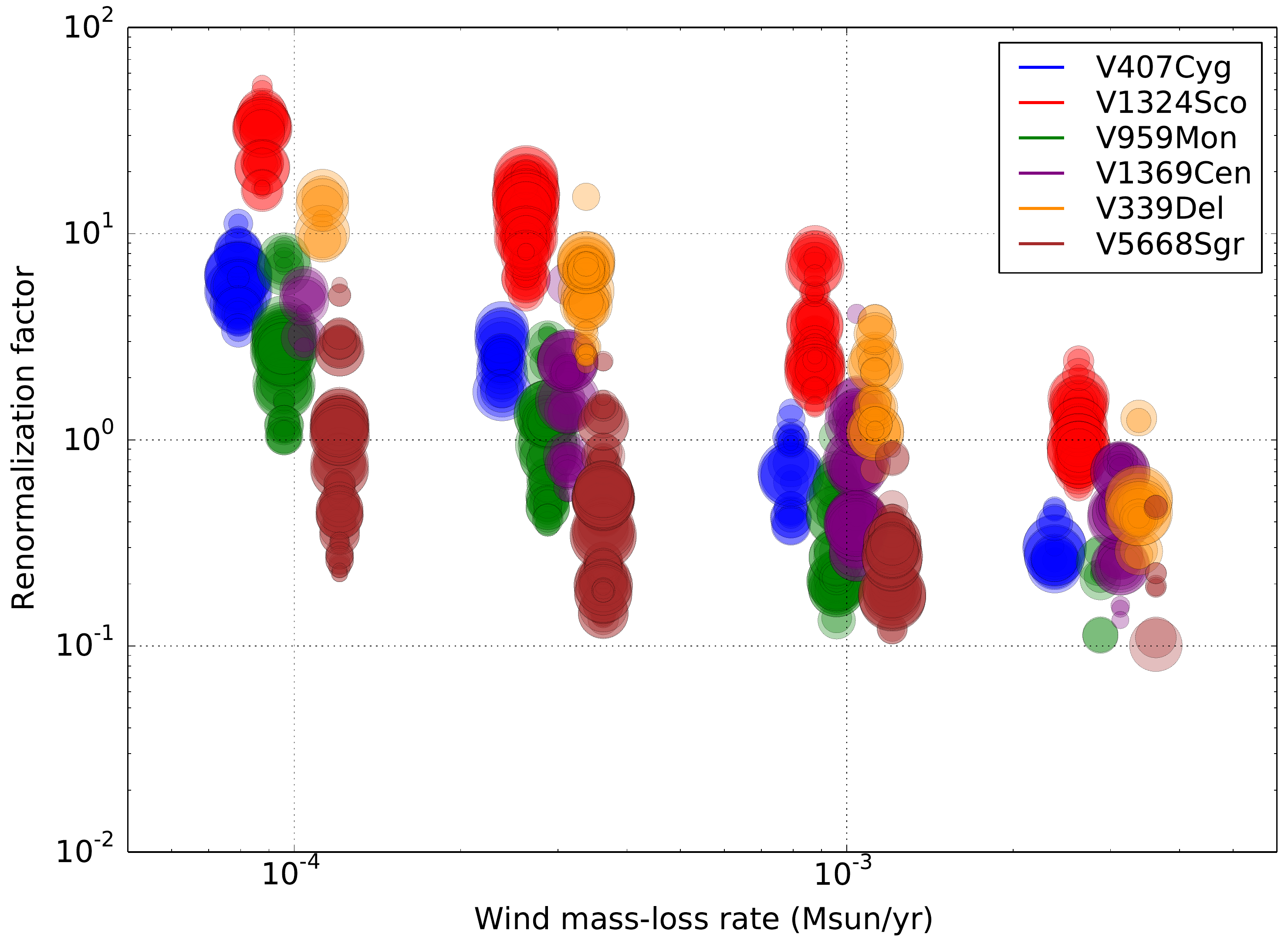}
\includegraphics[width=\columnwidth]{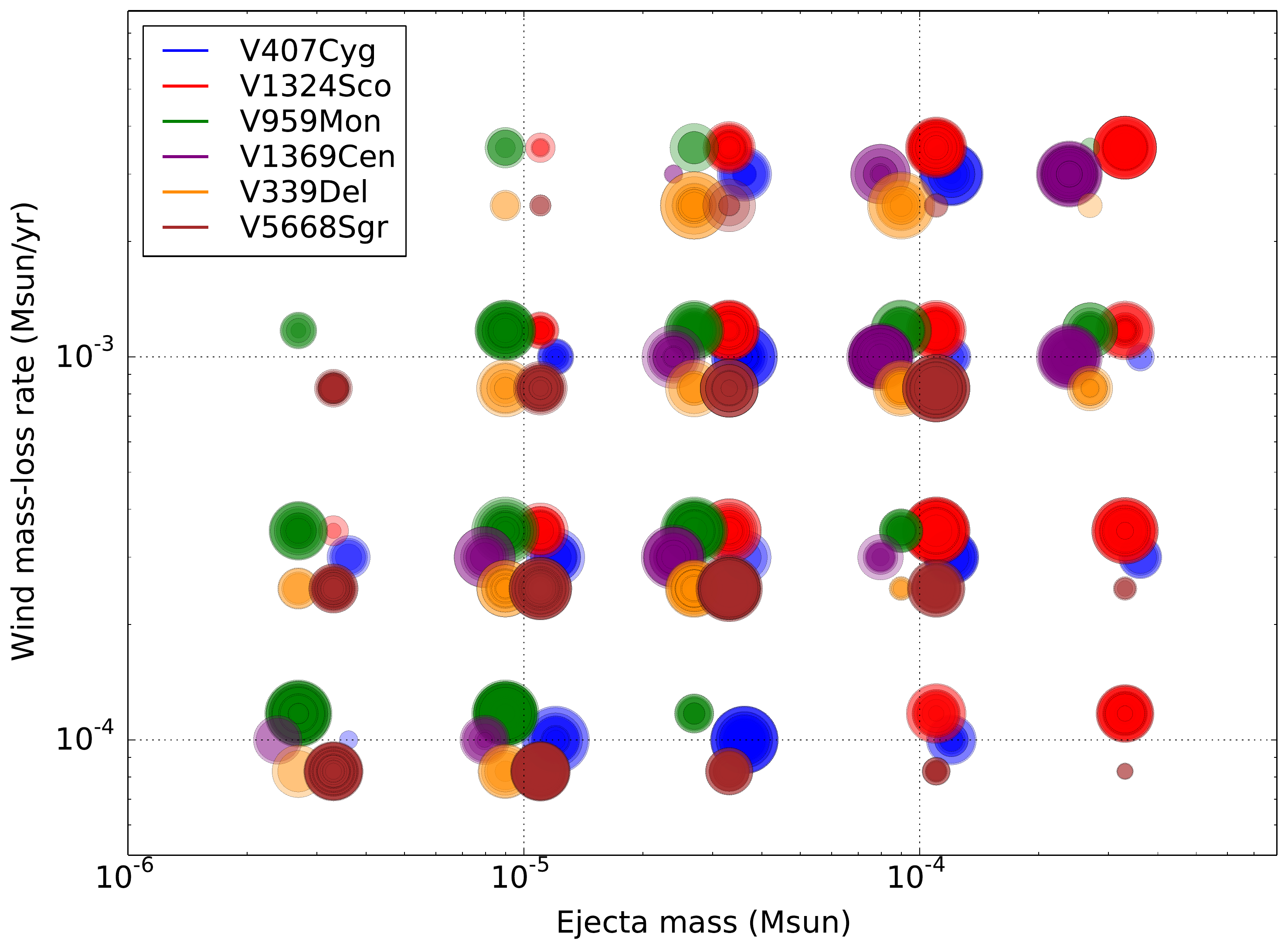}
\caption{Distribution of fit parameters for all models that are in the 68.3\% confidence interval. Each dot corresponds to a fitted model and has a size that is inversely proportional to the reduced $\chi^2$ and a color that indicates the nova observation to which it was fitted (several dots are superimposed for most parameter combinations). Dots for the different novae are slightly offset horizontally and vertically from the exact parameter values for readability.}
\label{fig_parspace}
\end{center}
\end{figure}

Figures \ref{fig_specfit_1} to \ref{fig_lcfit_2} reveal that a large number of models can account for the data, which stems mostly from large uncertainties in the measurements. Figure \ref{fig_parspace} shows for all novae the distribution of fit parameters for all models that are in the 68.3\% confidence interval, and this illustrates how poorly constrained most parameters are. The bottom panel shows that almost the full range of explored values for $M_{\rm ej}$ is likely, with only too massive ejecta $> 10^{-4}$\,M$_\odot$ being disfavoured for half of the novae. Similarly, the middle panel illustrates that most tested values for $\dot{M}_{\rm w}$ are likely, although most fits seem to favour $\dot{M}_{\rm w} \leq 10^{-3}$\,M$_\odot$\, yr$^{-1}$. The latter plot also shows two effects: firstly, there is a degeneracy between the wind mass-loss rate and the renormalization factor $f$, such that a stronger wind mass-loss rate is compensated by a lower injection fraction; secondly, the renormalization factor is correlated to the distance, with the highest values being obtained for the most distant nova, V1324 Sco, and the lowest values for the closest objects, V5668 Sgr and V959 Mon. The only parameter that seems to be rather well constrained is $v_{\rm w}$, whose best-fit values seem to be quite uniform among all 6 novae (which could have been expected because of the similarity of spectra and the discussion in Sect. \ref{pred_prop}). The top panel of Fig. \ref{fig_parspace} shows that the current observations of gamma-ray emitting novae favour wind velocities $< 2000$\,km\,s$^{-1}$. The need for relatively low wind velocities is strengthened by the fact that we neglected adiabatic heating downstream, which would tend to shift particle distributions to higher energies by a factor $\la 5$.

\begin{figure}[h]
\begin{center}
\includegraphics[width=\columnwidth]{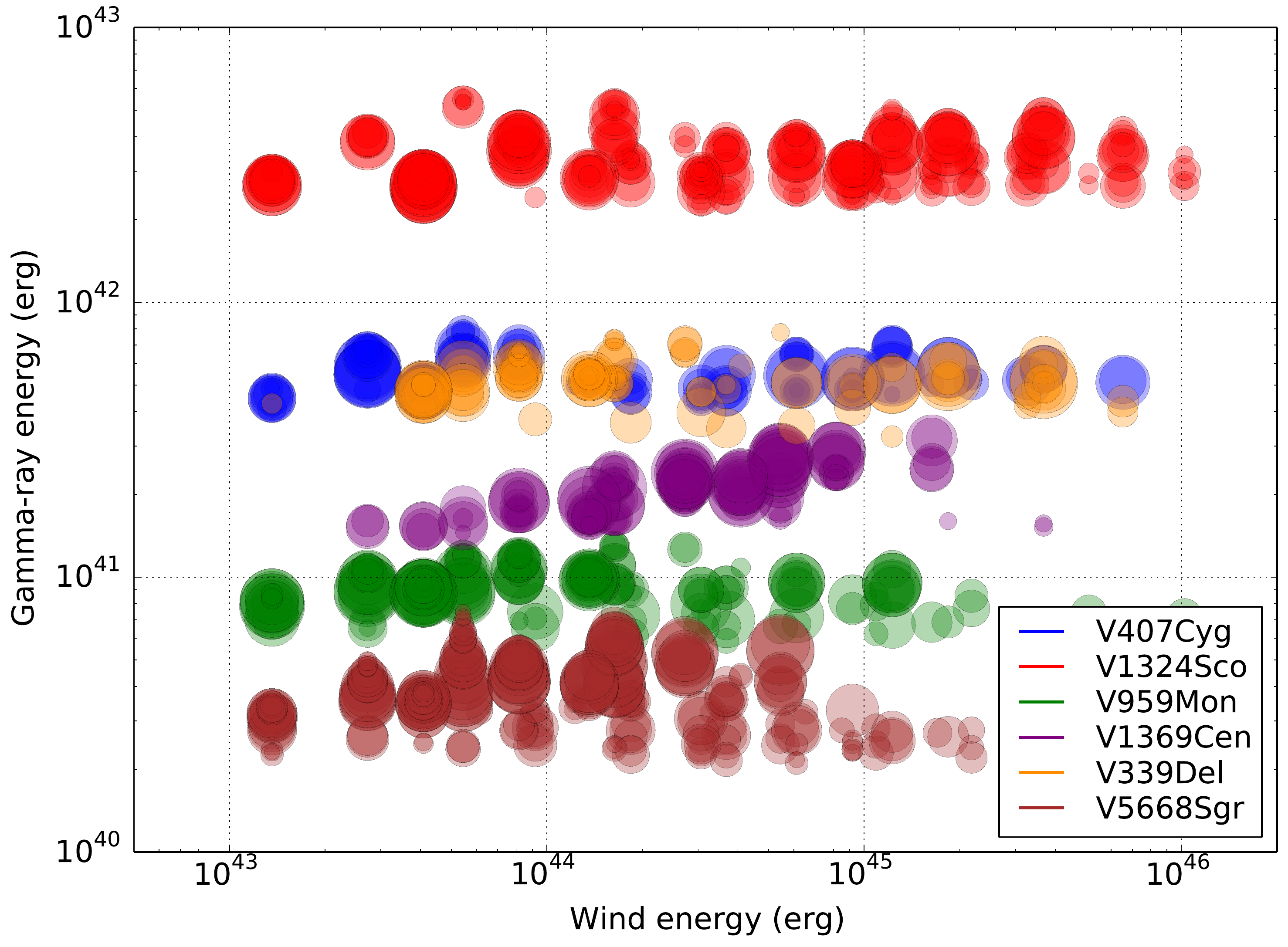}
\includegraphics[width=\columnwidth]{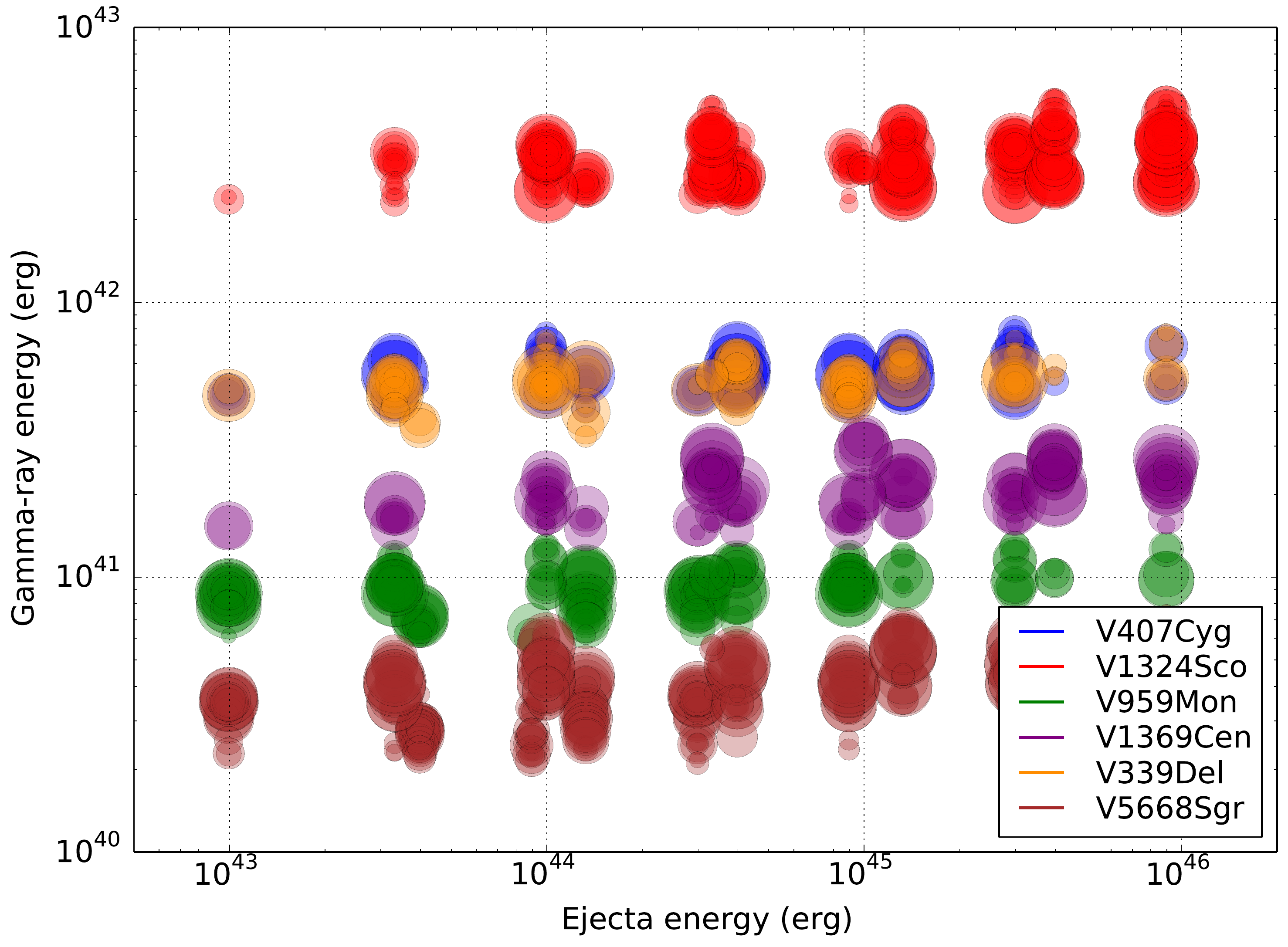}
\caption{Energy emitted in gamma-rays over the first 40 days as a function of the total wind energy (top panel) and ejecta energy (bottom panel), for all models that are in the 68.3\% confidence interval. Each dot corresponds to a fitted model and has a size that is inversely proportional to the reduced $\chi^2$ and a color that indicates the nova observation to which it was fitted.}
\label{fig_energetics}
\end{center}
\end{figure}

In terms of energetics, Fig. \ref{fig_energetics} illustrates that the gamma-ray output can vary by up to two orders of magnitude from one nova to another (assuming distance estimates are correct). Unfortunately, the large uncertainties in the gamma-ray measurements do not allow to strongly constrain the total kinetic energy released in the eruption. For each nova studied here, a given level of energy emitted in gamma-rays can be reproduced from widely different wind or ejecta energies, spanning two orders of magnitude or more. As a consequence, the efficiency of particle acceleration and gamma-ray production by novae cannot be strongly constrained from gamma-ray observations alone. 

Overall, gamma-ray observations of all six novae can be accounted for by about the same scenario: an ejecta of mass $10^{-5} - 10^{-4}$\,M$_\odot$ expands with a velocity 1000-2000\,km\,s$^{-1}$, followed within a day by a wind with mass-loss rate $10^{-4} - 10^{-3}$\,M$_\odot$\, yr$^{-1}$ and velocity $1000$\,km\,s$^{-1}$ and declining over a time scale of a few days; a fraction $10^{-4}$ of protons entering the reverse shock front are accelerated, diffusing across the shock in the Bohm limit, in a magnetic field with an energy density of 1\% of the upstream kinetic energy density, eventually reaching energies of about 100\,GeV and radiating from hadronic interactions in the dense downstream medium. Variations by factors of a few in ejecta mass and wind mass-loss rate from nova to nova result in gamma-ray luminosities spanning at least an order of magnitude in gamma-ray luminosity, which allows us to detect objects over a relatively wide range of distances.

\subsection{V407 Cyg: internal or external shocks ?}
\label{pred_v407}

Interestingly, this internal shock model seems to account reasonably well for the gamma-ray observations of V407 Cyg, which is not a classical nova but a nova in a symbiotic system. This calls into question the actual contribution of an external shock to the emission from such systems. In the model for V407 Cyg presented in \citet{Martin:2013a}, it was shown that an external shock propagating in a wind density profile cannot reproduce the observed light curve. Instead, it was proposed that matter accumulation in the orbital plane and around the white dwarf, as obtained in hydrodynamical simulations of the quiescent accretion phase, is required to get an early rise and rapid drop of the gamma-ray emission. The internal shock scenario explored here does not exclude an additional contribution from an external shock, but it certainly alleviates the constraints of early rise and rapid drop of the light curve. On the other hand, the rapid decline in the optical light curve and the early appearance of coronal lines and soft X-ray emission from the WD favour a low ejecta mass of $\sim 10^{-7} - 10^{-6}$\,M$_\odot$ \citep[see][and references therein]{Chomiuk:2012a}, which seems to contradict the best-fit ejecta masses of $\sim 10^{-5} - 10^{-4}$\,M$_\odot$ obtained here for V407 Cyg (see Fig. \ref{fig_parspace}).

\subsection{Prospects at other wavelengths}
\label{pred_other}

\begin{figure}[h]
\begin{center}
\includegraphics[width=\columnwidth]{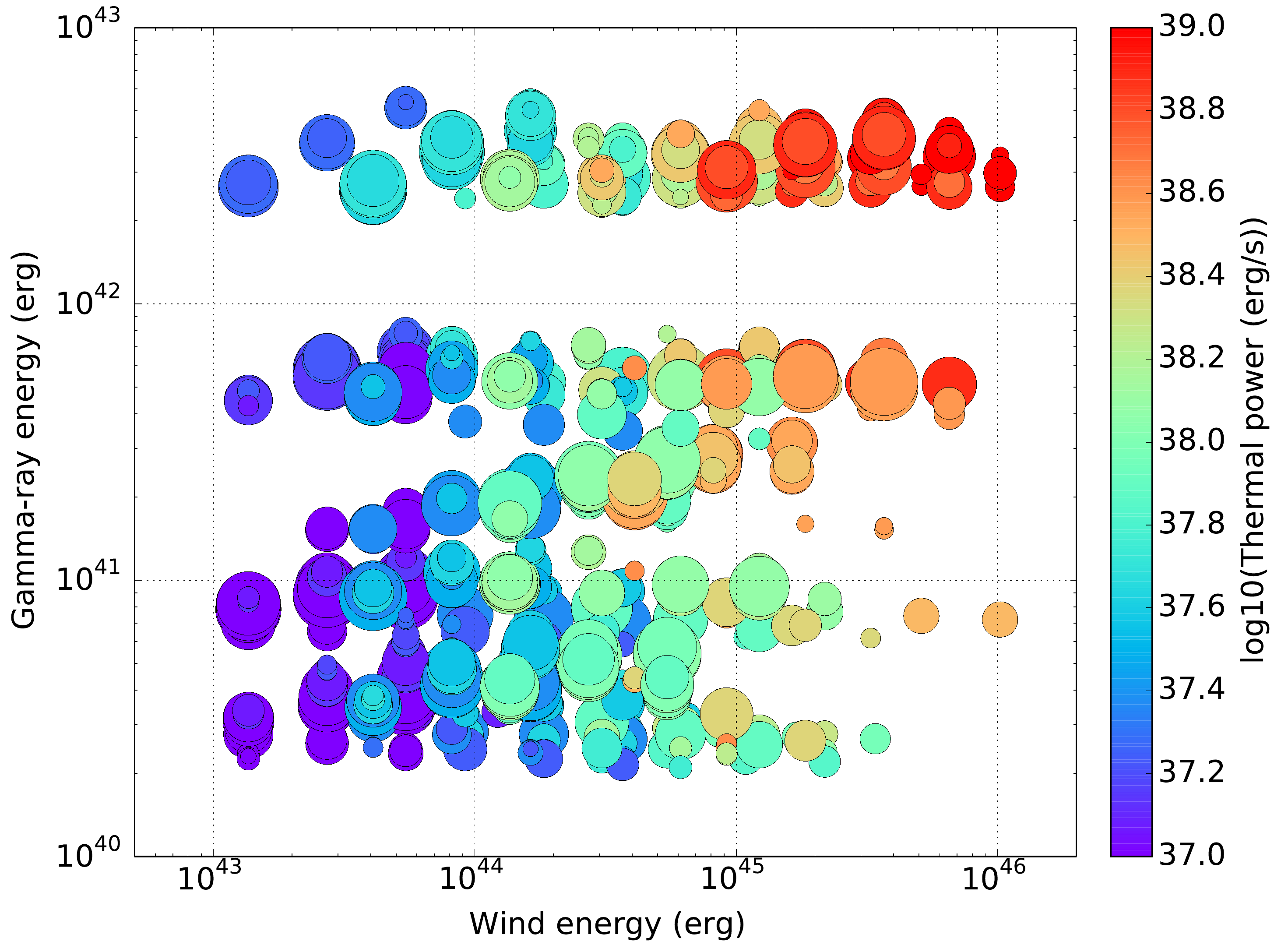}
\includegraphics[width=\columnwidth]{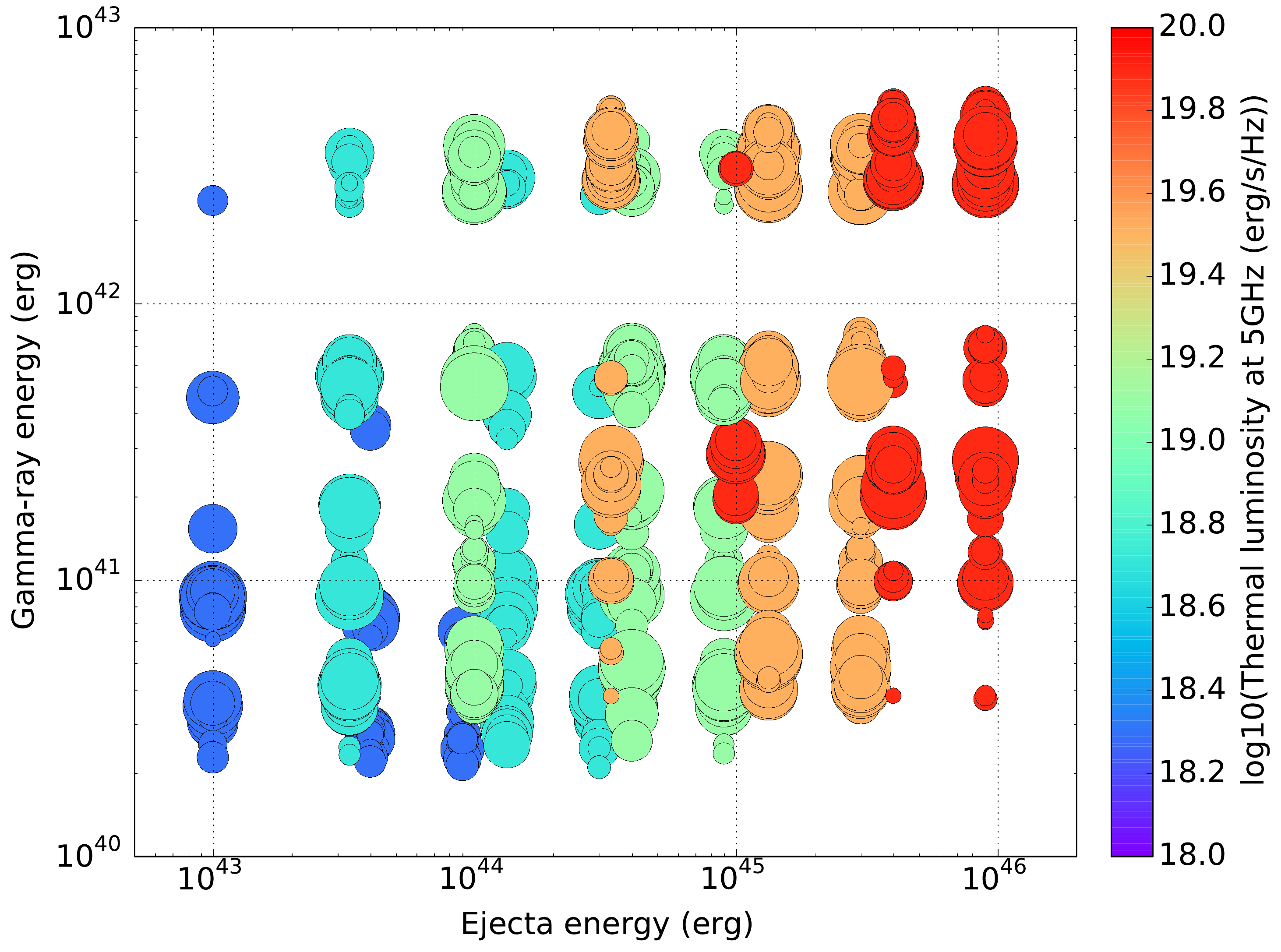}
\caption{Top panel: same as the top panel of Fig. \ref{fig_energetics}, except that the color indicates the peak thermal power dissipated at the reverse shock. Bottom panel: same as the bottom panel of Fig. \ref{fig_energetics}, except that the color indicates the peak thermal luminosity at 5GHz produced by the expanding ionized ejecta (undisturbed by internal shocks; see Sect. \ref{pred_other}).}
\label{fig_xradio}
\end{center}
\end{figure}

X-ray emission from the shocked material would most likely be a critical probe of the properties of the wind and ejecta. However, the X-ray emission from the shocked plasma may be absorbed and reprocessed to optical wavelengths, as pointed out in \citet{Metzger:2014a}. The top panel of Fig. \ref{fig_xradio} shows the peak thermal power, that is the maximum rate of conversion of mechanical energy into thermal energy at the reverse shock, corresponding to the best-fit model parameters obtained for the six novae \citep[computed using Eq. 28 from][for the radiative shock case]{Metzger:2014a}. The shock-related thermal power varies by two orders of magnitude over the parameter space constrained by the gamma-ray observations, and this will most likely have observational consequences. In particular, thermal powers much in excess of $10^{38}$\,erg\,s$^{-1}$, which is the typical (Eddington) luminosity of a nova, are probably excluded in most cases because they would give rise to over-luminous novae either in X-rays or in optical. This sets constraints on the wind energetics and would disfavour large wind velocities and combinations of large ejecta masses and large wind mass-loss rates. Conversely, our model shows that gamma-ray observations can be reproduced from low ejecta masses $\sim10^{-5}$\,M$_\odot$, low wind mass-loss rates $\sim10^{-4}$\,M$_\odot$\, yr$^{-1}$ and velocities $\sim1000$\,km\,s$^{-1}$, which would produce shock-related thermal emission about an order of magnitude below the canonical $10^{38}$\,erg\,s$^{-1}$. In either case, the velocities of the different ejecta/outflow components may be probed from the profile and evolution of emission lines \citep{Cassatella:2004a}.

\begin{figure}[h]
\begin{center}
\includegraphics[width=\columnwidth]{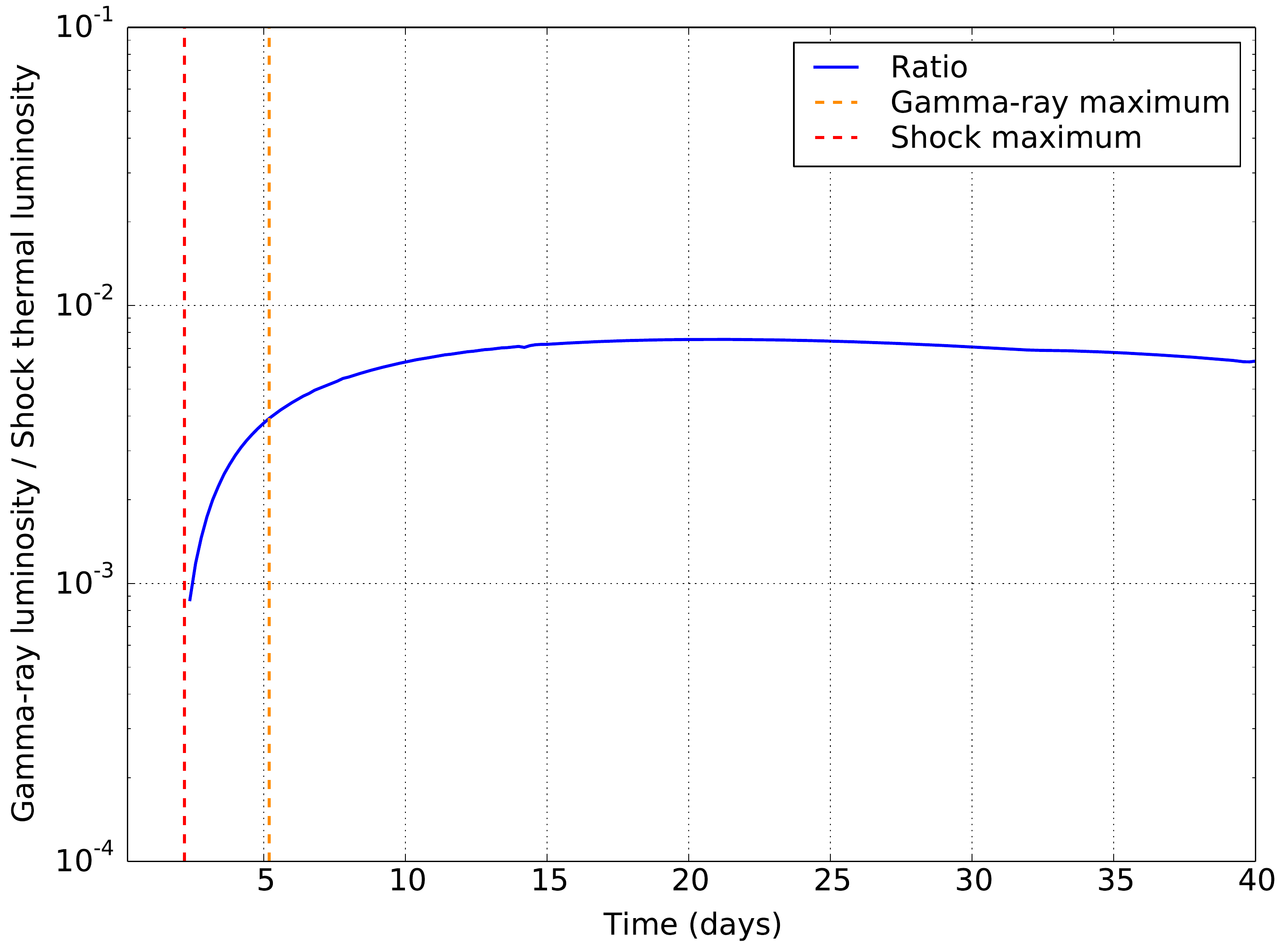}
\caption{Ratio of the gamma-ray luminosity to the reverse shock thermal power as a function of time, for a model with parameters $(M_{\rm ej},v_{\rm ej},v_{\rm in},\dot{M}_{\rm w},v_{\rm w},D_{\rm w},t_{\rm w})=(10^{-4},1000,10,3 \times 10^{-3},2000,10,2)$ and a proton injection fraction of $4 \times 10^{-5}$. The red and orange vertical dashed lines mark the peak of the shock thermal power and gamma-ray luminosity, respectively.}
\label{fig_ratio_gamth}
\end{center}
\end{figure}

\citet{Li:2017a} recently presented a convincing case of correlation between optical and gamma-ray emission from ASASSN-16ma, a nova with super-Eddington luminosity, thus lending support to the scenario put forward by \citet{Metzger:2014a}. \citet{Li:2017a} observed a nearly constant gamma-ray-to-optical flux ratio over 10 days past gamma-ray maximum and we compared their observations to a selected model of our grid. Using $(M_{\rm ej},v_{\rm ej},v_{\rm in},\dot{M}_{\rm w},v_{\rm w},D_{\rm w},t_{\rm w})=(10^{-4},1000,10,3 \times 10^{-3},2000,10,2)$ and a proton injection fraction of $4 \times 10^{-5}$, our model yields shock and gamma-ray luminosities with maximum values of $10^{39}$ and $2 \times 10^{36}$\,erg\,s$^{-1}$, respectively, which translates into fluxes of $5 \times 10^{-7}$ and $9 \times 10^{-10}$ \,erg\,s$^{-1}$\,cm$^{-2}$ \citep[assuming a distance of 4.2\,kpc, as used in][]{Li:2017a}. This is in very good agreement with the measured peak optical flux, expected to approximately trace the total shock power in the scenario of \citet{Metzger:2014a}, and with the peak gamma-ray flux measured with the \textit{Fermi}-LAT. Also consistent with the observations, the model predicts a ratio of the gamma-ray luminosity to the shock thermal power that is nearly constant after the gamma-ray maximum (see Fig. \ref{fig_ratio_gamth}; the ratio varies by less than a factor of 2 over a few weeks). The predicted ratio is, however, a factor of $\sim 2$ above the observed value, a discrepancy that may be solved by adopting for instance a slightly different density profile for the ejecta or a different time evolution for the wind properties.

Radio emission is also a powerful diagnostic of the mass and dynamics of the material ejected by a nova. In the classical picture, the expanding ejected envelope is photoionized and heated at temperature of $ \sim 1-4 \times 10^4$\,K by radiation from stable burning of the remaining hydrogen at the surface of the WD \citep{Cunningham:2015a}; this gives rise to thermal bremsstrahlung emission that is routinely detected at $\sim 1-10$\,GHz frequencies up to years after the eruption. This emission has characteristic light curves and spectra, resulting from the frequency-dependent transition from optically-thick to optically-thin radiation, and these can be used to estimate the mass, density and velocity structures of the envelope. The bottom panel of Fig. \ref{fig_xradio} shows the peak thermal emission at 5GHz corresponding to the best-fit model parameters obtained for the six novae \citep[computed following][assuming homologous expansion of a fully ionised hydrogen envelope at 10000\,K]{Bode:2008a}. The peak radio luminosity varies by nearly two orders of magnitude over the parameter space constrained by the gamma-ray observations, and this illustrates how combining radio and gamma-ray observations can reduce uncertainties on the parameters of the internal shock scenario. Using a similar model for the thermal emission, \citet{Finzell:2017a} derived a mass of $2.0 \pm 0.4 \times 10^{-5}$\,M$_\odot$ and velocity of $1150 \pm 40$\,km\,s$^{-1}$ for the ejecta of V1324 Sco, in agreement with our best-fit values for the gamma-ray emission model. 

We emphasize, however, that the radio luminosities presented in Fig. \ref{fig_xradio} do not take into account the effects of the nova wind. In the internal shock scenario, the dense cooled shell comprised between the forward and reverse shocks will shield the unshocked ejecta against ionization by the WD, at least for some time, thus affecting the ionization structure of the ejecta and potentially its thermal emission. This will also have an impact on the escape or not of synchrotron emission, another probe of the non-thermal phenomena, possibly detected in several systems \citep{Weston:2016a,Weston:2016b,Vlasov:2016a}. We will address these points elsewhere, but our first results show that the late thermal radio emission eventually scales with the mass of the envelope and reaches peak levels comparable to those obtained for a simple expanding envelope.

Last, the comparison of our model predictions to the \textit{Fermi}-LAT observations suggests that prospects for detecting novae at very high $\sim$TeV energies are currently rather limited if the six objects studied here provide a representative view on high-energy phenomena in novae. Existing Imaging Atmospheric Cherenkov Telescopes probably have too high energy thresholds and too low sensitivities to detect the high end of a nova spectrum \citep{Aliu:2012a,Ahnen:2015a}. The next-generation Cherenkov Telescope Array very-high-energy observatory might be able to detect $>100$\,GeV signals from novae if some of them can produce powerful high-velocity winds combined with a massive enough ejecta in order to generate a fast reverse shock. As described above, such conditions would most likely result in highly luminous early thermal emission, so targeting abnormally bright objects in optical or X-rays might be a good strategy to achieve very-high-energy detections of novae.

\section{Conclusions}
\label{conclu}

Numerical simulations of novae suggest that the sequence of mass ejection during an eruption may involve the slow expansion of an envelope followed by the launch of a faster wind. The interaction of both components would lead to the formation of a forward and reverse shock system, as supported by observational evidence in the radio, optical, and X-ray domains. Given the densities and velocities at stake, those shocks are expected to be strongly radiative over the first weeks, which would influence the bolometric luminosity of the nova, and to produce non-thermal emission, which could account for the gamma-ray emission detected from a handful of novae with the \textit{Fermi}-LAT instrument.

We present a model for particle acceleration and gamma-ray emission at internal shocks in novae. The dynamics of the radiative shock system in the nova ejecta is solved from the equation for momentum conservation, assuming spherical symmetry and neglecting effects of the companion star or interactions of the ejecta with pre-existing circumbinary material. From the resulting hydrodynamical quantities, particle acceleration is computed in the test particle approximation and assuming Bohm diffusion by solving a time-dependent transport equation for particle injection, acceleration, losses, and escape. 

For typical parameters, the $\ge 100$\,MeV gamma-ray emission arises predominantly from particles accelerated at the reverse shock and undergoing hadronic interactions in the dense cooling layer downstream of the shock. The maximum gamma-ray energy is driven by the wind velocity, while the wind mass-loss rate sets the gamma-ray luminosity. The emission rises within days after the onset of the wind, quickly reaches a maximum, and its subsequent decrease reflects mostly the time evolution of the wind properties. Departure from spherical symmetry, such as ejecta with different masses and velocities in the equatorial plane and polar directions, results in a flattening of the light curve and possibly on multiple peaks.

We fitted our model to the spectra and light curves measured with the \textit{Fermi}-LAT for six novae (V407 Cyg, V1324 Sco, V959 Mon, V339 Del, V1369 Cen, and V5668 Sgr), testing more than 2000 combinations of the ejecta and wind parameters. Because of the large uncertainties in the measurements, a large array of parameter values are able to reproduce the observations. The strongest constraints are obtained for the wind velocities, with values $\la 2000$\,km\,s$^{-1}$ favoured by the observed spectra. A typical model for all six novae is the following: an ejecta mass in the range $10^{-5} - 10^{-4}$\,M$_\odot$ expands with a velocity 1000-2000\,km\,s$^{-1}$, followed within a day by a wind with mass-loss rate $10^{-4} - 10^{-3}$\,M$_\odot$\, yr$^{-1}$ and velocity $\la 2000$\,km\,s$^{-1}$ and declining over a time scale of a few days; a fraction $\sim 10^{-4}$ of protons entering the reverse shock front are accelerated up to energies of about 100\,GeV and efficiently radiate from hadronic interactions. This typical model is able to account within a factor of 2 for the main features in the observations of the brightest nova ever detected in gamma-rays, ASASSN-16ma.

Gamma-ray observations hold potential for probing the mechanism of mass ejection in novae. A more comprehensive test of the internal shock hypothesis and tighter constraints on the parameters should be provided by modelling the associated thermal emissions, notably in radio and X-rays, and by clarifying whether synchrotron emission is expected. Radiation hydrodynamics calculations would certainly be useful to determine the ionization structure of the ejecta and compute in particular the reprocessing of X-rays and the opacity to optical and radio radiation.

\newpage
\newpage

\newpage
\newpage

\newpage
\newpage

\newpage
\newpage

\newpage
\newpage

\newpage
\newpage

\newpage
\newpage

\bibliographystyle{aa}
\bibliography{ClassicalNovae}

\begin{thebibliography}{43}
\expandafter\ifx\csname natexlab\endcsname\relax\def\natexlab#1{#1}\fi

\bibitem[{{Abdo} {et~al.}(2010){Abdo}, {Ackermann}, {Ajello}, {Atwood},
  {Baldini}, {Ballet}, {Barbiellini}, {Bastieri}, {Bechtol}, {Bellazzini}, \&
  et~al.}]{Abdo:2010a}
{Abdo}, A.~A., {Ackermann}, M., {Ajello}, M., {et~al.} 2010, Science, 329, 817

\bibitem[{{Ackermann} {et~al.}(2014){Ackermann}, {Ajello}, {Albert}, {Baldini},
  {Ballet}, {Barbiellini}, {Bastieri}, {Bellazzini}, {Bissaldi}, {Blandford},
  {Bloom}, {Bottacini}, {Brandt}, {Bregeon}, {Bruel}, {Buehler}, {Buson},
  {Caliandro}, {Cameron}, {Caragiulo}, {Caraveo}, {Cavazzuti}, {Charles},
  {Chekhtman}, {Cheung}, {Chiang}, {Chiaro}, {Ciprini}, {Claus},
  {Cohen-Tanugi}, {Conrad}, {Corbel}, {D'Ammando}, {de Angelis}, {den Hartog},
  {de Palma}, {Dermer}, {Desiante}, {Digel}, {Di Venere}, {do Couto e Silva},
  {Donato}, {Drell}, {Drlica-Wagner}, {Favuzzi}, {Ferrara}, {Focke},
  {Franckowiak}, {Fuhrmann}, {Fukazawa}, {Fusco}, {Gargano}, {Gasparrini},
  {Germani}, {Giglietto}, {Giordano}, {Giroletti}, {Glanzman}, {Godfrey},
  {Grenier}, {Grove}, {Guiriec}, {Hadasch}, {Harding}, {Hayashida}, {Hays},
  {Hewitt}, {Hill}, {Hou}, {Jean}, {Jogler}, {J{\'o}hannesson}, {Johnson},
  {Johnson}, {Kerr}, {Kn{\"o}dlseder}, {Kuss}, {Larsson}, {Latronico},
  {Lemoine-Goumard}, {Longo}, {Loparco}, {Lott}, {Lovellette}, {Lubrano},
  {Manfreda}, {Martin}, {Massaro}, {Mayer}, {Mazziotta}, {McEnery},
  {Michelson}, {Mitthumsiri}, {Mizuno}, {Monzani}, {Morselli}, {Moskalenko},
  {Murgia}, {Nemmen}, {Nuss}, {Ohsugi}, {Omodei}, {Orienti}, {Orlando},
  {Ormes}, {Paneque}, {Panetta}, {Perkins}, {Pesce-Rollins}, {Piron}, {Pivato},
  {Porter}, {Rain{\`o}}, {Rando}, {Razzano}, {Razzaque}, {Reimer}, {Reimer},
  {Reposeur}, {Saz Parkinson}, {Schaal}, {Schulz}, {Sgr{\`o}}, {Siskind},
  {Spandre}, {Spinelli}, {Stawarz}, {Suson}, {Takahashi}, {Tanaka}, {Thayer},
  {Thayer}, {Thompson}, {Tibaldo}, {Tinivella}, {Torres}, {Tosti}, {Troja},
  {Uchiyama}, {Vianello}, {Winer}, {Wolff}, {Wood}, {Wood}, {Wood},
  {Charbonnel}, {Corbet}, {De Gennaro Aquino}, {Edlin}, {Mason}, {Schwarz},
  {Shore}, {Starrfield}, {Teyssier}, \& {Fermi-LAT
  Collaboration}}]{Ackermann:2014a}
{Ackermann}, M., {Ajello}, M., {Albert}, A., {et~al.} 2014, Science, 345, 554

\bibitem[{{Ahnen} {et~al.}(2015){Ahnen}, {Ansoldi}, {Antonelli}, {Antoranz},
  {Babic}, {Banerjee}, {Bangale}, {Barres de Almeida}, {Barrio}, {Becerra
  Gonz{\'a}lez}, {Bednarek}, {Bernardini}, {Biasuzzi}, {Biland}, {Blanch},
  {Bonnefoy}, {Bonnoli}, {Borracci}, {Bretz}, {Carmona}, {Carosi},
  {Chatterjee}, {Clavero}, {Colin}, {Colombo}, {Contreras}, {Cortina},
  {Covino}, {Da Vela}, {Dazzi}, {De Angelis}, {De Caneva}, {De Lotto}, {de
  O{\~n}a Wilhelmi}, {Delgado Mendez}, {Di Pierro}, {Dominis Prester},
  {Dorner}, {Doro}, {Einecke}, {Eisenacher Glawion}, {Elsaesser},
  {Fern{\'a}ndez-Barral}, {Fidalgo}, {Fonseca}, {Font}, {Frantzen}, {Fruck},
  {Galindo}, {Garc{\'{\i}}a L{\'o}pez}, {Garczarczyk}, {Garrido Terrats},
  {Gaug}, {Giammaria}, {Godinovi{\'c}}, {Gonz{\'a}lez Mu{\~n}oz}, {Guberman},
  {Hanabata}, {Hayashida}, {Herrera}, {Hose}, {Hrupec}, {Hughes}, {Idec},
  {Kellermann}, {Kodani}, {Konno}, {Kubo}, {Kushida}, {La Barbera}, {Lelas},
  {Lewandowska}, {Lindfors}, {Lombardi}, {Longo}, {L{\'o}pez},
  {L{\'o}pez-Coto}, {L{\'o}pez-Oramas}, {Lorenz}, {Majumdar}, {Makariev},
  {Mallot}, {Maneva}, {Manganaro}, {Mannheim}, {Maraschi}, {Marcote},
  {Mariotti}, {Mart{\'{\i}}nez}, {Mazin}, {Menzel}, {Miranda}, {Mirzoyan},
  {Moralejo}, {Nakajima}, {Neustroev}, {Niedzwiecki}, {Nievas Rosillo},
  {Nilsson}, {Nishijima}, {Noda}, {Orito}, {Overkemping}, {Paiano}, {Palacio},
  {Palatiello}, {Paneque}, {Paoletti}, {Paredes}, {Paredes-Fortuny}, {Persic},
  {Poutanen}, {Prada Moroni}, {Prandini}, {Puljak}, {Reinthal}, {Rhode},
  {Rib{\'o}}, {Rico}, {Rodriguez Garcia}, {Saito}, {Saito}, {Satalecka},
  {Scapin}, {Schultz}, {Schweizer}, {Sillanp{\"a}{\"a}}, {Sitarek}, {Snidaric},
  {Sobczynska}, {Stamerra}, {Steinbring}, {Strzys}, {Takalo}, {Takami},
  {Tavecchio}, {Temnikov}, {Terzi{\'c}}, {Tescaro}, {Teshima}, {Thaele},
  {Torres}, {Toyama}, {Treves}, {Verguilov}, {Vovk}, {Will}, {Zanin},
  {Desiante}, \& {Hays}}]{Ahnen:2015a}
{Ahnen}, M.~L., {Ansoldi}, S., {Antonelli}, L.~A., {et~al.} 2015, \aap, 582,
  A67

\bibitem[{{Aliu} {et~al.}(2012){Aliu}, {Archambault}, {Arlen}, {Aune},
  {Beilicke}, {Benbow}, {Bouvier}, {Bradbury}, {Buckley}, {Bugaev}, {Byrum},
  {Cannon}, {Cesarini}, {Ciupik}, {Collins-Hughes}, {Connolly}, {Cui},
  {Decerprit}, {Dickherber}, {Duke}, {Dumm}, {Dwarkadas}, {Errando}, {Falcone},
  {Feng}, {Finley}, {Finnegan}, {Fortson}, {Furniss}, {Galante}, {Gall},
  {Godambe}, {Griffin}, {Grube}, {Gyuk}, {Hanna}, {Holder}, {Huan}, {Hughes},
  {Humensky}, {Kaaret}, {Karlsson}, {Kertzman}, {Khassen}, {Kieda},
  {Krawczynski}, {Krennrich}, {Lang}, {Lee}, {Maier}, {Majumdar}, {McArthur},
  {McCann}, {Millis}, {Moriarty}, {Mukherjee}, {Nu{\~n}ez}, {Ong}, {Orr},
  {Otte}, {Pandel}, {Park}, {Perkins}, {Pohl}, {Prokoph}, {Quinn}, {Ragan},
  {Reyes}, {Reynolds}, {Roache}, {Rose}, {Ruppel}, {Saxon}, {Schroedter},
  {Sembroski}, {Skole}, {Smith}, {Staszak}, {Telezhinsky}, {Te{\v s}i{\'c}},
  {Theiling}, {Thibadeau}, {Tsurusaki}, {Tyler}, {Varlotta}, {Vincent},
  {Vivier}, {Wakely}, {Ward}, {Weekes}, {Weinstein}, {Weisgarber}, {Welsing},
  {Williams}, \& {Zitzer}}]{Aliu:2012a}
{Aliu}, E., {Archambault}, S., {Arlen}, T., {et~al.} 2012, \apj, 754, 77

\bibitem[{{Banerjee} {et~al.}(2016){Banerjee}, {Srivastava}, {Ashok}, \&
  {Venkataraman}}]{Banerjee:2016a}
{Banerjee}, D.~P.~K., {Srivastava}, M.~K., {Ashok}, N.~M., \& {Venkataraman},
  V. 2016, \mnras, 455, L109

\bibitem[{{Bode} \& {Evans}(2008)}]{Bode:2008a}
{Bode}, M.~F. \& {Evans}, A. 2008, {Classical Novae}

\bibitem[{{Caprioli} \& {Spitkovsky}(2014)}]{Caprioli:2014c}
{Caprioli}, D. \& {Spitkovsky}, A. 2014, \apj, 794, 47

\bibitem[{{Cassatella} {et~al.}(2004){Cassatella}, {Lamers}, {Rossi},
  {Altamore}, \& {Gonz{\'a}lez-Riestra}}]{Cassatella:2004a}
{Cassatella}, A., {Lamers}, H.~J.~G.~L.~M., {Rossi}, C., {Altamore}, A., \&
  {Gonz{\'a}lez-Riestra}, R. 2004, \aap, 420, 571

\bibitem[{{Castor} {et~al.}(1975){Castor}, {McCray}, \&
  {Weaver}}]{Castor:1975a}
{Castor}, J., {McCray}, R., \& {Weaver}, R. 1975, \apjl, 200, L107

\bibitem[{{Cheung} {et~al.}(2015){Cheung}, {Jean}, {Fermi Large Area Telescope
  Collaboration}, \& {Shore}}]{Cheung:2015a}
{Cheung}, C.~C., {Jean}, P., {Fermi Large Area Telescope Collaboration}, \&
  {Shore}, S.~N. 2015, The Astronomer's Telegram, 7283

\bibitem[{{Cheung} {et~al.}(2016{\natexlab{a}}){Cheung}, {Jean}, {Shore}, \&
  {Fermi Large Area Telescope Collaboration}}]{Cheung:2016b}
{Cheung}, C.~C., {Jean}, P., {Shore}, S.~N., \& {Fermi Large Area Telescope
  Collaboration}. 2016{\natexlab{a}}, The Astronomer's Telegram, 9594

\bibitem[{{Cheung} {et~al.}(2016{\natexlab{b}}){Cheung}, {Jean}, {Shore},
  {Stawarz}, {Corbet}, {Kn{\"o}dlseder}, {Starrfield}, {Wood}, {Desiante},
  {Longo}, {Pivato}, \& {Wood}}]{Cheung:2016a}
{Cheung}, C.~C., {Jean}, P., {Shore}, S.~N., {et~al.} 2016{\natexlab{b}}, \apj,
  826, 142

\bibitem[{{Chomiuk} {et~al.}(2012){Chomiuk}, {Krauss}, {Rupen}, {Nelson},
  {Roy}, {Sokoloski}, {Mukai}, {Munari}, {Mioduszewski}, {Weston}, {O'Brien},
  {Eyres}, \& {Bode}}]{Chomiuk:2012a}
{Chomiuk}, L., {Krauss}, M.~I., {Rupen}, M.~P., {et~al.} 2012, \apj, 761, 173

\bibitem[{{Chomiuk} {et~al.}(2014){Chomiuk}, {Linford}, {Yang}, {O'Brien},
  {Paragi}, {Mioduszewski}, {Beswick}, {Cheung}, {Mukai}, {Nelson}, {Ribeiro},
  {Rupen}, {Sokoloski}, {Weston}, {Zheng}, {Bode}, {Eyres}, {Roy}, \&
  {Taylor}}]{Chomiuk:2014a}
{Chomiuk}, L., {Linford}, J.~D., {Yang}, J., {et~al.} 2014, \nat, 514, 339

\bibitem[{{Cunningham} {et~al.}(2015){Cunningham}, {Wolf}, \&
  {Bildsten}}]{Cunningham:2015a}
{Cunningham}, T., {Wolf}, W.~M., \& {Bildsten}, L. 2015, \apj, 803, 76

\bibitem[{{Finzell} {et~al.}(2017){Finzell}, {Chomiuk}, {Metzger}, {Walter},
  {Linford}, {Mukai}, {Nelson}, {Weston}, {Zheng}, {Sokoloski}, {Mioduszewski},
  {Rupen}, {Dong}, {Bohlsen}, {Buil}, {Prieto}, {Wagner}, {Bensby}, {Bond},
  {Sumi}, {Bennett}, {Abe}, {Koshimoto}, {Suzuki}, {P.}, {Tristram},
  {Christie}, {Natusch}, {McCormick}, {Yee}, \& {Gould}}]{Finzell:2017a}
{Finzell}, T., {Chomiuk}, L., {Metzger}, B.~D., {et~al.} 2017, ArXiv e-prints
  [\eprint[arXiv]{1701.03094}]

\bibitem[{{Finzell} {et~al.}(2015){Finzell}, {Chomiuk}, {Munari}, \&
  {Walter}}]{Finzell:2015a}
{Finzell}, T., {Chomiuk}, L., {Munari}, U., \& {Walter}, F.~M. 2015, \apj, 809,
  160

\bibitem[{{Franckowiak} {et~al.}(2017){Franckowiak}, {Jean}, {Wood}, {Cheung},
  \& C.C.}]{Franckowiak:2017a}
{Franckowiak}, A., {Jean}, P., {Wood}, M., {Cheung}, \& C.C., {Buson}, S. 2017,
  submitted to \aap

\bibitem[{{Gorbatskii}(1962)}]{Gorbatskii:1962a}
{Gorbatskii}, V.~G. 1962, \azh, 39, 198

\bibitem[{{Kato} \& {Hachisu}(1994)}]{Kato:1994a}
{Kato}, M. \& {Hachisu}, I. 1994, \apj, 437, 802

\bibitem[{{Li} \& {Chomiuk}(2016)}]{Li:2016b}
{Li}, K.-L. \& {Chomiuk}, L. 2016, The Astronomer's Telegram, 9699

\bibitem[{{Li} {et~al.}(2016){Li}, {Chomiuk}, \& {Strader}}]{Li:2016a}
{Li}, K.-L., {Chomiuk}, L., \& {Strader}, J. 2016, The Astronomer's Telegram,
  9736

\bibitem[{{Li} {et~al.}(2017){Li}, {Metzger}, {Chomiuk}, {Vurm}, {Strader},
  {Finzell}, {Beloborodov}, {Nelson}, {Shappee}, {Kochanek}, {Prieto}, {Kafka},
  {Holoien}, {Thompson}, {Luckas}, \& {Itoh}}]{Li:2017a}
{Li}, K.-L., {Metzger}, B.~D., {Chomiuk}, L., {et~al.} 2017, Accepted for
  publication in Nature Astronomy [\eprint[arXiv]{1709.00763}]

\bibitem[{{Linford} {et~al.}(2015){Linford}, {Ribeiro}, {Chomiuk}, {Nelson},
  {Sokoloski}, {Rupen}, {Mukai}, {O'Brien}, {Mioduszewski}, \&
  {Weston}}]{Linford:2015a}
{Linford}, J.~D., {Ribeiro}, V.~A.~R.~M., {Chomiuk}, L., {et~al.} 2015, \apj,
  805, 136

\bibitem[{{Martin} \& {Dubus}(2013)}]{Martin:2013a}
{Martin}, P. \& {Dubus}, G. 2013, \aap, 551, A37

\bibitem[{{Metzger} {et~al.}(2016){Metzger}, {Caprioli}, {Vurm}, {Beloborodov},
  {Bartos}, \& {Vlasov}}]{Metzger:2016a}
{Metzger}, B.~D., {Caprioli}, D., {Vurm}, I., {et~al.} 2016, \mnras, 457, 1786

\bibitem[{{Metzger} {et~al.}(2015){Metzger}, {Finzell}, {Vurm}, {Hasco{\"e}t},
  {Beloborodov}, \& {Chomiuk}}]{Metzger:2015a}
{Metzger}, B.~D., {Finzell}, T., {Vurm}, I., {et~al.} 2015, \mnras, 450, 2739

\bibitem[{{Metzger} {et~al.}(2014){Metzger}, {Hasco{\"e}t}, {Vurm},
  {Beloborodov}, {Chomiuk}, {Sokoloski}, \& {Nelson}}]{Metzger:2014a}
{Metzger}, B.~D., {Hasco{\"e}t}, R., {Vurm}, I., {et~al.} 2014, \mnras, 442,
  713

\bibitem[{{Mignone} {et~al.}(2012){Mignone}, {Zanni}, {Tzeferacos}, {van
  Straalen}, {Colella}, \& {Bodo}}]{Mignone:2012a}
{Mignone}, A., {Zanni}, C., {Tzeferacos}, P., {et~al.} 2012, \apjs, 198, 7

\bibitem[{{Munari} {et~al.}(1990){Munari}, {Margoni}, \&
  {Stagni}}]{Munari:1990a}
{Munari}, U., {Margoni}, R., \& {Stagni}, R. 1990, \mnras, 242, 653

\bibitem[{{Prialnik}(1986)}]{Prialnik:1986a}
{Prialnik}, D. 1986, \apj, 310, 222

\bibitem[{{Quataert} {et~al.}(2016){Quataert}, {Fern{\'a}ndez}, {Kasen},
  {Klion}, \& {Paxton}}]{Quataert:2016a}
{Quataert}, E., {Fern{\'a}ndez}, R., {Kasen}, D., {Klion}, H., \& {Paxton}, B.
  2016, \mnras, 458, 1214

\bibitem[{{Rybicki} \& {Lightman}(1986)}]{Rybicki:1986}
{Rybicki}, G.~B. \& {Lightman}, A.~P. 1986, {Radiative Processes in
  Astrophysics} ({Wiley-VCH})

\bibitem[{{Schaefer} {et~al.}(2014){Schaefer}, {Brummelaar}, {Gies},
  {Farrington}, {Kloppenborg}, {Chesneau}, {Monnier}, {Ridgway}, {Scott},
  {Tallon-Bosc}, {McAlister}, {Boyajian}, {Maestro}, {Mourard}, {Meilland},
  {Nardetto}, {Stee}, {Sturmann}, {Vargas}, {Baron}, {Ireland}, {Baines},
  {Che}, {Jones}, {Richardson}, {Roettenbacher}, {Sturmann}, {Turner},
  {Tuthill}, {van Belle}, {von Braun}, {Zavala}, {Banerjee}, {Ashok}, {Joshi},
  {Becker}, \& {Muirhead}}]{Schaefer:2014a}
{Schaefer}, G.~H., {Brummelaar}, T.~T., {Gies}, D.~R., {et~al.} 2014, \nat,
  515, 234

\bibitem[{{Schlickeiser}(2002)}]{Schlickeiser:2002}
{Schlickeiser}, R. 2002, {Cosmic Ray Astrophysics}, ed. {Schlickeiser, R.}
  (Springer)

\bibitem[{{Schure} {et~al.}(2009){Schure}, {Kosenko}, {Kaastra}, {Keppens}, \&
  {Vink}}]{Schure:2009a}
{Schure}, K.~M., {Kosenko}, D., {Kaastra}, J.~S., {Keppens}, R., \& {Vink}, J.
  2009, \aap, 508, 751

\bibitem[{{Shore} {et~al.}(2014){Shore}, {Schwarz}, {Walter}, {Crawford},
  {Woudt}, {Williams}, {Mason}, {Izzo}, {Page}, {Osborne}, {Ness},
  {Starrfield}, {Woodward}, {Vaisanen}, {Marang}, {Crause}, \&
  {Tyas}}]{Shore:2014a}
{Shore}, S.~N., {Schwarz}, G.~J., {Walter}, F.~M., {et~al.} 2014, The
  Astronomer's Telegram, 6413

\bibitem[{{Skopal} {et~al.}(2014){Skopal}, {Drechsel}, {Tarasova}, {Kato},
  {Fujii}, {Teyssier}, {Garde}, {Guarro}, {Edlin}, {Buil}, {Antao}, {Terry},
  {Lemoult}, {Charbonnel}, {Bohlsen}, {Favaro}, \& {Graham}}]{Skopal:2014a}
{Skopal}, A., {Drechsel}, H., {Tarasova}, T., {et~al.} 2014, \aap, 569, A112

\bibitem[{{Vlasov} {et~al.}(2016){Vlasov}, {Vurm}, \& {Metzger}}]{Vlasov:2016a}
{Vlasov}, A., {Vurm}, I., \& {Metzger}, B.~D. 2016, \mnras, 463, 394

\bibitem[{{V{\"o}lk} {et~al.}(2005){V{\"o}lk}, {Berezhko}, \&
  {Ksenofontov}}]{Volk:2005a}
{V{\"o}lk}, H.~J., {Berezhko}, E.~G., \& {Ksenofontov}, L.~T. 2005, \aap, 433,
  229

\bibitem[{{Vurm} \& {Metzger}(2016)}]{Vurm:2016a}
{Vurm}, I. \& {Metzger}, B.~D. 2016, ArXiv e-prints
  [\eprint[arXiv]{1611.04532}]

\bibitem[{{Weston} {et~al.}(2016{\natexlab{a}}){Weston}, {Sokoloski},
  {Chomiuk}, {Linford}, {Nelson}, {Mukai}, {Finzell}, {Mioduszewski}, {Rupen},
  \& {Walter}}]{Weston:2016b}
{Weston}, J.~H.~S., {Sokoloski}, J.~L., {Chomiuk}, L., {et~al.}
  2016{\natexlab{a}}, \mnras, 460, 2687

\bibitem[{{Weston} {et~al.}(2016{\natexlab{b}}){Weston}, {Sokoloski},
  {Metzger}, {Zheng}, {Chomiuk}, {Krauss}, {Linford}, {Nelson}, {Mioduszewski},
  {Rupen}, {Finzell}, \& {Mukai}}]{Weston:2016a}
{Weston}, J.~H.~S., {Sokoloski}, J.~L., {Metzger}, B.~D., {et~al.}
  2016{\natexlab{b}}, \mnras, 457, 887

\end{thebibliography}

\begin{acknowledgement}
We gratefully acknowledge use of the PLUTO code distributed at http://plutocode.ph.unito.it/. The authors acknowledge support from the {\em Programme National Hautes Energies} and from the {\em Centre National d'Etudes Spatiales}. PM and GD thank Brian Metzger for useful discussions on the subject, especially during the ``International Workshop on
Shocks and Particle Acceleration in Novae and Supernovae'' he organized in New York City in 2016.
\end{acknowledgement}

\begin{appendix}

\section{Adiabatic case}

\subsection{Shock propagation}
We detail here the case of an adiabatic interaction of the nova wind with the ejecta. The nova wind adds energy in the shocked wind region at a rate 
\begin{equation}
\fd{E}{t}=\fd{}{t}\left[ \frac{4}{3}\pi r_{\rm s}^3 \frac{P}{(\gamma-1)} \right]=\frac{1}{2}\dot{M}_wv_w^2-4\pi r_{\rm s}^2v_{\rm s}P
\label{pressure}
\end{equation}
where $\gamma$ is the adiabatic index of the gas. The pressure from the shocked wind then drives the shell of swept-up ejecta, assumed to be squeezed in a thin shell between the shocked wind and unshocked ejecta:
\begin{equation}
\fd{}{t}\left[M_{\rm ej,  s}v_{\rm s}\right]-\dot{M}_{\rm ej,s}v=4\pi r_s^2 P
\end{equation}
These equations are similar to those describing the evolution of a wind in an external medium \citep{Castor:1975a}. They allow for a self-similar solution with $P\propto t^\beta$ and $r_{\rm s}\propto t^\alpha$. Equation~\ref{pressure} imposes $\beta=1-3\alpha$. The solution, assuming $n_{\rm ej}\leq 3$, is
\begin{align}
\frac{r_{\rm s}}{r_{\rm ej}}={}&C_{\rm e}^{\frac{1}{n-5}}\left(\frac{t_{\rm w}}{t_{\rm E}}\right)^{\frac{1}{5-n}}\left(\frac{t}{t_{\rm w}}\right)^{\frac{3}{5-n}}~{\rm for}~t\ll t_{\rm w}\\
\frac{r_{\rm s}}{r_{\rm ej}}={}&C_{\rm l}^{\frac{1}{n-5}}\left(\frac{t_{\rm w}}{t_{\rm E}}\right)^{\frac{1}{5-n}}\left(\frac{t}{t_{\rm w}}\right)^{\frac{6-n}{5-n}}~{\rm for}~t\gg t_{\rm w}\label{alregime}
\end{align}
with the constants
\begin{align}
C_{\rm e}={}&\frac{6(7-2n)}{(5-n)^2} \left(\frac{1}{3(\gamma-1)}+\frac{3}{5-n}\right)\\
C_{\rm l} ={}&\frac{2(9-2n)}{(5-n)^2}\left(\frac{1}{3(\gamma-1)}+\frac{6-n}{5-n}\right)
\end{align}
and with
\begin{equation}
t_{\rm E}=\frac{M_{\rm ej}v_{\rm ej}^2}{\dot{M}_{\rm w} v_{\rm w}^2}=7\,\left(\frac{M_{\rm ej}}{10^{-5}\rm\, M_\odot}\right)\left(\frac{10^{-5}\rm\,M_\odot\,week^{-1}}{\dot{M}_{\rm w}}\right)\left(\frac{v_{\rm ej}}{v_{\rm w}}\right)^2\,{\rm days}
\end{equation}
the timescale over which the integrated kinetic power of the nova wind matches the kinetic energy of the ejecta. We focus on the limiting case $t\gg t_{\rm w}$ (Eq.~\ref{alregime}) as the more relevant one since we identify $t_{\rm w}$ with the delay between the initial ejection of material and the onset of the wind outflow. The shock has propagated throughout the ejecta when $r_{\rm s}=v_{\rm ej} t$, hence the time for the shock to catchup with the leading edge of the ejecta is
\begin{equation}
t_{\rm catch}\approx C_{\rm l}\, t_{\rm E}
\end{equation}
showing that the nova wind propagates adiabatically through the ejecta on the energy injection timescale $t_{\rm E}$. The constant $C_{\rm l}$ is $\approx 1.22$ for $n_{\rm ej}=0$ and $C_{\rm l}\approx 2.04$ for $n_{\rm ej}=2$ (with $\gamma=5/3$).

\subsection{Radiative losses}
Using  Eq.~\ref{alregime} to calculate $v_{\rm s}$, the Mach number of the shock for $t\gg t_{\rm w}$ is roughly
\begin{equation}
{\cal M}_{\rm fs}= \frac{v_{\rm s}-v}{c_s}\approx \left(\frac{v_{\rm ej}}{c_s}\right)\left(\frac{C_l^{\frac{1}{n-5}}}{5-n}\right) \left[\frac{t_{\rm energy}}{t_{\rm w}}\right]^{\frac{1}{n-5}} \left(\frac{t}{t_{\rm w}}\right)^{\frac{1}{5-n}}
\label{mach}
\end{equation}
where $c_s \ll v_{\rm ej}$ is the sound speed in the ejecta material. The shock is strong so the temperature of the shocked ejecta is 
\begin{equation}
kT=\frac{3}{16}\mu m_p (v_{\rm s}-v)^2=\frac{3}{16}\mu m_p c_s^2 {\cal M}_{\rm fs}^2
\label{temp}
\end{equation}
where $\mu$ is the mean molecular weight ($\mu=0.5$ for ionised hydrogen). The cooling timescale of the shocked ejecta is
\begin{equation}
\tau_{\rm cool} = \frac{(3/2) n k T}{4 n^2\Lambda} \approx \frac{3 k T}{8 n\Lambda_0 T^{1/2}} 
\end{equation}
with $n=\rho/(\mu m_p)$ the number density. $\Lambda=\Lambda_0 T^{1/2}= 2\times 10^{-27} T^{-27}$ is the cooling function assuming optically thin thermal bremsstrahlung emission. Using Eq.~\ref{alregime},\ref{mach}-\ref{temp} and the expression for $\rho$ (Eq.~\ref{rhoej}), we find that $\tau_{\rm cool}\leq t$ as long as 
\begin{equation}
t \leq  \left\{\frac{8 \Lambda_0(5-n)(3-n)}{\pi k^{1/2}(3\mu m_p)^{3/2}} \frac{M_{\rm ej}}{ v^4_{\rm ej}}\right\}^{\frac{5-n}{11-n}}  t_{\rm catch}^{\frac{n+1}{11-n}}
\end{equation}
which is about 156 days (resp. 42 days) for $n_{\rm ej}=0$ (resp. $n_{\rm ej}=2$), $t_{\rm catch}=7\rm\,days$, $M_{\rm ej}=10^{-5}\rm\,M_\odot$ and $v_{\rm ej}=1000\rm\,km\,s^{-1}$. Hence, the cooling timescale is much shorter than the ejecta crossing timescale. The shock into the ejecta is isothermal rather than adiabatic, provided that the radiation escapes.

Comparable conclusions are reached on the reverse shock into the nova wind. Approximating the Mach number of the reverse shock as
\begin{equation}
{\cal M}_{\rm rs}= \frac{v_{\rm w}-v_{\rm s}}{c_s}\approx \frac{v_w}{c_s}
\end{equation}
we find that $\tau_{\rm cool}\leq t$ for the reverse shock as long as 
\begin{equation}
t \leq  \left\{\frac{8 \Lambda_0(5-n)(3-n)}{\pi k^{1/2}(3\mu m_p)^{3/2}} \frac{\dot{M}_{\rm w}}{ v_{\rm w}^2 v^2_{\rm ej}}\right\}^{\frac{5-n}{7-n}}  t_{\rm catch}^{\frac{2}{7-n}}
\end{equation}
which is about 48 days (resp. 35 days) for $n_{\rm ej}=0$ (resp. $n_{\rm ej}=2$), $t_{\rm catch}=7\rm\,days$, $\dot{M}_{\rm w}=10^{-5}\rm\,M_\odot\,week^{-1}$, $v_{\rm w}=2000\rm\,km\,s^{-1}$ and $v_{\rm ej}=1000\rm\,km\,s^{-1}$. Again, the cooling timescale is shorter than the ejecta crossing timescale.

\section{Isothermal case}
We list here some analytical solutions of use in verifying the numerical integration of the equation of motion (Eq.~\ref{dynamics}).

\subsection{Analytical solution for $M_{\rm ej,s}\gg M_{\rm w,s}$\label{analyticalISO}}
If the mass of the swept-up nova wind material is negligible compared to the swept-up mass of ejecta material then Eq.~\ref{dynamics} simplifies to
\begin{equation}
\fd{}{t}\left[M_{\rm ej, s}v_{\rm s}\right]=\dot{M}_{\rm ej,s} v+\dot{M}_{\rm w,s}v_{\rm w}
\label{simpler}
\end{equation}
However, note that the assumption that $M_{\rm ej,s}\gg M_{\rm w,s}$ necessarily fails when $t\ga t_{\rm wind}$. As for the adiabatic case (Appendix A), power-law solutions of the form $r_{\rm s}\propto t^\alpha$ can be found for Eq.~\ref{simpler} provided that $v_{\rm w}\gg v_{\rm ej}$ and $n_{\rm ej}\leq2$
\begin{align}
\frac{r_{\rm s}}{r_{\rm ej}}={}&\left(\frac{4-n}{2}\right)^{\frac{1}{4-n}}\left(\frac{t_{\rm w}}{t_{\rm M}}\right)^{\frac{1}{4-n}}\left(\frac{t}{t_{\rm w}}\right)^{\frac{2}{4-n}}~{\rm for}~t\ll t_{\rm w}\\
\frac{r_{\rm s}}{r_{\rm ej}}={}&\left(\frac{4-n}{2}\right)^{\frac{1}{4-n}}\left(\frac{t_{\rm w}}{t_{\rm M}}\right)^{\frac{1}{4-n}}\left(\frac{t}{t_{\rm w}}\right)^{\frac{5-n}{4-n}}~{\rm for}~t\gg t_{\rm w}\label{lregime}
\end{align}
where $t_M$ defined in Eq.~\ref{tM}. Taking the second case (Eq.~\ref{lregime}), the time taken by the shock to cross all the ejecta is
\begin{equation}
t_{\rm catch}\approx \left(\frac{2}{4-n}\right)t_{\rm M}
\end{equation}
from which $r_{\rm catch}\approx \left(\tfrac{2}{4-n}\right) v_{\rm ej}t_{\rm M}$ and $v_{\rm catch}\approx \tfrac{5-n}{4-n}v_{\rm ej}$. In the isothermal case, the nova wind propagates through the ejecta on the momentum injection timescale $t_{\rm M}$.

\subsection{Analytical solution for $M_{\rm ej,s}=M_{\rm ej}$}
If the ejecta has been completely swept-up in the shell, that is when $t>t_{\rm catch}$, then Eq.~\ref{dynamics} simplifies to 
\begin{equation}
\fd{}{t}\left[\left(\dot{M}_{\rm w} \left(t - \frac{r_{\rm s}}{v_{\rm w}}\right)+M_{\rm ej}\right)v_{\rm s}\right]-\dot{M}_{\rm w} v_{\rm w} \left(1-\frac{v_{\rm s}}{v_{\rm w}}\right)=0
\end{equation}
This equation is integrated by changing variable to $u=\fd{}{t} \left[t-r_{\rm s}/v_{\rm w}\right]$, giving the result:
\begin{align}
r_{\rm s}={}&v_{\rm w} (t+t_{\rm W})\nonumber\\
&-v_{\rm w}\left(t_{\rm W}+t_{\rm catch}-\frac{r_{\rm catch}}{v_{\rm w}}\right)\left[1+\frac{2(t-t_{\rm catch})\left(1-\frac{v_{\rm catch}}{v_{\rm w}}\right)}{\left(t_{\rm W}+t_{\rm catch}-\frac{r_{\rm catch}}{v_{\rm w}}\right)}\right]^{1/2}
\end{align}
where $r_{\rm catch}$ and $v_{\rm catch}$ are the radius and the velocity of the shell at $t=t_{\rm catch}$, and $t_{\rm W}=M_{\rm ej}/\dot{M}_{\rm w}$. \citet{Gorbatskii:1962a} found a similar expression but for different initial conditions. 

\end{appendix}
\end{document}